\documentclass{amsart}
\usepackage{amssymb}
\usepackage{mathrsfs}
\usepackage{centernot}
\usepackage{MnSymbol}
\usepackage{color}
\usepackage{hyperref}

\usepackage{graphicx}
\usepackage{tikz}
\usetikzlibrary{positioning}
\usetikzlibrary{patterns}

\usepackage{hyperref}

\def\bC {\mathbf{C}}

\def\bF {\mathbf{F}}

\def\bH {\mathbf{H}}

\def\bN {\mathbf{N}}

\def\bR {\mathbf{R}}

\def\fD {\mathfrak{D}}
\def\fd {\mathfrak{d}}
\def\fH {\mathfrak{H}}
\def\fk {\mathfrak{k}}
\def\fK {\mathfrak{K}}
\def\fS {\mathfrak{S}}

\def\cA {\mathcal{A}}

\def\cC {\mathcal{C}}
\def\cD {\mathcal{D}}
\def\cE {\mathcal{E}}

\def\cH {\mathcal{H}}

\def\cK {\mathcal{K}}
\def\cL {\mathcal{L}}

\def\cN {\mathcal{N}}
\def\cO {\mathcal{O}}
\def\cP {\mathcal{P}}
\def\cQ {\mathcal{Q}}

\def\cS {\mathcal{S}}
\def\cT {\mathcal{T}}
\def\cU {\mathcal{U}}
\def\cV {\mathcal{V}}
\def\cW {\mathcal{W}}

\def\a {{\alpha}}
\def\b {{\beta}}
\def\g {{\gamma}}
\def\Ga {{\Gamma}}
\def\de {{\delta}}
\def\eps {{\epsilon}}
\def\th {{\theta}}

\def\lbd {{\lambda}}
\def\L {{\Lambda}}
\def\si {{\sigma}}

\def\om {{\omega}}
\def\Om {{\Omega}}

\def\d {{\partial}}
\def\grad {{\nabla}}
\def\Dlt {{\Delta}}

\def\rstr {{\big |}}
\def\indc {{\bf 1}}

\def\la {\langle}
\def\ra {\rangle}

\def\bu {{\bullet}}

\newcommand{\Div}{\operatorname{div}}
\newcommand{\Grad}{\operatorname{grad}}

\newcommand{\Dom}{\operatorname{Dom}}

\newcommand{\Supp}{\operatorname{supp}}

\newcommand{\Tr}{\operatorname{trace}}

\newcommand{\Dist}{\operatorname{dist}}

\newcommand{\Ker}{\operatorname{Ker}}
\newcommand{\IM}{\operatorname{Ran}}

\newcommand{\Lip}{\operatorname{Lip}}

\def\hb {{\hbar}}

\newcommand{\ba}{\begin{aligned}}
\newcommand{\ea}{\end{aligned}}

\newcommand{\be}{\begin{equation}}
\newcommand{\ee}{\end{equation}}

\newcommand{\lb}{\label}

\def\bu {$\bullet$}

\begin{document}

\title[QOT and Many-Body Problems]{Quantum Optimal Transport:\\ Quantum Couplings and Many-Body Problems}

\author[F. Golse]{Fran\c cois Golse}
\address[F.G.]{CMLS, \'Ecole polytechnique, IP Paris, 91128 Palaiseau Cedex, France}
\email{francois.golse@polytechnique.edu}

\date{\today}

\begin{abstract}
This text is a set of lecture notes for a 4.5-hour course given at the Erd\H os Center (R\'enyi Institute, Budapest) during the Summer School ``Optimal Transport on Quantum Structures''
(September 19th-23rd, 2023). Lecture I introduces the quantum analogue of the Wasserstein distance of exponent $2$ defined in [F. Golse, C. Mouhot, T. Paul: Comm. Math. Phys.
\textbf{343} (2016), 165--205], and in [F. Golse, T. Paul: Arch. Ration. Mech. Anal. \textbf{223} (2017) 57--94].  Lecture II discusses various applications of this quantum analogue of
the Wasserstein distance of exponent $2$, while Lecture III discusses several of its most important properties, such as the triangle inequality, and the Kantorovich duality in the quantum 
setting, together with some of their implications.
\end{abstract}



\maketitle


\section*{Introduction}


Optimal transport has become a thriving field of mathematical research in the last decade of the 20th century. Nowadays, it is used in a variety of subjects
which the founder of this theory (G. Monge, 1781) could obviously not have foreseen: statistics, machine learning, probability theory, fluid mechanics, besides 
more classical applications such as the calculus of variations, partial differential equations, geometry. In his book \cite{VillaniAMS}, C. Villani speaks of a
``revival'' of optimal transport following Y. Brenier's remarkable paper \cite{Brenier}.

Among the many applications of optimal transport is a quite fascinating observation by Dobrushin \cite{Dobru}, who realized that one special kind of approximation
used in classical nonequilibrium statistical mechanics, namely the time-dependent mean-field limit, can be proved rigorously by using a notion of distance on 
the set of probability measures which is defined in terms of optimal transport.

The same mean-field limit is routinely used in quantum dynamics, but, until relatively recently, the methods of proof used in the rigorous justification of this
limit were radically different in the quantum and in the classical settings. This state of affairs was slightly disturbing, since one would expect that it should be
possible to take the classical limit of quantum mechanics and the mean-field limit of both classical and quantum dynamics in any order --- in other words,
one expects that both limits should be represented by a commutative diagram.

There have been some attempts in that direction at the beginning of the 2000s \cite{GraffiMartPulvi,PulviPezzo}, and, slightly later, this suggested the natural
idea of ``lifting'' Dobrushin's approach \cite{Dobru} to the quantum setting. Since Dobrushin used an optimal transport metric to compare $N$-particle densities
and their mean-field limits, it became a very natural motivation for defining an analogous optimal transport ``metric'' for the purpose of comparing quantum states.
Ideally, this ``metric'' should converge to the metric used by Dobrushin in the classical limit. 

I started working on this problem with C. Mouhot and T. Paul, after a visit to C. Mouhot in Cambridge in September 2014. At this point, we realized that mean-field
quantum dynamics satisfies an analogous continuous dependence on the initial data defined in terms of a quantum optimal transport problem as in Dobrushin's
analysis. Our initial contribution \cite{FGMouPaul} was followed by a series of works involving other collaborations \cite{FGPaulARMA,CaglioFGPaul,
CagliotiFGPaulSNS,FGJinPaul,FGPaulJMPA,FGPaulJFA,FGPaulM3AS}, and exploring both our quantum analogue of optimal transport and its applications to 
various problems in quantum dynamics. It is precisely this approach to quantum optimal transport which is presented in these lectures.

But there are other, possibly (most likely?) unrelated approaches to quantum optimal transport. 

One such approach, due to E. Carlen and J. Maas is based on the Benamou-Brenier formula (Theorem 8.1 in \cite{VillaniAMS}): see \cite{CarlenMaas}. Another
approach, based on the notion of quantum channel and closer in spirit to the one presented in this course, yet different, is due to G. De Palma and D. Trevisan
\cite{DePalmaTrev} --- see also Dario Trevisan's course \cite{DT} in this volume. Some of these different approaches to a theory of quantum optimal transport 
(see in particular Eric Carlen's lectures \cite{EC}) will be presented in this school. Still another approach to the problem of quantum optimal transport is the very 
early reference \cite{ZycSlo} (see also the beautiful book \cite{BZ} by I. Bengtsson and K. \.Zyczkowski, and especially section 7.7 there).

Since this text is a set of lecture notes, several quizzes/exercises are proposed to the reader. Some of these exercises review classical material which the reader
is expected to master before going further; some others discuss natural extensions of the material presented in this course. In any case, solving these exercises
is strongly recommended in order to gain familiarity with the notions presented in these notes.

I am indebted to several colleagues for the mathematics discussed in these lectures, in the first place to C. Mouhot, T. Paul, E. Caglioti, and S. Jin, with whom
I had the pleasure to work on various problems related to quantum optimal transport. The observation inequality for the Schr\"odinger equation discussed in
Lecture II comes from questions posed by C. Bardos. I owe my first acquaintance with Dobrushin's remarkable paper \cite{Dobru} to M. Pulvirenti, who gave
a most lucid account of it in a lecture at \'Ecole normale sup\'erieure in 1997. I also benefited from numerous discussions on optimal transport with Y. Brenier.
Most of these lectures are based on the numerous analogies between classical and quantum optimal transport. However, some very fundamental properties 
of classical optimal transport may fail to have quantum analogues. I am very grateful to D. Serre, who kindly showed me an example where the quantum and
the classical theory significantly differ (see Quiz 31 in Lecture III).

Finally, I wish to express my gratitude to the organizers of this 2022 Summer School on Optimal Transport on Quantum Structures at the Erd\H os Center,
J. Maas, S. Rademacher, T. Titkos and D. Virosztek for their kind invitation, and more generally to the R\'enyi Institute for its most enjoyable hospitality.

\bigskip
\newpage

\centerline{\sc Table of Contents}

\bigskip
\noindent
\textbf{Lecture I:} Extending the Wasserstein distance of exponent $2$ to density operators.

\noindent
\textbf{Lecture II:} Applying the quantum Wasserstein pseudometric to particle dynamics.

\noindent
\textbf{Lecture III:} Triangle inequalities and optimal transport in the quantum setting.


\section{Lecture I: Extending the Wasserstein Distance of Exponent $2$\\ to Density Operators}


Our general purpose is to extend optimal transport (Wasserstein) distances, defined on Borel probability measures on phase space, i.e. $\bR^d\times\bR^d$, to their quantum 
analogue, i.e. to density operators on the Hilbert space $L^2(\bR^d)$.

\bigskip
In this first lecture, we

\smallskip
\noindent
\bu recall some fundamental results on classical optimal transport (section \ref{SS-ClassOT}),

\noindent
\bu recall some material on trace-class and Hilbert-Schmidt operators (section \ref{SS-DensOp}),

\noindent
\bu introduce one first noncommutative extension of optimal transport (section \ref{SS-ConnesDist}),

\noindent
\bu present our quantum extension of the Wasserstein metric $\cW_2$ (section \ref{SS-QuantW2}), and

\noindent
\bu discuss some basic estimates and examples of computations (section \ref{SS-QOTCheap}).

\subsection{A Crash-Course on Classical Optimal Transport}\lb{SS-ClassOT}


Before embarking on a description of a quantum analogue of the Wasserstein distance of exponent $2$ for density operators, we need to recall some fundamental notions
and results pertaining to the the classical theory of optimal transport. There are many excellent reference textbooks on optimal transport, such as \cite{VillaniAMS, AmbrosioGS, 
VillaniOTON, Santambrogio, FigalliGlaudo}, where the interested reader will find the proofs of all the statements in this section --- together with many fascinating applications
of optimal transport in various areas of mathematics. Of course, Alessio Figalli's course \cite{AF} in this school is strongly recommended as a general introduction to optimal 
transport.

\subsubsection{The Monge and the Kantorovich Problems}

Optimal transport grew from Monge's celebrated ``M\'emoire\footnote{``D\'eblai'' and ``remblai'' are technical terms for earthwork in French. ``D\'eblai'' means excavation,
whereas ``remblai'' is the French word for embankment or backfill. A first version of Monge's memoir was read on February 7th 1776 at the Acad\'emie des sciences, followed by 
a second version, read on March 27th 1781, and finally published in 1784.} sur la th\'eorie des d\'eblais et des remblais''. 

In modern mathematical terminology, Monge's problem can be stated as follows.

\smallskip
\noindent
\textbf{Monge's problem.} For all $\mu,\nu\in\cP_1(\bR^n)$, find $T:\,\bR^n\to\bR^n$ measurable such that $T\#\mu=\nu$ and 
$$
\ba
\int_{\bR^n}|T(x)-x|\mu(dx)
\\
=\inf\left\{\int_{\bR^n}|F(x)-x|\mu(dx)\text{ with }F:\,\bR^n\to\bR^n\text{ measurable s.t. }F\#\mu=\nu\right\}&\,,
\ea
$$
where $\cP(\bR^n)$ designates the set of all Borel probability measures on $\bR^n$, while
$$
\cP_k(\bR^n):=\left\{\mu\in\cP(\bR^n)\text{ s.t. }\int_{\bR^n}|x|^k\mu(dx)<\infty\right\}\,,
$$
and $T\#\mu$ is the push-forward of the measure $\mu$ by the transformation $T$, defined by the formula
$$
T\#\mu(B):=\mu(T^{-1}B)\,,\quad\text{ for all Borel }B\subset\bR^n\,.
$$

\smallskip
With such a general formulation, Monge's problem does not always have at least one solution. (For example, set $n=1$, choose $\mu:=\de_0$ and $\nu:=\tfrac12(\de_{+1}+\de_{-1})$.
Then there does not exist any map $T:\,\bR\to\bR$ such that $T\#\mu=\nu$.)

However, if $\mu$ is absolutely continuous with respect to the Lebesgue measure of $\bR^n$, Monge's problem always has at least one solution. This was proved more than 200 years 
after Monge's first version of his memoir, in 1979,  by Sudakov \cite{Sudakov} (see also section 6 of \cite{AmbrosioLN}, where a gap in Sudakov's original argument is fixed).

\smallskip
Before Sudakov's proof, Kantorovich proposed in \cite{Kantoro1942} a relaxed version of Monge's problem, for which the existence of a solution is elementary.

\smallskip
\noindent
\textbf{The Kantorovich relaxation of Monge's problem.} For all $\mu,\nu\in\cP_1(\bR^n)$, find
$$
\cW_1(\mu,\nu):=\min_{\rho\in\cC(\mu,\nu)}\iint_{\bR^n\times\bR^n}|x-y|\rho(dxdy)\,,
$$
where $\cC(\mu,\nu)$ is the set of ``couplings'', or ``transport plans'' between $\mu$ and $\nu$, defined as follows:
$$
\cC(\mu,\nu):=\left\{\rho\!\in\!\cP(\bR^{2n})\left\vert\ba\rho(A\!\times\!\bR^n)\!=\!\mu(A),\\ \rho(\bR^n\!\times\! A)\!=\!\nu(A),\ea\right.\quad\text{ for all Borel }A\subset \bR^n\right\}\,.
$$
Observe that $\mu\otimes\nu\in\cC(\mu,\nu)$, so that $\cC(\mu,\nu)\not=\varnothing$.

That Kantorovich's problem is indeed a relaxation of Monge's problem is explained by the following elementary observation. 

\noindent
\textbf{Remark.} If $T:\,\bR^n\to\bR^n$ is a measurable map such that $T\#\mu=\nu$, the probability measure $\rho(dxdy)=\mu(dx)\de_{T(x)}(dy)$ belongs to $\cC(\mu,\nu)$ (this
is trivial). Moreover, for all $\mu,\nu\in\cP_1(\bR^n)$, one has
$$
\cW_1(\mu,\nu)=\inf\left\{\int_{\bR^n}|F(x)-x|\mu(dx)\text{ with }F:\,\bR^n\to\bR^n\text{ measurable s.t. }F\#\mu=\nu\right\}
$$
provided that $\mu$ has no atom (this is not obvious: see Theorem 2.1 in \cite{AmbrosioLN}).

\smallskip
\noindent
\textbf{Quiz 1.} Prove the existence of a solution to the Kantorovich problem. (Hint: if $\mu,\nu\in\cP_1(\bR^n)$, then $\cC(\mu,\nu)\subset\cP_1(\bR^n\times\bR^n)$, and
$\cC(\mu,\nu)$ is weakly relatively compact in $\cP(\bR^n\times\bR^n)$ by Prokhorov's Theorem.)

\smallskip
The main result in \cite{Kantoro1942} is the following equivalent variational formula for $\cW_1(\mu,\nu)$, which can be deduced from the definition above by convex duality,
applying the Fenchel-Moreau-Rockafellar duality theorem (see Theorem 1.12 in \cite{BrezisFA}).

\noindent
\textbf{Kantorovich(-Rubinstein) duality.} For all $\mu,\nu\in\cP_1(\bR^n)$,
$$
\cW_1(\mu,\nu)=\sup_{\chi\in\Lip(\bR^n,\bR)\atop \Lip(\chi)\le 1}\left|\int_{\bR^n}\chi(z)\mu(dz)-\int_{\bR^n}\chi(z)\nu(dz)\right|\,.
$$

For a proof of Kantorovich duality, see chapter 1 of \cite{VillaniAMS}.

\smallskip
The Kantorovich duality formula for $\cW_1$ has several important applications to the topology of $\cP_1(\bR^n)$, listed below.

\smallskip
\noindent
\textbf{Consequences of the Kantorovich(-Rubinstein) duality.}

\noindent
(1) The functional $\cW_1$ is a metric on $\cP_1(\bR^n)$.

\noindent
(2) Let $\mu\in\cP_1(\bR^n)$ and $\mu_j$ be a sequence of elements of $\cP_1(\bR^n)$. Then the three conditions below are equivalent

(a) $\cW_1(\mu_j,\mu)\to 0$ as $j\to\infty$,

(b) $\mu_j\to\mu$ weakly in the sense of probability measures as $j\to\infty$ and
$$
\lim_{R\to\infty}\sup_{j\ge 1}\int_{|x|>R}|x|\mu_j(dx)=0\,,
$$

(c) $\mu_j\to\mu$ weakly in the sense of probability measures as $j\to\infty$ and
$$
\lim_{j\to\infty}\int_{\bR^n}|x|\mu_j(dx)=\int_{\bR^n}|x|\mu(dx)\,.
$$

Statement (2) is Theorem 7.12 in \cite{VillaniAMS}. As for (1), the Kantorovich(-Rubinstein) duality formula obviously implies that $\cW_1$ is nonnegative, symmetric in both
its arguments, and satisfies the triangle inequality. Finally, if $\cW_1(\mu,\nu)=0$, then
$$
\int_{\bR^n}\psi(z)\mu(dz)=\int_{\bR^n}\psi(z)\nu(dz)
$$
for each $\psi\in C^\infty_c(\bR^n)$, since $\phi:=\psi/(1+\|\Grad\psi\|_{L^\infty(\bR^n)})$ satisfies $\Lip(\phi)\le 1$. Hence $\mu=\nu$ (viewing $\mu$ and $\nu$ as distributions
of order $0$ on $\bR^n$). 

The functional $\cW_1$ is usually referred to as the Wasserstein distance of exponent $1$, and sometimes as the Monge-Kantorovich or the Kantorovich-Rubinstein distance.

\subsubsection{The Wasserstein Distance of Exponent $2$}

In Monge's own words ``Le prix du transport d'une mol\'ecule [est], toute choses \'egales d'ailleurs, proportionnel \`a son poids \& \`a l'espace qu'on lui fait parcourir'' (all else
being equal, the cost of transport for one molecule is proportional to its weight and to the distance over which it is transported).

But one could assume instead that the cost of transport is proportional to some power of the distance. In this section, we consider the special case where the cost of transport 
is proportional to the \textit{square} distance. In that case, the Monge and the Kantorovich problems are as follows.

\smallskip
\noindent
\textbf{Monge's problem.} For all $\mu,\nu\in\cP_2(\bR^n)$, find $T:\,\bR^n\to\bR^n$ measurable such that $T\#\mu=\nu$ and 
$$
\ba
\int_{\bR^n}|T(x)-x|^2\mu(dx)
\\
=\inf\left\{\int_{\bR^n}|F(x)-x|^2\mu(dx)\text{ with }F:\,\bR^n\to\bR^n\text{ measurable s.t. }F\#\mu=\nu\right\}&\,.
\ea
$$
\textbf{The Kantorovich problem.} For all $\mu,\nu\in\cP_2(\bR^n)$, find
$$
\cW_2(\mu,\nu):=\left(\min_{\rho\in\cC(\mu,\nu)}\iint_{\bR^n\times\bR^n}|x-y|^2\rho(dxdy)\right)^{1/2}\,.
$$
The existence of a solution to the Kantorovich problem is proved by exactly the same weak compactness argument as in Quiz 1 above.

\smallskip
\noindent
\textbf{Remark.} For each $p\in(1,+\infty)$, there exists an analogue of the functional $\cW_2$, denoted $\cW_p$, in the case where the cost of transport is proportional to the
$p$th power of the distance. We have chosen to restrict our attention to the cases $p=1$ and $p=2$, for which ``quantum'' analogues have been defined.

\smallskip 
The Kantorovich duality in that case is, at first sight, different from the case of exponent $1$.

\smallskip
\noindent
\textbf{Kantorovich duality for $\cW_2$.} For all $\mu,\nu\in\cP_2(\bR^n)$,
$$
\cW_2(\mu,\nu)^2=\sup_{a(x)+b(y)\le|x-y|^2\atop a,b\in C_b(\bR^n)}\int_{\bR^n}a(z)\mu(dz)+\int_{\bR^n}b(z)\nu(dz)\,.
$$
Moreover, there exist two l.s.c. proper convex functions $\a,\b$ such that $\a\in L^1(\bR^n,\mu)$ and $\b\in L^1(\bR^n,\nu)$, satisfying\footnote{A convex function 
$\a:\,\bR^n\to\bR\cup\{+\infty\}$ is said to be proper if there exists at least one point $x\in\bR^n$ such that $\a(x)<+\infty$. Its Legendre dual is 
$$\a^*(p):=\sup_{x\in\bR^n}(p\cdot x-\a(x))\,,\qquad p\in\bR^n\,.$$ If $\a$ is a l.s.c. proper convex function on $\bR^n$, then the function $\a^*$ is also l.s.c. proper 
and convex on $\bR^n$, and one has $\a^{**}=\a$. See section 1.4 and Theorem 1.11 in \cite{BrezisFA}.} $\a^*=\b$ and $\b^*=\a$, and such that
$$
\cW_2(\mu,\nu)^2=\int_{\bR^n}(|x|^2-2\a(x))\mu(dx)+\int_{\bR^n}(|y|^2-2\b(y))\nu(dy)\,.
$$

The first equality is Theorem 1.3 in \cite{VillaniAMS}, while the existence of an optimal pair $\a,\b$ is Theorem 2.9 in \cite{VillaniAMS}.

\smallskip
Perhaps the most important consequence of the Kantorovich duality for $\cW_2$ is the structure of optimal couplings for the Kantorovich problem.

\smallskip
\noindent
\textbf{Optimal couplings.} 

\noindent
(a) (Knott-Smith Theorem.) Let $\mu,\nu\in\cP_2(\bR^n)$. A probability measure $\rho\in\cC(\mu,\nu)$ is an optimal coupling for $\cW_2$ if and only if there exists 
$\Phi:\,\bR^n\to\bR\cup\{+\infty\}$, l.s.c. proper and convex, such that
$$
\Supp(\rho)\subset\text{graph}(\d\Phi)\,,
$$
where $\d\Phi$ is the subdifferential\footnote{If $\Phi:\,\bR^n\to\bR\cup\{+\infty\}$ is a proper convex function, its subdifferential at $x\in\bR^n$ is the set
$$\d\Phi(x):=\{\xi\in\bR^n\text{ s.t. }\Phi(y)\ge\Phi(x)+\xi\cdot(y-x)\,,\quad y\in\bR^n\}\,.$$ See Example 2.1.4 in \cite{BrezisOpMM}. The subdifferential of
a proper convex function is an example of monotone operator: if $x_1,x_2\in\bR^n$ and if $\xi_1\in\d\Phi(x_1)$ and $\xi_2\in\d\Phi(x_2)$, then 
$$(\xi_2-\xi_1)\cdot(x_2-x_1)\ge 0\,.$$ One can check that (1) $\Phi$ is differentiable at $x\in\bR^n$ if and only if $\d\Phi(x)$ contains a single element,
which is $\Grad\Phi(x)$, and (2) if a l.s.c. proper convex function $\Phi$ on $\bR^n$ is strictly convex in a neighborhood of 
$x\in\bR^n$, its Legendre dual $\Phi^*$ is differentiable on $\d\Phi(x)$, and $\Grad\Phi^*(\xi)=x$ for all $\xi\in\d\Phi(x)$.} of $\Phi$.

\noindent
(b) (Brenier's Theorem.) Let $\mu,\nu\in\cP_2(\bR^n)$. If $\mu(S)=0$ for each Borel set $S\subset\bR^n$ of Hausdorff dimension $\cH-\text{dim}(S)\le n-1$, there 
exists a unique optimal coupling for $\cW_2$, of the form
$$
\rho(dxdy)=\mu(dx)\de_{\Grad\Phi(x)}(dy)
$$
with $\Phi:\,\bR^n\to\bR\cup\{+\infty\}$ convex and such that $\Phi\in L^1(\bR^n,\mu)$. 

\smallskip
See Theorem 2.12 in \cite{VillaniAMS}, and more generally chapter 2 of \cite{VillaniAMS} for a proof of the Knott-Smith and the Brenier theorems.

Since $\Phi\in L^1(\bR^n,\mu)$, one has $\mu(\Phi^{-1}(\{+\infty\}))=0$, on the other hand $\Phi$ is locally Lipschitz continuous on the interior of its domain 
$\Dom(\Phi):=\bR^n\setminus\Phi^{-1}(\{+\infty\})$. Then (1) $\Dom(\Phi)$ is a convex subset of $\bR^n$, and hence $\d\Dom(\Phi)$ has Hausdorff dimension
$\le n-1$, and (2) the set of points $x$ in the interior of $\Dom(\Phi)$ such that $\Phi$ is not differentiable at $x$ has Hausdorff dimension $\le n-1$ (see
\cite{AlbertiAmbrosioCannarsa}). Since
$$
(\Grad\Phi)^{-1}(B):=\{x\in\bR^n\text{ s.t. }\Phi\text{ is differentiable at }x\text{ and }\Grad\Phi(x)\in B\}\,,
$$
and since $\Grad\Phi(x)$ exists for $\mu$-a.e. $x\in\bR^n$, the push-forward measure $(\Grad\Phi)_\#\mu$ is well-defined.

\smallskip
We shall conclude this section with a brief list of the most important topological properties of $\cW_2$.

\smallskip
\noindent
\textbf{Properties of $\cW_2$.}

\noindent
(1) The functional $\cW_2$ is a metric on $\cP_2(\bR^n)$.

In particular
$$
\cW_2(\mu,\nu)=0\iff\mu=\nu\,;
$$
the optimal coupling in that case is $\rho(dxdy)\!=\!\mu(dx)\de_{x}(dy)$, and the Brenier transport map is the identity, which is the gradient of the convex function 
$x\mapsto\tfrac12|x|^2$.

\noindent
(2) Let $\mu\in\cP_2(\bR^n)$ and $\mu_j$ be a sequence of elements of $\cP_2(\bR^n)$. Then the two conditions below are equivalent:

\noindent
(a) $\cW_2(\mu_j,\mu)\to 0$ as $j\to\infty$,

\noindent
(b) $\mu_j\to\mu$ weakly in the sense of probability measures, and
$$
\lim_{R\to\infty}\sup_{j\ge 1}\int_{|x|>R}|x|^2\mu_j(dx)=0\,.
$$

\smallskip
Statement (1) is Theorem 7.3 in \cite{VillaniAMS}, while statement (2) is Theorem 7.12 in \cite{VillaniAMS}. The proof of the triangle inequality is definitely nontrivial, 
at variance with the case of $\cW_1$, for which the triangle inequality follows from the expression of $\cW_1$ using the Kantorovich duality. 

\smallskip
Here is a nontrivial example where $\cW_2(\mu,\nu)$ can be computed explicitly. See \cite{GivensShortt} for a proof of the formula below.

\smallskip
\noindent
\textbf{Example.} Let $G_1,G_2$ be Gaussian laws on $\bR^n$ with means $m_1,m_2$ and covariance matrices $A_1,A_2$. Then
$$
\cW_2(G_1,G_2)^2=|m_1-m_2|^2+\Tr\left(A_1+A_2-2\left(\sqrt{A_1}A_2\sqrt{A_1}\right)^\frac12\right)
$$

\smallskip
\noindent
\textbf{Quiz 2.} For all $m_1,m_2\in\bR^n$, compute $\cW_p(\delta_{m_1},\delta_{m_2})$ for $p=1$ and $p=2$. Does there exist optimal transport map(s) in both cases?

\smallskip
In most reference textbooks on optimal transport, the proof of the triangle inequality for $\cW_p$ with $1<p<\infty$ is based on a nontrivial construction referred to as ``glueing''
couplings having a common marginal (Lemma 7.6 in \cite{VillaniAMS}, or Lemma 5.3.2 in \cite{AmbrosioGS}). This procedure is itself based on the notion of ``disintegration'' of a 
probability measure on a Cartesian product with respect to one of its marginals (obviously related to the notion of conditional probability): see Theorem 5.3.1 in \cite{AmbrosioGS}.
There is an alternative to the approach based on the disintegration theorem, which uses instead the Hahn-Banach theorem: see Exercise 7.9 in \cite{VillaniAMS}.

In the case where there exist optimal transport maps (e.g. when the sets of Hausdorff codimension $\ge 1$ are negligible for the probability measures considered in the triangle 
inequality, according to the Brenier theorem), the proof of the triangle inequality is very simple: see Lemma 5.3 in \cite{Santambrogio}. In the setting considered here (where the
underlying metric space is $\bR^n$ with its canonical Euclidean distance), one can always reduce the triangle inequality for general probability measures to this simple case by
an approximation argument (Lemma 5.2 in \cite{Santambrogio}). 

However, we shall see later that the very notion of a transport map in the quantum setting remains to be clarified. Likewise, the possibility of ``glueing'' couplings with a common
marginal in the quantum setting seems to be an open problem at the time of this writing. 

For these reasons, a third approach to the triangle inequality for $\cW_2$, entirely based on the Kantorovich duality formula, is proposed in the exercise below\footnote{I came up
with this proof during the week of the summer school at the R\'enyi Institute. The same method applies to the Wasserstein distance of exponent $p$ for all $p\in(1,\infty)$, but the
computations (which I shall publish elsewhere) are more involved than in the case $p=2$, which is the only one of interest here. I have not seen this proof in any of the reference 
textbooks on optimal transport that I have been using, and I do not know whether it is original.}. Perhaps this could be useful later, for instance in the quantum setting.

\smallskip
\noindent
\textbf{Quiz 3.} Let $\mu,\nu,\rho\in\cP_2(\bR^n)$, and let $A,\Ga$ be optimal functions in the Kantorovich duality formula for $\cW_2(\mu,\rho)$ such that $A\in L^1(\bR^n,\mu)$
and $\Ga\in L^1(\bR^n,\rho)$, i.e.
$$
\cW_2(\mu,\rho)^2=\int_{\bR^n}A(x)\mu(dx)+\int_{\bR^n}\Ga(z)\rho(dz)\,.
$$
while $\a$ and $\g$ defined by $\a(x):=\tfrac12(|x|^2-A(x))$ and $\g(z):=\tfrac12(|z|^2-\Ga(z))$ are l.s.c. convex functions such that $\a^*=\g$ and $\g^*=\a$. For each $\eta>0$,
set
$$
B_\eta(y):=(1+\tfrac1\eta)\inf_{z\in\bR^n}\left(|y-z|^2-\tfrac{\Ga(z)}{1+\frac1\eta}\right)\,,\qquad y\in\bR^n\,.
$$

\noindent
(1) Prove that the function 
$$
y\mapsto\tfrac12\left(|y|^2-\tfrac{B_\eta(y)}{1+\tfrac1\eta}\right)
$$
is the Legendre transform of a l.s.c. proper convex function to be computed in terms of $\Ga$.

\noindent
(2) Prove that $A(x)-B_\eta(y)\le(1+\eta)|x-y|^2$ for all $x,y\in\bR^d$.

\noindent
(3) Prove that $B_\eta\in L^1(\bR^n,\nu)$ for each $\eta>0$.

\noindent
(4) Prove that, for each $\eta>0$
$$
\int_{\bR^n}A(x)\mu(dx)-\int_{\bR^n}B_\eta(y)\nu(dy)\le(1+\eta)\cW_2(\mu,\nu)^2\,.
$$
(5) Prove that, for each $\eta>0$
$$
\int_{\bR^n}B_\eta(y)\nu(dy)+\int_{\bR^n}\Ga(z)\rho(dz)\le(1+\tfrac1\eta)\cW_2(\nu,\rho)^2\,.
$$
(6) Prove that, for each $\eta>0$
$$
\cW_2(\mu,\rho)^2\le(1+\eta)\cW_2(\mu,\nu)^2+(1+\tfrac1\eta)\cW_2(\nu,\rho)^2\,.
$$
(7) Prove that $\cW_2$ satisfies the triangle inequality.

\subsection{Density Operators in Quantum Mechanics}\lb{SS-DensOp}


In classical mechanics, the state of a point particle is completely defined by its position $q\in\bR^d$ and its momentum $p\in\bR^d$. The point particle phase space is therefore
$\bR^d\times\bR^d$, and it is natural to study the probability of finding a point particle in subsets of the phase space. This is precisely the statistical formalism used by Maxwell
and Boltzmann in the kinetic theory of gases. 

In quantum mechanics, the analogous formalism involves density operators, a special class of operators on the Hilbert space $L^2(\bR^d)$. Before studying density operators, 
we need to recall some fundamental notions in the theory of operators on Hilbert spaces.

\subsubsection{Trace of an Operator}


Let $\fH$ be a complex, separable Hilbert space with inner product denoted by $(\cdot\,|\,\cdot)_\fH$, and let $\cL(\fH)$ designate the algebra of bounded operators on the Hilbert 
space $\fH$.

\smallskip
\noindent
\textbf{Positive operators.} An operator $T\in\cL(\fH)$ is said to be positive if
$$
T=T^*\,,\quad\text{ and }\quad(x|Tx)_\fH\ge 0\,,\qquad x\in\fH\,.
$$
Equivalently, if $T\in\cL(\fH)$, then
$$
T=T^*\ge 0\iff\text{ there exists }S\in\cL(\fH)\text{ s.t. }T=S^*S\,.
$$

\smallskip
\noindent
\textbf{Trace of a positive operator.} For $T\in\cL(\fH)$ such that $T=T^*\ge 0$, we define
$$
\Tr_\fH(T):=\sum_{j\ge 1}(e_j|Te_j)_\fH\in[0,+\infty]\text{ for all Hilbert basis }(e_j)_{j\ge 0}\text{ of }\fH\,.
$$
One easily checks that, if this sum is finite for \textit{one} Hilbert basis $(e_j)_{j\ge 0}$ of $\fH$, it is finite for \textit{all} Hilbert basis of $\fH$, since the (infinite dimensional) 
``transition matrix'' from any Hilbert basis of $\fH$ to $(e_j)_{j\ge 0}$ is a unitary operator on $\fH$.

The definition of the trace follows the definition of the Lebesgue integral on the real line: first we define trace of any self-adjoint positive operator, as we define the integral of 
any measurable positive function. Then we define the analogue of the Lebesgue space $L^1$, and extend the trace to this space by linearity, exactly in the same way as the 
integral is extended from the set of measurable positive functions to the Lebesgue space $L^1$.

\smallskip
\noindent
\textbf{Trace-class operators.} The set of trace-class operators is
$$
\cL^1(\fH):=\{T\!\in\!\cL(\fH)\text{ s.t. }\|T\|_1\!:=\!\Tr_\fH(|T|)<\infty\}\,,
$$
where $|T|:=\sqrt{T^*T}$ for each $T\in\cL(\fH)$. 

The following properties of $\cL^1(\fH)$ are well known.

\smallskip
\noindent
\textbf{Properties of trace-class operators.}

\noindent
(a) Trace-class operators are compact:
$$
\cL^1(\fH)\subset\cK(\fH)\,,
$$
where $\cK(\fH)$ is the set of compact operators on $\fH$, i.e. the operator-norm closure of the set of finite rank operators on $\fH$. (One easily checks that $\cK(\fH)$ 
is a two-sided ideal of $\cL(\fH)$, stable by the involution $T\mapsto T^*$.)

\noindent
(b) The set of trace-class operators $\cL^1(\fH)$ is a two-sided ideal of $\cL(\fH)$, which is stable by the involution $T\mapsto T^*$:
$$
\ba
A\in\cL(\fH)\text{ and }T\in\cL^1(\fH)&\implies AT\text{ and }TA\in\cL^1(\fH)\,,
\\
T\in\cL^1(\fH)&\implies T^*\in\cL^1(\fH)\,.
\ea
$$
(c) The trace, which is defined on the set of positive trace-class operators on $\fH$, extends as a linear functional on $\cL^1(\fH)$ satisfying the properties
$$
\Tr(T^*)=\overline{\Tr(T)}\,,\quad\Tr(AT)=\Tr(TA)\,,\quad\text{ and }|\Tr(AT)|\le\|A\|\|T\|_1\,,
$$
for all $T\in\cL^1(\fH)$ and all $A\in\cL(\fH)$.

\noindent
(d) The $\bC$-linear space $\cL^1(\fH)$ is a Banach space for the trace norm $T\mapsto\|T\|_1$; besides\footnote{The  topological dual of a real or complex normed linear 
space $E$, i.e. the set of continuous linear functionals on $E$ (with real or complex values), is denoted by $E'$.}
$$
\cK(\fH)'=\cL^1(\fH)\,,\quad\text{ and }\quad\cL^1(\fH)'=\cL(\fH)\,.
$$
In both equalities, the duality is defined by the trace, i.e.
$$
\la T,K\ra_{\cL^1(\fH),\cK(\fH)}:=\Tr_\fH(TK)\,,\qquad \la A,T\ra_{\cL(\fH),\cL^1(\fH)}:=\Tr_\fH(AT)\,.
$$

\smallskip
\noindent
\textbf{Partial Trace.} The construction of the trace explained before is strikingly similar to the construction of the Lebesgue integral. In the present section, we are going
to study an analogue of the Fubini theorem.

For each $T\in\cL^1(\fH_1\otimes\fH_2)$, one defines $T_1=\Tr_2(T)\in\cL^1(\fH_1)$ by the formula
$$
\Tr_{\fH_1}(T_1A)=\Tr_{\fH_1\otimes\fH_2}(T(A\otimes I_{\fH_2}))\,,\qquad\text{ for all }A\in\cL(\fH)\,.
$$
There is a similar definition of $\Tr_1(T)\in\cL^1(\fH_2)$.

\smallskip
Observe indeed that the map
$$
\cK(\fH_1)\ni A\mapsto\Tr_{\fH_1\otimes\fH_2}(T(A\otimes I_{\fH_2}))\in\bC
$$
is a norm-continuous linear functional on $\cK(\fH_1)$, and is therefore represented by a unique trace-class operator $T_1$ on $\fH_1$. That the defining identity for 
$\Tr_2(T)$ holds for all $A\in\cL(\fH_1)$ follows from the density of finite rank operators in $\cL^1(\fH)$ for the trace-norm.

\smallskip
\noindent
\textbf{Remark.} The analogue of this construction in the context of integration is the following form of the Fubini theorem: if $f\in L^1(\bR^m\times\bR^n)$, the function 
$y\mapsto f(x,y)$ belongs to $L^1(\bR^n)$ for a.e. $x\in\bR^m$, the function
$$
F:\,x\mapsto\int_{\bR^n}f(x,y)dy
$$
belongs to $L^1(\bR^m)$, and 
$$
\int_{\bR^m\times\bR^n}f(x,y)dxdy=\int_{\bR^m}F(x)dx\,.
$$
The function $F$ in this statement is easily seen to be the analogue of $\Tr_2(T)$ for $T\in\cL^1(\fH_1\otimes\fH_2)$.

\subsubsection{Hilbert-Schmidt Operators}


An operator $T\in\cL(\fH)$ is said to be a Hilbert-Schmidt operator if
$$
\Tr_\fH(T^*T)<\infty\,.
$$
The Hilbert-Schmidt class is the set of Hilbert-Schmidt operators:
$$
\cL^2(\fH):=\{T\in\cL(\fH)\text{ s.t. }\Tr_\fH(T^*T)<\infty\}\,.
$$

The Hilbert-Schmidt class $\cL^2(\fH)$ is a Hilbert space for the inner product
$$
(T_1|T_2)_2:=\Tr_\fH(T_1^*T_2)
$$
defining the Hilbert-Schmidt norm 
$$
\|T\|_2:=\sqrt{\Tr_\fH(T^*T)}\,.
$$
Moreover
$$
\cL^1(\fH)\subset\cL^2(\fH)\subset\cK(\fH)\subset\cL(\fH)
$$
with continuous inclusions, and
$$
\|T\|\le\|T\|_2\le\|T\|_1\,,\qquad T\in\cL^1(\fH)\,.
$$
Besides $\cL^2(\fH)$ is a two-sided ideal of $\cL(\fH)$.

\smallskip
If $T=T^*\in \cL^2(\fH)$, then $T$ is a compact operator, so that there exists $(e_j)_{j\ge 1}$, a Hilbert basis of $\fH$ and $(\tau_j)_{j\ge 1}\in\ell^2(\bN^*;\bR)$ such that 
$$
T=\sum_{j\ge 1}\tau_jP_j\,,\quad\|T\|_2^2=\sum_{j\ge 1}\tau_j^2\quad\text{ with }P_j\phi:=(e_j|\phi)_\fH e_j\,.
$$

Besides, in the case where $\fH=L^2(\bR^d)$, one has therefore
\be\lb{OpIntT}
T\phi(x)=\int_{\bR^d}t(x,y)\phi(y)dy\,,\qquad\text{ for all }\phi\in\fH\,,
\ee
where
$$
t(x,y):=\sum_{j\ge 1}\tau_je_j(x)\overline{e_j(y)}\,.
$$
In particular
\be\lb{HSNormT}
\|T\|_2^2=\sum_{j\ge 1}\tau_j^2=\iint_{\bR^d\times\bR^d}|t(x,y)|^2dxdy\,.
\ee

Conversely, an integral operator $T$ of the form \eqref{OpIntT} belongs to $\cL^2(\fH)$ if and only if $t\in L^2(\bR^d\times\bR^d)$, and the first left-hand side in \eqref{HSNormT} 
is equal to the last right-hand side of in \eqref{HSNormT}.

\smallskip
\noindent
The interested reader will discover additional important properties of trace-class and Hilbert-Schmidt operators in the next two exercises.

\smallskip
\noindent
\textbf{Quiz 4.} In this exercise, $\fH:=L^2(\bR^d)$.

\noindent
(1) Prove that any $T\in L^1(\fH)$ can be put in the form $T=T_1T_2$ with $T_1,T_2\in\cL^2(\fH)$, and that $\|T\|_1\le\|T_1\|_2\|T_2\|_2$.

\noindent
(2) For all $T\in L^1(\fH)$, can one find $T_1,T_2\in\cL^2(\fH)$ such that 
$$
T=T_1T_2\quad\text{ and }\quad\|T\|_1=\|T_1\|_2\|T_2\|_2\,?
$$

\noindent
(3) Prove that for each $T\in L^1(\fH)$, there exists $t\equiv t(x,y)$ such that $z\mapsto t(x+z,x)$ belongs to $C_b(\bR^d_z;L^1(\bR^d_x))$, 
which is an integral kernel for $T$, in the sense that
$$
T\phi(x)=\int_{\bR^d}t(x,y)\phi(y)dy\,,\qquad\text{ for all }\phi\in\fH\,,
$$
and that
$$
\Tr_\fH(T)=\int_{\bR^d}t(x,x)dx\,.
$$

\smallskip
Question (3) suggests that $t(x,y)$ is the analogue for $T$ of the entries of a matrix if the infinite dimensional Hilbert space $\fH$ is replaced with $\bC^n$. One might
therefore believe that an integral operator is trace-class if the restriction of its integral kernel to the diagonal is summable. This is not the case, as shown by the next
example.

\smallskip
\noindent
\textbf{Quiz 5.} Consider the Volterra operator $V$ defined on $L^2([0,1])$ by the formula
$$
V\phi(x):=\int_0^x\phi(y)dy\,,\qquad\phi\in L^2([0,1])\,.
$$
(1) Prove that $V$ is the operator defined by the integral kernel $v(x,y)=\indc_{0\le y\le x}$. 

\noindent
(2) Is $V$ a Hilbert-Schmidt operator on $L^2([0,1])$? 

\noindent
(3) Does the function $x\mapsto v(x,x)$ belong to $L^1([0,1])$? 

\noindent
(4) Is $V$ a trace-class operator on $L^2([0,1])$?

\noindent
(5) What are the eigenvalues of $V$?

\noindent
(6) What is the spectral radius of $V$?

\smallskip
This last exercise shows the importance of the continuity condition in question (3) of Quiz 3. Any trace-class operator $T$ on $L^2(\bR^d)$, being a Hilbert-Schmidt operator, 
is an integral operator defined by a unique integral kernel in $L^2(\bR^d\times\bR^d)$. Since the diagonal is a Lebesgue-negligible set in $\bR^d\times\bR^d$, the restriction 
to the diagonal of this integral kernel is a priori not a well defined function. Yet, any trace-class operator $T$ on $L^2(\bR^d)$ has an integral kernel satisfying the continuity
condition of (3) in Quiz 3, which is a representative of the unique $L^2(\bR^d\times\bR^d)$ integral kernel of $T$ viewed as a Hilbert-Schmidt operator on $L^2(\bR^d)$, and
the trace of $T$ is indeed the integral of the restriction of this particular integral kernel to the diagonal, which is a well-defined element of $L^1(\bR^d)$.

\subsubsection{Density Operators}


As mentioned above, density operators on $L^2(\bR^d)$ are the quantum analogue of Borel probability measures on the single-particle phase space $\bR^d\times\bR^d$. 
The positivity of a probability measure $\mu$ on $\bR^d\times\bR^d$ becomes the positivity of an operator $R$ on $L^2(\bR^d)$, while the analogue of the normalization 
condition $\mu(\bR^d\times\bR^d)=1$ in the quantum setting is the condition $\Tr_\fH(R)=1$. 

Throughout this section, we set $\fH=L^2(\bR^d)$.

\smallskip
\noindent
\textbf{(Quantum) Density operators.} A density operator on $\fH$ is an element of
$$
\cD(\fH):=\{T\in \cL(\fH)\text{ s.t. }T=T^*\ge 0\text{ and }\Tr_\fH(T)=1\}\subset\cL^1(\fH)\,.
$$
As explained above, $\cD(\fH)$ is the quantum analogue of $\cP(\bR^d\times\bR^d)$.

\smallskip
When dealing with computations involving quantum states, it will be especially convenient to use the notation involving bras and kets, which is recalled below.

\smallskip
\noindent
\textbf{Dirac bra-ket notation.} For $\phi,\psi\in\fH$, we denote by $|\psi\ra$ the vector $\psi$, while
$$
\la\phi|\text{ denotes the linear functional }\psi\mapsto\int_{\bR^d}\overline{\phi(x)}\psi(x)dx=\la\phi|\psi\ra\,.
$$
With this notation, one easily checks that
$$
\psi\in\fH\text{ and }\|\psi\|_\fH=1\implies|\psi\ra\la\psi|=\text{ orthogonal projection on }\bC\psi\,.
$$

\smallskip
\noindent
\textbf{Example: Schr\"odinger's coherent state.} For $q,p\in\bR^d$, set
$$
|q,p\ra(x):=(2\pi\hb)^{-d/4}\exp\left(-\tfrac1{2\hb}|x-q|^2\right)\exp\left(\tfrac{i}{\hb}p\cdot(x-\tfrac{q}2)\right)\,.
$$
One easily checks that $\|\,|q,p\ra\,\|_\fH=1$ so that $|q,p\ra\la q,p|\in\cD(\fH)$. 

The density operator $|q,p\ra\la q,p|$ defined above is a quantum analogue (and by no means the only one) of the probability measure $\de_{(q,p)}\in\cP(\bR^d\times\bR^d)$.

\smallskip
Since density operators are the quantum analogue of phase space (Borel) probability measures, a natural problem is that of comparing two density operators by some
procedure which corresponds to a comparison between phase space (Borel) probability measures in the classical limit of quantum mechanics. In the sequel, we shall
consider an important example, involving Schr\"odinger's coherent states, for which explicit computations are very easy.

\smallskip
\noindent
\textbf{Key example.} For $(q_1,p_1)\not=(q_2,p_2)\in\bR^d\times\bR^d$, set 
$$
R_1:=|q_1,p_1\ra\la q_1,p_1|\quad\text{ and }\quad R_2=|q_2,p_2\ra\la q_2,p_2|\,.
$$
Then $R_1-R_2$ is a self-adjoint, rank-$2$ operator, such that $\Tr(R_1-R_2)=0$, and hence there exists $\lbd>0$ and an orthonormal basis $(e,f)$ of $\IM(R_1-R_2)$ 
for which
$$
R_1-R_2=\lbd|e\ra\la e|-\lbd|f\ra\la f|\,.
$$
Therefore
$$
\|R_1-R_2\|_1=2\lbd\quad\text{ and }\quad\|R_1-R_2\|_2=\sqrt{2}\lbd\,,
$$
so that 
$$
\|R_1-R_2\|_1=\sqrt{2}\|R_1-R_2\|_2\,.
$$
On the other hand, $R_1-R_2$ is the Hilbert-Schmidt integral operator with integral kernel
$$
\ba
r(x,y):=&(2\pi\hb)^{-d/2}\exp\left(-\tfrac1{2\hb}(|x-q_1|^2+|y-q_1|^2)\right)\exp\left(\tfrac{i}{\hb}p_1\cdot(x-y)\right)
\\
&-(2\pi\hb)^{-d/2}\exp\left(-\tfrac1{2\hb}(|x-q_2|^2+|y-q_2|^2)\right)\exp\left(\tfrac{i}{\hb}p_2\cdot(x-y)\right)\,,
\ea
$$
so that
$$
\|R_1-R_2\|_2^2=\iint_{\bR^d\times\bR^d}|r(x,y)|^2dxdy= 2(1-e^{-(|q_1-q_2|^2+|p_1-p_2|^2)/2\hb})\,.
$$
Thus
$$
\|R_1-R_2\|_1= 2\sqrt{1-e^{-(|q_1-q_2|^2+|p_1-p_2|^2)/2\hb}}\to 2\quad\text{ as }\hb\to 0^+
$$
since $(q_1,p_1)\not=(q_2-p_2)$. In other words, passing to the limit as $\hb\to 0$,
$$
\|\,|q_1,p_1\ra\la q_1,p_1|-|q_2,p_2\ra\la q_2,p_2|\,\|_1\to\|\de_{(q_1,p_1)}-\de_{(q_2,p_2)}\|_{TV}
=\left\{\ba{}&2&&\text{ if }(q_1,p_1)\not=(q_2,p_2)\,,\\&0&&\text{ if }(q_1,p_1)=(q_2,p_2)\,,\ea\right.
$$
In particular, $\|\cdot\|_1$ fails to discriminate between density operators concentrating on phase space points at a distance $\gg O(\hb^{1/2})$ of each other  in the classical limit.

\subsection{The Connes Distance in Noncommutative Geometry}\lb{SS-ConnesDist}


It is well known that one of the main differences between quantum and classical mechanics is that the product of phase space coordinates of a point particle in classical mechanics
is a commutative operation, whereas the quantum analogues of these quantities are operators on a Hilbert space, and their product is in general noncommutative. For example,
think of a point particle in space dimension $1$, and let $q,p\in\bR$ designate respectively its position and momentum in classical mechanics. Obviously
$$
pq-qp=0\,.
$$
In quantum mechanics, the real-valued functions $(q,p)\mapsto q$ and $(q,p)\mapsto p$ are replaced with (unbounded) operators $\hat q$ and $\hat p$ on $L^2(\bR)$ satisfying
the canonical commutation relation (CCR)
$$
[\hat p,\hat q]=\hat p\hat q-\hat q\hat p=-i\hbar\text{Id}\,.
$$

To the best of our knowledge, the first attempt at extending the notions of optimal transport to the noncommutative setting is due to Connes \cite{Connes1989}. We shall briefly
describe his work on this topic in this section.

Let $\cA$ be a unital $C^*$-algebra. (We recall that a $C^*$-algebra is a complex Banach algebra endowed with a linear involution $x\mapsto x^*$ such that $(\a x)^*=\bar\a x^*$
for all $\a\in\bC$ and $\|x^*x\|=\|x\|^2$ for all $x$ in the algebra. The latter condition is known as ``the $C^*$ identity''. This is a very strong condition, which connects the norm
with the algebraic structure. For instance, it implies that $\|x\|^2$ is the spectral radius of $x^*x$ for each $x$ in the algebra.) Here are a few examples of $C^*$-algebras:

\noindent
(a) $\cA=C(X,\bC)$ with $X$ compact; in this example
$$
f^*(x):=\overline{f(x)}\text{ for all }x\in X\,,\qquad\|f\|:=\sup_{x\in X}|f(x)|\,.
$$
This is the prototype of a unital commutative $C^*$-algebra, the unit being the constant function $x\mapsto 1$.

\noindent
(b) $\cA=\cL(\fH)$ where $\fH$ is a separable complex Hilbert space; in this example, the involution is $T\mapsto T^*$ where $T^*$ is the adjoint of the operator $T$ on $\fH$,
while $\|T\|$ is the operator-norm of $T$, i.e. $\|T\|:=\sup\{\|Tx\|_\fH\,:\,\|x\|_\fH=1\}$. This is a unital $C^*$-algebra, with unit $\text{Id}_\fH$, and it is not separable for the norm
topology.

\noindent
(c) $\cA=\cK(\fH)$, the set of compact operators on an infinite dimensional, separable complex Hilbert space $\fH$; this is a non unital $C^*$-subalgebra of $\cL(\fH)$; in fact
$\cK(\fH)$ is the only norm-closed two-sided ideal of $\cL(\fH)$.

\noindent
(d) $\cA=\cL(\fH)/\cK(\fH)$ where $\fH$ is an infinite dimensional, separable complex Hilbert space; this a unital $C^*$-algebra, which is simple --- meaning that its only closed 
two-sided ideals are $\{0\}$ and $\cA$ --- and known as the Calkin algebra; obviously $\cA$ is not of the form $\cL(\fH_1)$ for any separable Hilbert space.

\smallskip
A \textit{state} on a $C^*$-algebra $\cA$ is a positive linear functional on $\cA$ --- positive meaning that $\om(a^*a)\ge 0$ --- of norm $1$ --- meaning that 
$$
\|\om\|:=\sup\{|\om(x)|\,:\,x\in\cA\text{ and }\|x\|\le 1\}=1\,.
$$
It is a classical exercise\footnote{Here is a hint for the interested reader to prove the direct implication. If $\om$ is a positive linear functional on $\cA$, check that $(x,y)\mapsto\om(x^*y)$ 
is a positive sesquilinear form on $\cA$, and use the Cauchy-Schwarz inequality to check that $\om(x^*)=\overline{\om(x)}$ and that $\om$ is continuous on $\cA$ with norm $\|\om\|=\om(1)$.} 
to check that, if $\om$ is continuous linear functional on a unital $C^*$-algebra $\cA$, then
$$
\om\text{ is positive }\iff\|\om\|=\om(1)\,.
$$

\smallskip
\noindent
\textbf{Examples of states}

\noindent
\bu $\cA=C(X;\bC)$ with $X$ compact, and $\om(f)=f(x_0)=\la\de_{x_0},f\ra$, for some $x_0\in X$;

\noindent
\bu $\cA=\cL(\fH)$ and $\om(A):=\la\psi|A|\psi\ra$ for some $\psi\in\fH$ with $\|\psi\|_\fH=1$;

\noindent
\bu $\cA=\cL(\fH)$ and $\om(A):=\Tr_\fH(RA)$ for some $R\in\cD(\fH)$.

\smallskip
Let $(\fH,D)$ be a Fredholm module on $\cA$, meaning that

\noindent
(a) there is a $^*$-linear representation\footnote{I.e. $\pi$ is a morphism of algebras from $\cA$ to $\cL(\fH)$ satisfying the condition $\pi(a^*)=\pi(a)^*$)} $\pi$ of $\cA$ 
in $\fH$,

\noindent
(b) $D=D^*$ is a self-adjoint unbounded operator on $\fH$ such that 
$$
(I+D^2)^{-1}\in\cK(\fH)\,,
$$
(c) $\{a\in\cA\text{ s.t. }[D,\pi(a)]\in\cL(\fH)\}$ is norm-dense in $\cA$.

\smallskip
\noindent
\textbf{Theorem (Connes \cite{Connes1989}).} Assume that
$$
\{a\in\cA\text{ s.t. }\|[D,\pi(a)]\|_\fH\le 1\}/\bC 1\text{ is bounded.}
$$
Then, the following formula metrizes the set of states on $\cA$:
$$
\Dist_C(\om_1,\om_2):=\sup\{|\om_1(a)-\om_2(a)|\text{ s.t. }\|[D,\pi(a)]\|_\fH\le 1\}\,.
$$

\bigskip
There is an obvious similarity between Connes'  definition and the Kantorovich(-Rubinstein) duality for $\cW_1$: it clearly suggests to think of Connes' distance as the
analogue of the metric $\cW_1$ in noncommutative geometry.

It is interesting to see how Connes' definition can be applied to a commutative setting, corresponding to example (a) above of a $C^*$-algebra.

\smallskip
\noindent
\textbf{Example 1: the Dirac operator as  a Fredholm module.} Set $\cA=C(M)$ where $M$ is a compact spin Riemannian manifold (see for instance \cite{BerGetzVergne}), 
with Riemannian metric $g$. Let $S$ be the spinor bundle on $M$, and set $\fH:=L^2(M;S)$, the Hilbert space of $L^2$ sections of $S$. Let $\cA$ act on $\fH$ by scalar 
multiplication --- to avoid unnecessary complications in the notation, we write $a\xi$ instead of $\pi(a)\xi$ everywhere in this example. Finally, let $D$ be the Dirac operator 
on $M$. 

Then, the geodesic distance $\Dist_g$ on $M$ satisfies the following property:
$$
\Dist_g(x,y)=\sup\{|a(x)-a(y)|\,:\,a\in C(M)\text{ s.t. }\|[D,a]\|\le 1\}=\Dist_C(\de_x,\de_y)\,.
$$
\begin{proof} Denote by $\g(v)\zeta$ the Clifford multiplication of $\zeta\in S_x$ by $v\in T_xM$. Since $D$ is a differential operator of order $1$, one easily checks that
$$
([D,a]\xi)_x\!=\!\g((\text{grad }a)_x)\xi_x\,,\quad\xi\in\fH\,,
$$
for all $a\in C^1(M)$, so that
$$
\|[D,a]\|\!=\!\|\text{grad }a\|_{L^\infty(M)}\,.
$$
Hence
$$
\Dist_C(\de_x,\de_y)=\sup_{\Lip(a)\le 1}|a(x)-a(y)|\,,
$$
and
$$
\Dist_g(x,y)\le\Dist_C(\de_x,\de_y)\le\Dist_g(x,y)\,.
$$
The upper bound is obvious by definition of the Lipschitz constant; as for the lower bound, it suffices to pick the function $a(z):=\Dist_g(z,y)$.
\end{proof}

\smallskip
\noindent                                                 
\textbf{Example 2: the word length on a discrete group as a Fredholm module.} Let $\Gamma$ be a discrete group, with reduced $C^*$-algebra $C^*_{red}(\Gamma)$ 
defined as the $C^*$-algebra generated by the left regular representation $\lbd$ on $\fH:=\ell^2(\Gamma)$. (We recall that $(\lbd(g)\xi)_h:=\xi_{g^{-1}h}$ for each 
$\xi=(\xi_h)_{h\in\Gamma}\in\ell^2(\Gamma)$.) Let $L:\,\Gamma\to\bR_+$ be a length function (for instance the 
word length with respect to a system of generators of $\Gamma$). In other words, we assume that
$$
L(1)=0\,,\quad L(g^{-1})L(g)\,,\quad L(gh)\le L(g)+L(h)\,,\qquad g,h\in\Gamma\,.                                                                        
$$
Assume that $L(g)\to+\infty$  as $g\to\infty$, and set                     
$$
D\xi:=(L(g)\xi_g)_{g\in\Gamma}\quad\text{ for all }\xi=(\xi_g)_{g\in\Gamma}\in\ell^2(\Gamma)\,.
$$ 
Then $(\fH,D)$ is an unbounded  Fredholm module on $C^*_{red}(\Gamma)$. As in the preceding example, we seek to compute $\|[D,a]\|$ for $a\in C^*_{red}(\Gamma)$.
One finds that
$$
\|[D,\lbd(g)]\|=L(g)\,.
$$
(Indeed, one has 
$$
(\lbd(g)D\lbd(g^{-1})\xi)_h=(D\lbd(g^{-1})\xi)_{g^{-1}h}=L(g^{-1}h)(\lbd(g^{-1})\xi)_{g^{-1}h}=L(g^{-1}h)\xi_h\,,
$$
so that
$$
((\lbd(g)D\lbd(g^{-1})-D)\xi)_h=(L(g^{-1}h)-L(h))\xi_h\,,
$$
and
$$
\|[\lbd(g),D]\|=\|\lbd(g)D\lbd(g^{-1})-D\|=\sup_{h\in\Gamma}|L(g^{-1}h-L(h)|=L(g)\,.
$$

\smallskip
Connes' distance in noncommutative geometry is defined via an analogue of the Kantorovich(-Rubinstein) duality formula. Whether there exists a formula involving a notion
of noncommutative coupling or transport plan seems to be an open question at the time of this writing --- see however an interesting contribution to this problem by D'Andrea 
and Martinetti \cite{DAndreaMartinetti}. 

In spite of its great interest, the Connes distance is not exactly the quantum analogue of the Wasserstein distance  which we are looking for. The first example presented above
suggests that the Connes distance is a noncommutative analogue of the classical distance on the space of positions, and not of a phase space distance. Moreover, the Connes
distance is clearly an analogue of the Monge, or Wasserstein distance of exponent $1$, instead of the Wasserstein distance of exponent $2$, for which the optimal transport
problem seems to have more structure (in particular by the Knott-Smith or the Brenier theorems).

\subsection{A Quantum Analogue of $\cW_2$.}\lb{SS-QuantW2}


This section gathers together analogues of the Wasserstein distance $\cW_2$ in the quantum setting introduced in \cite{FGMouPaul,FGPaulARMA}. The presentation given
here is closer in spirit to \cite{FGPaulJFA}.

Other approaches to the problem of generalizing $\cW_2$ to the quantum setting have been proposed by other authors: see for instance \cite{DePalmaTrev} and the earlier
reference \cite{ZycSlo}.

\subsubsection{Transport Cost}

The first step in extending the Wasserstein distance $\cW_2$ to the quantum setting is obviously to find some appropriate definition of the transport cost.

In classical mechanics, the phase space coordinates of a point particle are its position $q\in\bR^d$ and its momentum $p\in\bR^d$.

In quantum mechanics, these coordinates, or functions thereof, must be replaced with appropriate operators. This procedure --- associating operators on a Hilbert space
to functions of the classical phase space coordinates --- is called ``quantization''.

The simplest quantization procedure

\noindent
(i) associates to each function of the position variable only a multiplication operator
$$
a(q)\to\text{ multiplication by }a(y)\text{ in }L^2(\bR^d_y)\,;
$$
(ii) associates to the classical momentum variable $p$ the momentum operator
$$
p\mapsto-i\hb\grad_y\text{ viewed as an unbounded self-adjoint operator on }L^2(\bR^d_y)\,.
$$

The classical transport cost is a function on the Cartesian product of the phase space with itself which is the square Euclidean distance from the phase space point $(x,\xi)$
to the phase space point $(q,p)$ in $\bR^d\times\bR^d$, i.e.
$$
|x-q|^2+|\xi-p|^2\,.
$$

The most natural thing to do is to quantize this expression in the variables $(q,p)$ to measure the cost of transporting a particle from the classical phase space point $(x,\xi)$
to a quantum state with position $y$ and momentum $-i\hb\grad_y$ (whatever it means). This leads to the

\noindent
\textbf{Classical-to-quantum} transport cost, which is an operator on $L^2(\bR^d_y)$:
$$
c_\hb(x,\xi):=|x-y|^2+|\xi+i\hb\grad_y|^2\,.
$$
This is ``the'' quantization\footnote{In truth, there is more than one quantization procedure; we shall see later in this lecture a notion of ``Toeplitz quantization'', which in this 
case would give a slightly different result, with a difference of order $O(\hb)$.} in $(y,\eta)$ of $(q,p)\mapsto|x-q|^2+|\xi-p|^2$.

One can also quantize this expression in both variables $(x,\xi)$ and $(q,p)$ to measure the cost of transporting a quantum particle from a quantum state with position $x$ 
and momentum $-i\hb\grad_x$ to a quantum state with position $y$ and momentum $-i\hb\grad_y$. This leads to the

\noindent
\textbf{Quantum-to-quantum} transport cost, which is an operator on $L^2(\bR^d_x\times\bR^d_y)$:
$$
C_\hb:=|x-y|^2-\hb^2(\grad_x-\grad_y)\cdot(\grad_x-\grad_y)\,.
$$
This is ``the'' quantization of the function $(x,\xi,q,p)\mapsto |x-q|^2+|\xi-p|^2$. 

\smallskip
Notice that $c_\hb(x,\xi)$ is the Hamiltonian of the harmonic oscillator, shifted in phase space by $(x,\xi)$, while $C_\hb$ is the Hamiltonian of a harmonic oscillator in the
variable $x-y$.

\smallskip
Observe that
$$
c_\hb(x,\xi)\ge d\hb I_{\fH}\,,\quad\text{ for all }x,\xi\in\bR^d\,,
$$
while
$$
C_\hb\ge 2d\hb I_{\fH\otimes\fH}\,,
$$
where $\fH=L^2(\bR^d)$. These lower bounds are implied by Heisenberg's uncertainty inequalities.

The exercise below gathers together several important facts related to the Hamiltonian of the harmonic oscillator $y^2-\hb^2\d_y^2$ on the real line.

\smallskip
\noindent
\textbf{Quiz 6.} In this exercise $\fH=L^2(\bR)$. Set 
$$
\om(x):=\pi^{-1/4}e^{-x^2/2}\,,
$$
and
$$
a:=\tfrac1{\sqrt{2}}(x+\d_x)\,,\quad a^*:=\tfrac1{\sqrt{2}}(x-\d_x)\,.
$$
(1) Find $\Ker(a)$, and prove that $\Ker(a)\subset\fH$. Compute the commutator 
$$
[a,a^*]=aa^*-a^*a
$$
together with the operators $aa^*$ and $a^*a$.

Set $\cV:=\{\psi\in H^1(\bR)\text{ s.t. }y\mapsto y\psi(y)\in L^2(\bR)\}$. 

\noindent
(2) Find
$$
\inf_{\psi\in\cV,\,\|\psi\|_{L^2}=1}\int_\bR\overline{\psi(y)}(y^2-\hbar^2\partial_y^2)\psi(y)dy\,.
$$
(3) Find
$$
\inf_{\psi\in\cV,\,\|\psi\|_{L^2}=1}\left(\int_\bR|y|^2|\psi(y)|^2dy\right)^{1/2}\left(\int_\bR|\hbar\d_y\psi(y)|^2dy\right)^{1/2}\,.
$$
(Hint: change $\hbar$ in $\eps\hbar$ in question (2) where $\eps>0$ is arbitrary, and conclude by a minimization argument in $\eps>0$.) 

Question (3) leads to an inequality which is a mathematical formulation of Heisenberg's uncertainty principle (for its physical interpretation, see for instance \S 16 in \cite{LL6}
or chapter I.C.3 in \cite{CTDL}.

\noindent
(4) Find the spectrum of $a^*a$. (Hint: find $\Ker(a^*a)$ by using question (1). Then, argue as in the solution of the following classical exercise in algebra: if $\cA$ is a unital 
algebra with unit denoted by $1$, and if $a,b\in\cA$, then $1-ab$ is invertible in $\cA$ iff $1-ba$ is invertible in $\cA$. To solve this exercise, the idea is to guess a formula
relating $(1-ab)^{-1}$ and $(1-ba)^{-1}$, which can be done easily by writing $(1-x)^{-1}$ as a formal series in powers of $x$.)

\noindent
(5) Compute $a^*af_n$ for each integer $n\ge 0$, where $f_n:=(a^*)^n\om$.

\noindent
(6) Set $\phi_n=f_n/\sqrt{n!}$. Prove that $(\phi_n)_{n\ge 0}$ is an orthonormal system of $\fH$.

\noindent
(7) Prove that  $\phi_n=(-1)^n\om H_n/\sqrt{2^nn!}$ for each integer $n\ge 0$, where
$$
H_n=(-1)^n\om^{-2}\d_x^n\om^2
$$
is the $n$-th Hermite polynomial in the so-called ``physical form''. What is the leading coefficient in $H_n$?

\noindent
(8) Prove that the orthonormal system $(\phi_n)_{n\ge 0}$ is complete in $\fH$.

\noindent
(9) Consider the Fourier transform scaled as follows:
$$
\bF\psi(\xi):=\tfrac1{\sqrt{2\pi}}\int_\bR\psi(x)e^{-i\xi x}dx\,.
$$
Compute $\bF(\phi_n)$ for each integer $n\ge 0$.

\subsubsection{Finite Energy Density Operators}

Before going further in the definition of a quantum analogue of the Wasserstein distance $\cW_2$, we need to define the analogue of $\cP_2(\bR^d\times\bR^d)$.

Using the basic quantization procedure given in the previous section, shows that the phase space Euclidean norm is transformed into the Hamiltonian of a quantum harmonic 
oscillator:
$$
\underbrace{|q|^2+|p|^2}_{\text{phase space}\atop\text{Euclidean norm}}\to\underbrace{|x|^2-\hb^2\Dlt_x}_{\text{harmonic oscillator}}\,.
$$
This suggests the following definition of a quantum analogue to $\cP_2(\bR^d\times\bR^d)$ (the set of Borel probability measures on phase space with finite second order moments):
$$
\cD_2(\fH)\!:=\{R\in\cD(\fH)\text{ s.t. }\Tr_\fH(R^{\frac12}(|x|^2-\hb^2\Dlt_x)R^{\frac12})<\infty\}\,.
$$
In other words, $\cD_2(\fH)$ is the set of density operators with finite energy for the quantum harmonic oscillator.

If $\{\psi_n\!\in\! L^2(\bR^d,|x|^2dx)\cap H^1(\bR^d)\}$ is an orthonormal system in $\fH\!=\!L^2(\bR^d)$, then
$$
R\!=\!\sum_{n\ge 1}\!\rho_n|\psi_n\ra\la\psi_n|\!\in\!\cD_2(\fH)\!\iff\!
\left\{\ba{}&\rho_n\ge 0\text{ and }\sum_{n\ge 1}\rho_n=1\,,\\ &\sum_{n\ge 1}\rho_n(\|x\psi_n\|_\fH^2\!\!+\!\hb^2\|\grad\psi_n\|^2_\fH)\!<\!\infty\,.\ea\right.
$$
The very simple verification of this statement is left to the reader.

\subsubsection{Couplings}

We already know the notion of coupling between two probability densities $\mu$ and $\nu$ belonging to $\cP(\bR^d\times\bR^d)$: the set of such couplings is denoted by
$\cC(\mu,\nu)$.

Similarly, we define the set of \textbf{couplings of $2$ quantum density} operators $R,S\in\cD(\fH)$:
$$
\ba
\cC(R,S):=\{T\in\cD(\fH\otimes\fH)\text{ s.t. }\Tr_\fH(T(A\otimes I+I\otimes B))=\Tr_\fH(RA+SB)\,,
\\
\quad\text{ for all }A,B\in\cL(\fH)\}&\,.
\ea
$$
The condition involving the test operators $A,B\in\cL(\fH)$ can be equivalently replaced by conditions on partial traces:
$$
\Tr_2(T)=R\quad\text{ and }\quad\Tr_1(T)=S\,.
$$

Finally, we define a notion of \textbf{coupling of a classical probability density $f$ and a quantum density operator $R$}. Let $f(x,\xi)$ be a probability density on $\bR^{2d}$
and let $R\in\cD(\fH)$. A coupling of $f$ and $R$ is a measurable\footnote{Since $f\in L^1(\bR^d\times\bR^d)$ and $\Tr_\fH(Q(x,\xi))=\|Q(x,\xi)\|_1=f(x,\xi)<\infty$ for a.e.
$(x,\xi)\in\bR^d\times\bR^d$, one has $Q(x,\xi)\ni\cL^1(\fH)$ for a.e. $(x,\xi)\in\bR^d\times\bR^d$. Since $\cL^1(\fH)$ is separable, the map $Q$ is weakly measurable if
and only if it is strongly measurable by the Pettis Theorem (see Theorem 2 in chapter II of \cite{DiestelUhl}). In other words, there are no measurability issues with $Q$.} 
operator-valued map $Q$ defined on $\bR^d\times\bR^d$, with values in $\cL(\fH)$, satisfying the following properties:
$$
\ba
\bR^{2d}\ni(x,\xi)\mapsto Q(x,\xi)=Q(x,\xi)^*\in\cL(\fH)\text{ s.t. }Q(x,\xi)\ge 0\text{ a.e.}
\\
\Tr_\fH(Q(x,\xi))=f(x,\xi)\text{ a.e.,}\qquad\text{ and }\int_{\bR^{2d}}Q(x,\xi)dxd\xi=R\,.
\ea
$$
In this case again, the set of couplings of $f$ with $R$ will be denoted by $\cC(f,R)$.

\smallskip
Here are trivial examples of couplings (usually not the most clever couplings in optimal transport).

\noindent
\textbf{Examples of couplings.} 

\noindent
(a) For all $R,S\in\cD(\fH)$, the tensor product $R\otimes S$ belongs to $\cC(R,S)$.

\noindent
(b) For each $f$ probability density on $\bR^{2d}$, and each density operator $R$ on $\fH$, the operator-valued map
$$
fR=f\otimes_\bC R:\,(x,\xi)\mapsto f(x,\xi)R
$$
belongs to $\cC(f,R)$. 

\smallskip
In particular
$$
\cC(R,S)\not=\varnothing\quad\text{ and }\quad\cC(f,R)\not=\varnothing\,.
$$
(We have already observed that the set of couplings of two probability measures is never empty, since it always contains the tensor product of these two measures.)

\subsubsection{Extending the Wasserstein Distance to $\fD:=\cP_2(\bR^d\times\bR^d)\cup\cD_2(\fH)$}

We are now ready to define the most important object in these lectures, namely the extension of the Wasserstein $\cW_2$ distance to the (disjoint) union of the sets
of classical and quantum densities.

First we define the (disjoint) union of the set of (classical) Borel probability measures on phase space with finite second order moments, and of the set of finite energy
(quantum) density operators:
$$
\fD:=\cP_2(\bR^d\times\bR^d)\cup\cD_2(\fH)\,.
$$
\textbf{Definition of $\fd$ on $\fD\times\fD$.}

\noindent
(1) For each $\mu,\nu\in\cP_2(\bR^{2d})$, set 
$$
\fd(\mu,\nu):=\cW_2(\mu,\nu)\,.
$$
(2) For each phase space probability density $f$ such that $f(x,\xi)dxd\xi\in\cP_2(\bR^{2d})$ and each $R\in\cD_2(\fH)$, set
$$
\fd(f,R):=\inf_{Q\in\cC(f,R)}\left(\int_{\bR^{2d}}\Tr_\fH(Q(x,\xi)^\frac12 c_\hb(x,\xi)Q(x,\xi)^\frac12)dxd\xi\right)^\frac12\,.
$$
(3) For each $R,S\in\cD_2(\fH)$, set
$$
\fd(R,S):=\inf_{T\in\cC(R,S)}\left(\Tr_{\fH\otimes\fH}(T^\frac12 C_\hb T^\frac12)\right)^\frac12\,.
$$
\textbf{Remark.} For all $f(x,\xi)dxd\xi\in\cP_2(\bR^{2d})$ and all $R,S\in\cD_2(\fH)$, one has
$$
\fd(f,R)\ge\sqrt{d\hb}\qquad\text{ and }\qquad\fd(R,S)\ge\sqrt{2d\hb}\,.
$$
In particular $\fd(R,R)>0$, so that $\fd$ is not a bona fide metric on $\fD$.

\smallskip
\noindent
\textbf{Quiz 7.} Let $\fH$ be a (complex) separable Hilbert space.

\noindent
(1) Let $A,B\in\cL(\fH)$. Prove that
$$
A^*B+B^*A\le |A|^2+|B|^2\,.
$$
(We recall that $|A|^2:=A^*A$.)

\noindent
(2) Prove that, for all $\eps>0$ and all $A,B\in\cL(\fH)$, one has
$$
A^*B+B^*A\le \eps |A|^2+\tfrac1\eps |B|^2\,.
$$
(3) Prove that, for all $\eps>0$ and all $A,B\in\cL(\fH)$, one has
$$
|A+B|^2\le(1+\eps)|A|^2+(1+\tfrac1\eps)|B|^2\,.
$$
(4) Let $A,B$ be Hilbert-Schmidt operators on $\fH$. Deduce from (2) that 
$$
|\Tr_\fH(B^*A)|\le\sqrt{\Tr_\fH(|A|^2)}\sqrt{\Tr_\fH(|B|^2)}\,.
$$
(5) Let $R=R^*\ge 0$ be a trace-class operator on $\fH$. Prove that
$$
|\Tr_\fH(B^*AR)|\le\sqrt{\Tr_\fH(|A|^2R)}\sqrt{\Tr_\fH(|B|^2R)}\,.
$$
(6) Let $f$ be a convex function on $\bR$ (in particular $f\in C(\bR)$), let $R\in\cD(\fH)$, and let $A=A^*\in\cL(\fH)$. Prove that
$$
f(\Tr_\fH(AR))\le\Tr_\fH(f(A)R)\,.
$$
(Hint: prove that, for each $z,m\in\bR$ and for all $\lbd\in[f'_g(m),f'_d(m)]$, one has the inequality $f(z)\ge f(m)+\lbd(z-m)$. Using the spectral measure $(\xi|E(dz)\xi)$ of $A$, where 
$\xi\in\fH$ and $E$ is the spectral decomposition\footnote{See for instance chapter 12 in \cite{RudinFA}.} of $A$, prove that $f(A)\ge f(m)I+\lbd(A-mI)$. Conclude by choosing $m$ 
appropriately.)

\noindent
(7) How should one modify (1) if $A$ and $B$ are unbounded operators on $\fH$?

\noindent
(8) Prove that
$$
\fd(f,R)+\fd(R,S)<\infty
$$
for each probability density $f$ with finite second order moments on $\bR^d\times\bR^d$, and all $R,S\in\cD_2(\fH)$.

\smallskip
In practice, the ``pseudometric'' $\fd$ is not easy to compute, except in a few cases. But before discussing these cases, we need to return to the question of ``quantization''
--- i.e. associating an operator on $L^2(\bR^d)$ to a function on phase space (i.e. $\bR^d\times\bR^d$).

\subsubsection{Toeplitz Operators and Husimi Transform.}

We begin with the definition of a quantum analogue of the Dirac mass at the phase space point $(q,p)\in\bR^d\times\bR^d$. 

\noindent
\textbf{Gaussian wave packet (Schr\"odinger coherent state).} For all $q,p\in\bR^d$, set
$$
|q,p\ra(x):=(\pi\hb)^{-d/4}\exp(-\tfrac1{2\hb}|x-q|^2)\exp(\tfrac{i}{\hb}p\cdot(x-\tfrac{q}2))\,.
$$
This is a plane wave oscillating at frequency $|p|/\hb$ in the direction $p/|p|$, modulated by a Gaussian envelope of width $O(\sqrt{\hb})$, centered at the position $q$.
Therefore, the oscillating profile of the wave function $|q,p\ra$ encodes the momentum vector $p$, while the envelope of the oscillations encodes the position $q\in\bR^d$.

\begin{figure}

\begin{center}

\includegraphics[width=8cm]{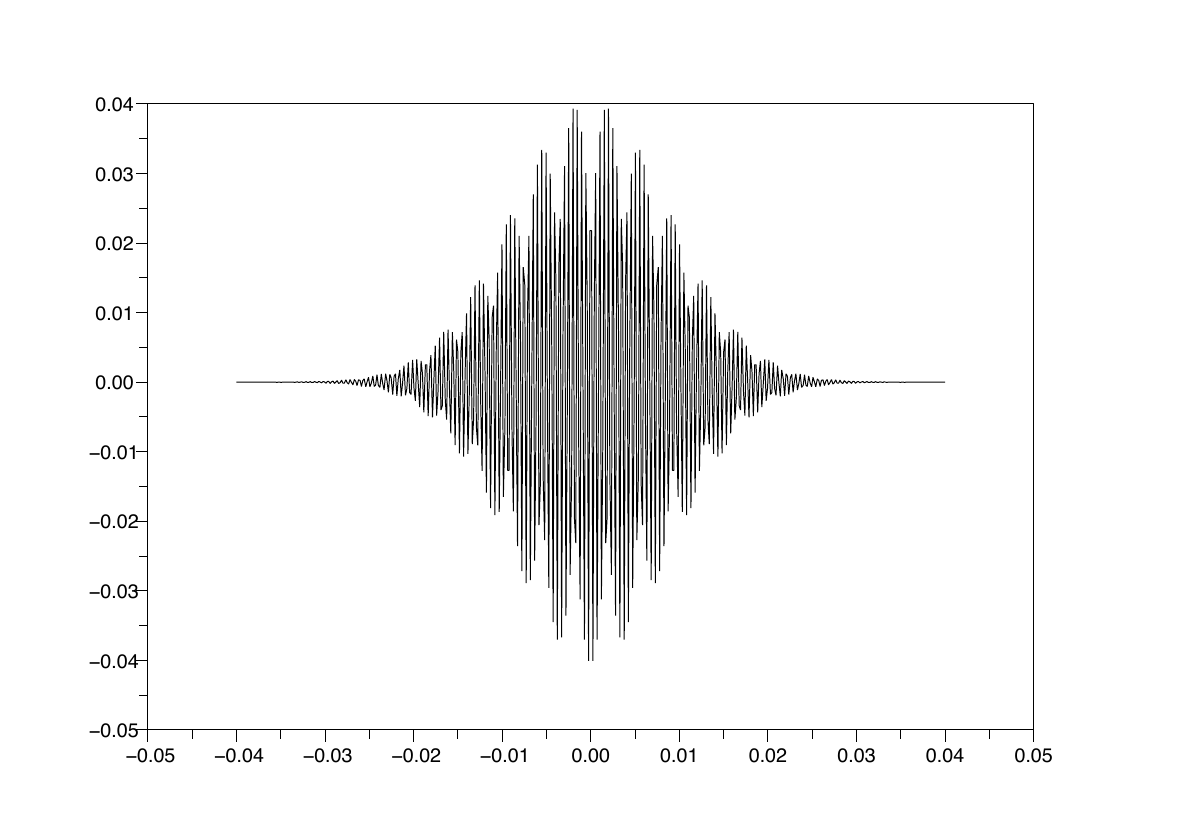}

\caption{Oscillating structure of a Gaussian wave-packet}

\end{center}

\end{figure}

\begin{figure}

\begin{center}

\includegraphics[width=8cm]{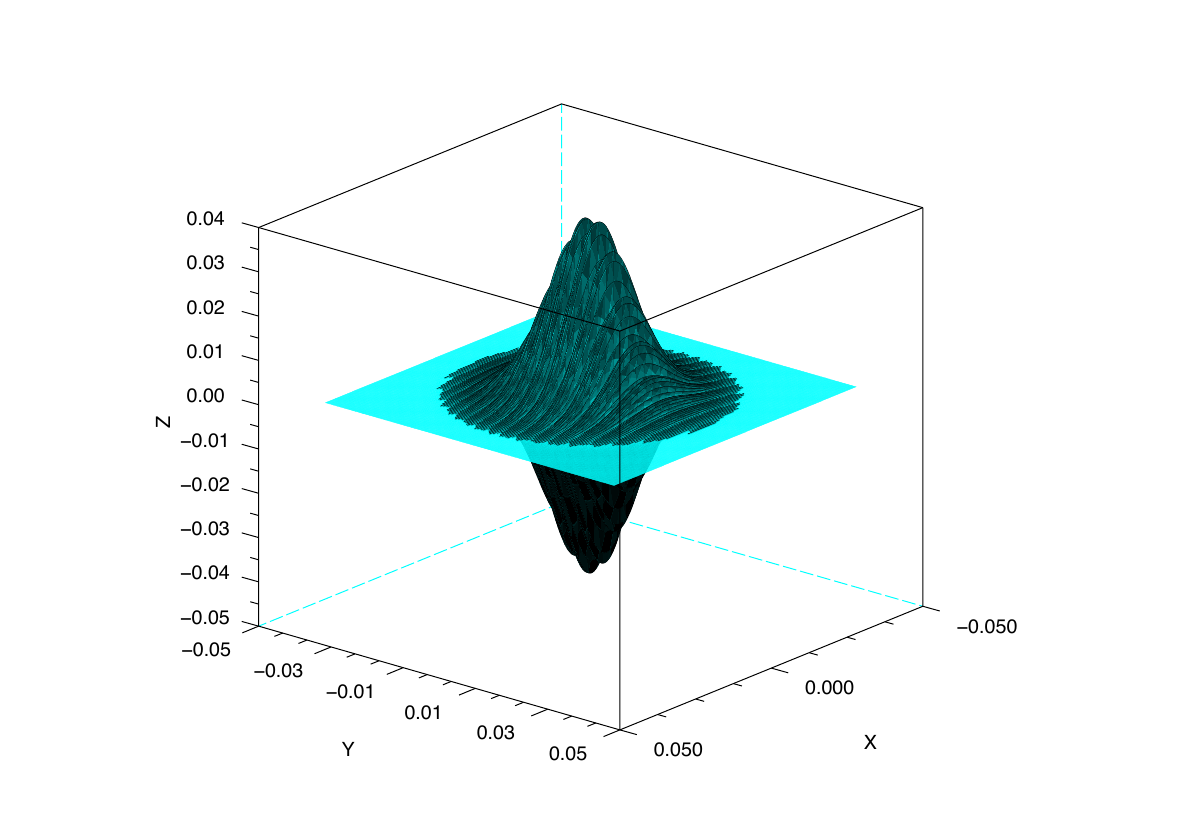}

\caption{With $\hb=8\cdot 10^{-5}$, plot of $Z=$ real part of the coherent state centered at $q=(0,0)$ with momentum $p=(1,0)$ with space variable $(X,Y)\in\bR^2$}

\end{center}

\end{figure}

\noindent
\textbf{Toeplitz map.} To $m$, a (complex) Radon measure on the phase space $\bR^d\times\bR^d$, one associates the operator
$$
\cT[m]:=\int_{\bR^d}|q,p\ra\la q,p|m(dqdp)\,.
$$
The form-domain of $\cT[m]$ is the set of $\phi\in\fH$ such that the function $(q,p)\mapsto\la q,p|\psi\ra$ belongs to $L^2(\bR^{2d};m)$.

Observe that $|q,p\ra\la q,p|$ is a self-adjoint positive operator (specifically, a rank-$1$ projection in $\fH$). Therefore, if the measure $m$ is real-valued, the operator 
$\cT[m]$ (with some appropriate domain) is expected to be self-adjoint, and if $m$ is a positive measure on phase space, the associated Toeplitz operator $\cT[m]$
is expected to be a positive operator on $\fH$ (possibly unbounded, with some appropriate domain).

\smallskip
\noindent
\textbf{Basic properties of the Toeplitz map.}

\noindent
(1) The set of Schr\"odinger coherent states is a resolution of the identity:
$$
\cT[1]=\int_{\bR^{2d}}|q,p\ra\la q,p|dqdp=(2\pi\hb)^dI_\fH\,.
$$
(2) The image by the Toeplitz map $\cT$ of the set of Borel phase space probability measure on $\bR^d\times\bR^d$ is included in $\cD(L^2(\bR^d))$:
$$
m\in\cP(\bR^d\times\bR^d)\implies\cT[m]\in\cD(\fH)\,,\quad\text{ with }\fH=L^2(\bR^d)\,.
$$
(3) One has 
$$
\cT[q]=(2\pi\hb)^dx\text{ (position operator),}
$$
while 
$$
\cT[p]=(2\pi\hb)^d(-i\hb\grad_x)\text{ (momentum operator).}
$$
(4) If $f$ is a quadratic form on $\bR^d$, then
$$
\left\{
\ba
\cT[f(q)]&=(2\pi\hb)^d\left(f(x)+\tfrac14\hb(\Dlt f)I_\fH\right)\,,\text{ and}
\\
\cT[f(p)]&=(2\pi\hb)^d\left(f(-i\hb\grad_x)+\tfrac14\hb(\Dlt f)I_\fH\right)\,.
\ea
\right.
$$

\smallskip
\noindent
\textbf{Quiz 8.} Prove the statements (1)-(4) above. (Hint: recall the formulas for moments of order $\le 2$ of Gaussian distributions, together with the oscillating
integrals
$$
\tfrac1{2\pi}\int_{\bR^d}e^{ip\cdot(x-y)}dp=\de_0(x-y)\,,\qquad\tfrac1{2\pi}\int_{\bR^d}p_je^{ip\cdot(x-y)}dp=-i\d_{x_j}\de_0(x-y)\,,
$$
and
$$
\tfrac1{2\pi}\int_{\bR^d}p_jp_ke^{ip\cdot(x-y)}dp=-\d_{x_j}\d_{x_k}\de_0(x-y)\,,\qquad j,k=1,\ldots,d
$$
--- which are to be understood in the sense of tempered distributions. For instance, the first formula above is equivalent to the Fourier inversion formula on
the set of tempered distributions.)

\smallskip
The Toeplitz map associates an operator on $\fH=L^2(\bR^d)$ to a function on the phase space $\bR^d\times\bR^d$. Conversely, given an operator on $\fH$, 
we seek to associate a function on phase space. There are various ways of doing this, one of which is the Husimi transform.

\noindent
\textbf{Husimi Transform.} To $T\in\cL(\fH)$, one associates its Husimi transform
$$
\cH[T](q,p):=\tfrac1{(2\pi\hb)^d}\la q,p|T|q,p\ra\,.
$$
This definition can be extended to all unbounded operators on $T$ such that the Gaussian wave packet $|q,p\ra$ belongs to the form domain of $T$ for each
$q,p\in\bR^d$.

\smallskip
\noindent
\textbf{Basic properties of the Husimi transform.}

\noindent
(1) For each $T\in\cL(\fH)$,
$$
T=T^*\implies\cH[T](q,p)\in\bR\,,\text{ and }T\ge 0\implies\cH[T]\ge 0\,.
$$
(2) The Husimi transform is an ``almost inverse'' of the Toeplitz map: for each $m\in\cP(\bR^d\times\bR^d)$
$$
\cH[\cT[m]]=e^{\frac\hb{2}\Dlt_{q,p}}m\,,
$$
since
$$
\la q,p|q',p'\ra=e^{-\frac1{4\hb}(|q-q'|^2+|p-p'|^2)}e^{-\frac{i}\hb(p\cdot q'-q\cdot p')}\,.
$$
(3) One has
$$
\cH[I]=(2\pi\hb)^{-d}\,,
$$
while
$$
\left\{
\ba
{}&\cH[f(x)](q,p)=(2\pi\hb)^{-d}(I+\tfrac14\hb\Dlt)f(q)\,,\text{ and }
\\
&\cH[f(-i\hb\grad_x)](q,p)=(2\pi\hb)^{-d}(I+\tfrac14\hb\Dlt)f(p)\,.
\ea
\right.
$$
(4) One has
$$
\Tr_\fH(R^*\cT[f])=(2\pi\hb)^d\iint_{\bR^d\times\bR^d}\overline{\cH[R](q,p)}f(q,p)dqdp\,.
$$

\smallskip
\noindent
\textbf{Quiz 9.} Prove the statements (1)-(4) above.

\smallskip
An important property of the Husimi transform is that it is a one-to-one transformation. In other words, the Husimi transform of an operator specifies it completely.

\newpage
\noindent
\textbf{Quiz 10.} Set $\fH:=L^2(\bR^d)$.

\noindent
(1) Let $R\in\cD_2(\fH)$. Prove that $\cH[R]$ is a probability density, and compute
$$
\iint_{\bR^d\times\bR^d}(|q|^2+|p|^2)\cH[R](q,p)dqdp\,.
$$
(2) Let $R,S\in\cD_2(\fH)$, and assume that $\cH[R]=\cH[S]$. Prove that $R=S$. (Hint: let $r\equiv r(y,y')$ be an integral kernel of $R$. Set
$$
J(x,\xi)=\iint_{\bR^d\times\bR^d}r(y,y')e^{-(|y|^2+|y'|^2)/2\hb}e^{x\cdot(y+y')-i\xi\cdot(y-y')/\hb}dydy'\,.
$$
Prove that $J$ extends as a holomorphic function on $\bC^d\times\bC^d$, and therefore is uniquely determined by its restriction to $\bR^d\times\bR^d$. Conclude 
by (a) computing the formula relating $\cH[R]$ to $J$, and (b) by computing the integral kernel $r$ of $R$ in terms of $J$.)

\smallskip
Here is another way of associating a function on the phase space $\bR^d\times\bR^d$ to an operator on $L^2(\bR^d)$.

\noindent
\textbf{Quiz 11.} Set $\fH:=L^2(\bR^d)$. To each $A\in\cL^1(\fH)$ with integral kernel $a\equiv a(x,y)$ such that $z\mapsto a(x+z,x)$ belongs to $C_b(\bR^d;L^1(\bR^d))$
(see Quiz 4), we associate its Wigner transform
$$
W[A](x,\xi):=\tfrac1{(2\pi)^d}\int_{\bR^d}a(x+\tfrac12\hb y,x-\tfrac12\hb y)e^{-i\xi\cdot y}dy
$$
(where the integral above is to be understood as the partial Fourier transform of the continuous bounded function $y\mapsto a(x+\tfrac12\hb y,x-\tfrac12\hb y)$ with
values in $L^1(\bR^d_x)$, which is therefore a tempered distribution).

\noindent
(1) Prove that 
$$
\overline{W[A]}=W[A^*]\,,
$$
and that, for each $A,B\in\cL^1(\fH)$
$$
\Tr_\fH(A^*B)=(2\pi\hb)^d\iint_{\bR^d\times\bR^d}\overline{W[A](x,\xi)}W[B](x,\xi)dxd\xi\,,
$$
and
$$
\|A\|_2=(2\pi\hb)^{d/2}\|W[A]\|_{L^2(\bR^d\times\bR^d)}\,.
$$
Prove that the Wigner transform has a unique extension to $\cL^2(\fH)$.

\noindent
(2) Prove that, for each $A\in\cL^1(\fH)$,
$$
\Tr_\fH(A)=\iint_{\bR^d\times\bR^d}W[A](x,\xi)dxd\xi\,.
$$

\noindent
(3) Let $t\mapsto R(t)$ be a time-dependent density operator, solution of the von Neumann equation
$$
i\hb\d_tR(t)=[-\tfrac12\hb^2\Dlt+V,R(t)]\,.
$$
Prove that $W[R(t)]$ is a solution of the Wigner equation
$$
(\d_t+\xi\cdot\grad_x)W[R(t)](x,\xi)+\Theta[V]W[R(t)](x,\xi)=0\,,
$$
where $\Theta[V]$ is the linear operator with distribution kernel
$$
\tfrac1{(2\pi)^d}\int_{\mathbf R^d}\tfrac1{i\hb}(V(x+\tfrac12\hb y)-V(x-\tfrac12\hb y))e^{iy\cdot(\eta-\xi)}dy\,.
$$
Prove that 
$$
\Theta[V]=-\grad V(x)\cdot\grad_\xi
$$
in the case where $V$ is a polynomial of degree $2$.

In that case (for a quadratic potential $V$), the Wigner equation coincides with the classical Liouville equation on the phase space $\bR^d\times\bR^d$
$$
\d_tW[R(t)](x,\xi)+\{\tfrac12|\xi|^2+V(x),W[R(t)](x,\xi)\}=0\,.
$$
(4) However, one cannot think of $W[A]$ as a distribution function as in the kinetic theory of gases. Indeed
$$
0\le A=A^*\in\cL^1(\fH)\text{ does not imply }W[A]\ge 0\,.
$$
To see this, compute $W[|\psi\ra\la\psi|](0,0)$ where $\psi(x)=\sqrt{2}\pi^{-1/4}xe^{-x^2/2}$ for $x\in\bR$.

\noindent
(5) The relation between the Wigner and the Husimi transform is given by the following formula: for each $A\in \cL^1(\fH)$, one has
$$
\cH[A](x,\xi)=\exp(\tfrac{\hb}4\Dlt_{x,\xi})W[A](x,\xi)\,.
$$
(6) Let $\psi\in L^2(\bR)$ satisfy $\|\psi\|_{L^2(\bR)}=1$ and $W[|\psi\ra\la\psi|]\ge 0$. Prove that there exist $q_0,p_0\in\bR^d$ and $u\in\bC$ with
$|u|=1$ such that $\psi=u|q_0,p_0\ra$. (Hint: prove that 
$$
F(z):=\int_\bR\psi(x)e^{-\frac12x^2-zx}dx
$$
defines an entire function on $\bC$, that 
$$
0<|F(z)|^2\le Ce^{\Re(z)^2}\,,\qquad z\in\bC\,,
$$
for some constant $C>0$. Conclude by Hadamard's theorem\footnote{See for instance chapter 5.3.2 in \cite{Ahlfors}.} that $F(z)=e^{g(z)}$ where $g$ is a polynomial of degree $2$.)

\subsubsection{Explicit Computations/Estimates}

We have gathered together in this section several useful explicit computations, or bounds, on the ``pseudometric'' $\fd$.

\smallskip
\noindent
\textbf{Theorem 1.}

\noindent
(1) For all $f,g$ probability densities on $\bR^{2d}$ with finite 2nd order moments,
$$
\ba
\fd(\cT[f],\cT[g])^2&\!\le\!\cW_2(f,g)^2\!+\!2d\hb,\qquad&&\fd(\cT[f],\cT[f])=\sqrt{2d\hb}\,,
\\
\fd(f,\cT[g])^2&\!\le\!\cW_2(f,g)^2\!+\!d\hb,\qquad&&\quad\,\,\,\fd(f,\cT[f])=\sqrt{d\hb}\,.
\ea
$$
(2) For all $R,S\in\cD_2(\fH)$ and all probability density $f$ on $\bR^{2d}$ with finite 2nd order moments,
$$
\ba
\cW_2(\cH[R],\cH[S])^2&\le\fd(R,S)^2+2d\hb\,,
\\
\cW_2(f,\cH[R])^2&\le\fd(f,R)^2+d\hb\,.
\ea
$$
(3) Moreover, if $\text{rank}(R)=1$, then
$$
\ba
\fd(R,S)&=\Tr_{\fH\otimes\fH}((R\otimes S)^\frac12 C_\hb(R\otimes S)^\frac12)^\frac12\,,\quad\text{ and }
\\
\fd(f,R)&=\left(\int_{\bR^{2d}}f(x,\xi)\Tr_\fH(R^\frac12 c_\hb(x,\xi)R^\frac12)dxd\xi\right)^\frac12\,.
\ea
$$

\smallskip
\noindent
\textbf{Remark.} The second inequality in (1) can be recast as
$$
\fd(f,\cT[g])^2\le\fd(f,g)^2+\fd(g,\cT[g])^2
$$
since
$$
\fd(g,\cT[g])=\sqrt{d\hb}=\min\fd\,.
$$
This suggests that

\noindent
(1) ``the segment $[g,\cT[g]]$ is orthogonal to the set of classical densities'', and

\noindent
(2) the ``angle'' $\th$ between the ``segment'' $[g,f]$ and the ``segment'' $[g,\cT(g)]$ in $\fD_2$ is acute.

\smallskip
Therefore, one could think of $\cP_2(\bR^d\times\bR^d)$ as a limit set (for the classical limit $\hb\to 0$) --- or boundary --- of $\fD$, and that the set $\cD(\fH)$ of quantum densities 
lies on the ``concave'' side of the set of classical densities.

\begin{figure}

\begin{center}

\includegraphics[width=8cm]{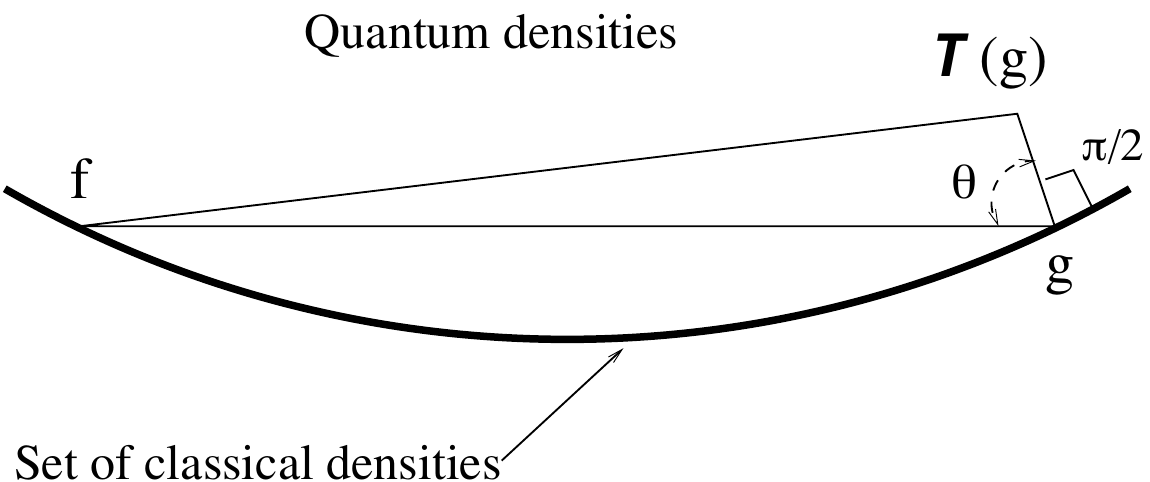}

\caption{A geometric interpretation of Theorem 1 (1).}

\end{center}

\end{figure}

\smallskip
\noindent
\textit{Proof of Theorem 1.} 

\noindent
\underline{Proof of (1).} Let $\grad\Phi$ (with $\Phi$ convex) be the Brenier map pushing $f$ to $g$. The optimal coupling of $f$ and $g$ for $\cW_2$ is 
$$
\Lambda:=f(x,\xi)\de_{\grad\Phi(x,\xi)}(dyd\eta)dxd\xi\,.
$$
Hence
$$
\cT[\Lambda]\!\in\!\cC(\cT[f],\!\cT[g])\quad\text{ and }\quad(x,\xi)\!\mapsto\!f(x,\xi)\cT[\de_{\grad\Phi(x,\xi)}]\!\in\!\cC(f,\!\cT[g])\,.
$$
On the other hand (see Appendix B, and especially formulas (52)-(53), in \cite{FGMouPaul})
$$
\ba
\cH[C_\hb](q,p,q',p')&=(2\pi\hb)^{-2d}(|q-q'|^2+|p-p'|^2+2d\hb)\,,
\\
\cH[c_\hb(x,\xi)](q,p)&=(2\pi\hb)^{-d}(|x-q|^2+|\xi-p|^2+d\hb)\,.
\ea
$$
Therefore
$$
\ba
\fd(\cT[f],\cT[g])^2\le&\Tr_{\fH\otimes\fH}(\cT[\Lambda]^\frac12 C_\hb\cT[\Lambda]^\frac12)
\\
=&\int_{\bR^{4d}}\underbrace{(|q-q'|^2+|p-p'|^2+2d\hb)}_{=(2\pi\hb)^{2d}\overline{\cH[C_\hb](q,p)}}\Lambda(dqdpdq'dp')
\\	\\
=&\cW_2(f,g)^2+2d\hb\,,
\ea
$$
and
$$
\ba
\fd(f,\cT[g])^2\le&\int_{\bR^{2d}}\Tr_\fH(\cT[\de_{\grad\Phi(x,\xi)}]^\frac12c_\hb(x,\xi)\cT[\de_{\grad\Phi(x,\xi)}]^\frac12)f(x,\xi)dxd\xi
\\
=&\int_{\bR^{2d}}\underbrace{(|(x,\xi)-\grad\Phi(x,\xi)|^2+d\hb)}_{=(2\pi\hb)^d\overline{\cH[c_\hb(x,\xi)](\grad\Phi(x,\xi))}}f(x,\xi)dxd\xi
\\	\\
=&\cW_2(f,g)^2+d\hb\,.
\ea
$$
\underline{Proof of (2).} Pick sequences $a_n,b_n\in C_b(\bR^{2d};\bR)$ such that
$$
\ba
a_n(q,p)+b_n(q',p')\le|q-q'|^2+|p-p'|^2\,,\quad\text{ and}
\\
\underbrace{\int_{\bR^{2d}}a_n(q,p)\cH[R](q,p)dqdp}_{=(2\pi\hb)^{-d}\Tr_\fH(\cT[a_n]R)}+\underbrace{\int_{\bR^{2d}}b_n(q',p')\cH[S](q',p')dq'dp'}_{=(2\pi\hb)^{-d}\Tr_\fH(\cT[b_n]S)}
\\	\\
\to\cW_2(\cH[R],\cH[S])^2
\ea
$$
as $n\to\infty$. That such sequences exist is a consequence of the Kantorovich duality formula for the Wasserstein distance $\cW_2$.

On the other hand, for each $T\in\cC(R,S)$, one has
$$
\ba
(2\pi\hb)^{-d}\Tr_\fH(\cT[a_n]R)+(2\pi\hb)^{-d}\Tr_\fH(\cT[b_n]S)
\\
=(2\pi\hb)^{-d}\Tr_{\fH\otimes\fH}(T^\frac12(\cT[a_n]\otimes I+I\otimes\cT[b_n])T^\frac12)
\\
=(2\pi\hb)^{-2d}\Tr_{\fH\otimes\fH}(T^\frac12\cT[a_n\otimes 1+1\otimes b_n]T^\frac12)
\\
\le(2\pi\hb)^{-2d}\Tr_{\fH\otimes\fH}(T^\frac12\cT[|q\!-\!q'|^2\!+\!|p\!-\!p'|^2]T^\frac12)&\,.
\ea
$$
Now, one has (see the basic properties of the Toeplitz map and Quiz 8 above)
$$
\cT[|q\!-\!q'|^2\!+\!|p\!-\!p'|^2]=(2\pi\hb)^{2d}(C_\hb+2d\hb I_{\fH\otimes\fH})\,.
$$
Thus, for all $T\in\cC(R,S)$, one has
$$
\ba
\cW_2(\cH[R],\cH[S])^2=&\lim_{n\to\infty}(2\pi\hb)^{-d}\left(\Tr_\fH(\cT[a_n]R)+\Tr_\fH(\cT[b_n]S)\right)
\\
\le&\Tr_{\fH\otimes\fH}(T^\frac12(C_\hb+2d\hb I_{\fH\otimes\fH})T^\frac12)
\\
=&\Tr_{\fH\otimes\fH}(T^\frac12C_\hb T^\frac12)+2d\hb\,.
\ea
$$
Minimizing the r.h.s. in $T\in\cC(R,S)$ leads to
$$
\cW_2(\cH[R],\cH[S])^2\le\fd(R,S)^2+2d\hb\,.
$$
\underline{Proof of (3).} We begin with a question of a rather fundamental nature in quantum mechanics.

\noindent
\textbf{Question.} What is the structure of couplings for rank-$1$ density operators?

\smallskip
This question is answered by the following lemma.

\smallskip
\noindent
\textbf{Lemma 2.}  Let $R\in\cD(\fH)$. Then
$$
\text{rank}(R)=1\implies\left\{\ba{}&\cC(f,R)=\{fR\}\,,\quad f\!\in\!\cP(\bR^d\!\times\!\bR^d)\,,\\ &\cC(R,S)=\{R\otimes S\}\,,\qquad S\in\cD(\fH)\,.\ea\right.
$$

\smallskip
Obviously Lemma 2 implies statement (3) in Theorem 1.

\rightline{$\Box$}

\smallskip
\noindent
\textit{Proof of Lemma 2.} Since $\text{rank}(R)=1$, it is of the form $R=|\phi\ra\la\phi|$, with $\|\phi\|_\fH=1$. We shall prove the second statement in the lemma. Let $Q\in\cC(R,S)$. Then
$$
\ba
\Tr(((I-R)\otimes I)Q((I-R)\otimes I))=\Tr(Q((I-R)^2\otimes I))
\\
=\Tr(Q((I-R)\otimes I))=\Tr(R(I-R))=0&\,,
\ea
$$
Since
$$
((I-R)\otimes I)Q((I-R)\otimes I)\ge 0\,,
$$
one has
$$
((I-R)\otimes I)Q((I-R)\otimes I)=0\,.
$$
Next, we deduce from the Cauchy-Schwarz inequality that
$$
\ba
|\la\psi_1\otimes\psi_2|(R\otimes I)Q((I-R)\otimes I)\psi'_1\otimes\psi'_2\ra|^2
\\
\le\la\psi'_1\otimes\psi'_2|((I-R)\otimes I)Q((I-R)\otimes I)\psi'_1\otimes\psi'_2\ra
\\
\times\la\psi_1\otimes\psi_2|(R\otimes I)Q(R\otimes I)\psi_1\otimes\psi_2\ra&\,.
\ea
$$
Hence
$$
\ba
(R\otimes I)Q((I-R)\otimes I)=0&=((R\otimes I)Q((I-R)\otimes I))^*
\\
&=((I-R)\otimes I)Q(R\otimes I)\,,
\ea
$$
Since we already know that $((I-R)\otimes I)Q((I-R)\otimes I)=0$, this implies that
$$
Q=(R\otimes I)Q(R\otimes I)\,.
$$
Therefore $Q=R\otimes T$, where
$$
\la\psi|T|\psi'\ra:=\la\phi\otimes\psi|Q|\phi\otimes\psi'\ra\,.
$$
Finally, $T=S$, since, for all $A\in\cL(\fH)$, one has
$$
\Tr(SA)=\Tr(Q(I\otimes A))=\Tr((R\otimes T)(I\otimes A))=\Tr(TA)\,.
$$
This proves the second statement in the lemma.

\rightline{$\Box$}

\smallskip
\noindent
\textbf{Quiz 12.} Complete the proof of Lemma 2: prove that $\cC(f,R)=\{f\otimes_\bC R\}$, in the case where $f$ is a probability density on $\bR^d\times\bR^d$ and 
$R$ is a rank-$1$ density operator on $\fH=L^2(\bR^d)$.

\smallskip
\noindent
\textbf{Remark.} It is well known that, for all $\mu\in\cP(\bR^d\times\bR^d)$ and all $(q,p)\in\bR^d\times\bR^d$, the set of couplings of $\mu$ with $\de_{(q,p)}$ contains
only one element:
$$
\cC(\mu,\de_{(q,p)})=\{\mu\otimes\de_{(q,p)}\}\,.
$$
Lemma 2 suggests that \textbf{all} rank-$1$ density operators are quantum analogues of the Dirac mass in phase space. Thus the Schr\"odinger equation governing 
the evolution of the wave function $\psi(t,x)$ --- or the von Neumann equation specialized to $|\psi(t,\cdot)\ra\la\psi(t,\cdot)|$ ---  is the quantum analogue of Newton's 
second law of motion in classical mechanics, which can be viewed as the equation governing $\de_{(q(t),p(t))}$, where $q(t)$ and $p(t)$ are respectively the position 
and the momentum of a moving classical particle. This analogy also explains why the wave function is a purely quantum object, which has no classical analogue. 
Indeed, if a classical analogue of the wave function existed, it could be thought of as a ``square root'' of the phase space Dirac measure $\de_{(q(t),p(t))}$.

\smallskip
\noindent
\textbf{Quiz 13: another proof of Theorem 1 (2).}

\noindent
(1) Prove that 
$$
\cT[|q-q'|^2+|p-p'|^2]=(2\pi\hb)^{2d}(C_\hb+2d\hb I_{\fH\otimes\fH})\,.
$$
(2) For each $T\in\cC(R,S)$, prove that
$$
\ba
\Tr_{\fH\otimes\fH}\left(T^\frac12C_\hb T^\frac12\right)+2d\hb\ge&\tfrac1{(2\pi\hb)^d}\Tr_{\fH\otimes\fH}\left(T\cT\left[\tfrac{|q-q'|^2+|p-p'|^2}{1+\eps|q-q'|^2+\eps|p-p'|^2}\right]\right)\
\\
=&\int_{\bR^{4d}}\cH[T](q,p,q',p')\tfrac{|q-q'|^2+|p-p'|^2}{1+\eps|q-q'|^2+\eps|p-p'|^2}dqdpdq'dp'\,.
\ea
$$
(3) Conclude by monotone convergence in the right-hand side of the inequality above as $\eps\to 0^+$, after observing that $\cH[T]$ is a coupling of $\cH[R]$ and $\cH[S]$.

\subsection{Quantum Optimal Transport is Cheaper!}\lb{SS-QOTCheap}

To conclude this first lecture, we shall study an example where $\fd$ can be computed essentially explicitly,  with interesting implications on the structure of optimal
couplings. The material in this section --- together with the somewhat provocative title --- is taken from \cite{CaglioFGPaul}.

We begin with a simple lemma, which can be viewed as an amplification of statement (1) in Theorem 1.

\noindent
\textbf{Lemma 3.} For $\rho_1,\rho_2\in\cP_2(\bR^d\times\bR^d)$ with optimal coupling $\Pi$ for $\cW_2$, one has
$$
\ba
\fd(\cT[\rho_1],\cT[\rho_2])^2=\cW_2(\rho_1,\rho_2)^2+2d\hb
\\
\iff\cT[\Pi]\in\cC(\cT[\rho_1],\cT[\rho_2])\text{ is a quantum optimal coupling for }\fd&\,.
\ea
$$

The proof of this lemma is left to the reader as an easy exercise (see the proof of Theorem 1).

\smallskip
Here is an example of this kind of situation. With $d=1$ and $0<a<b$, set 
$$
\mu:=\tfrac12(\de_{(+a,0)}+\de_{(-a,0)})\,\text{ and }\,\nu:=\tfrac12(\de_{(+b,0)}+\de_{(-b,0)})\in\cP_2(\bR\times\bR)\,.
$$
In other words, we have equal masses (1/2), but different positions --- since $0<a<b$.

\noindent
\textbf{Proposition 4.} One has
$$
\fd(\cT[\mu],\cT[\nu])^2=\cW_2(\mu,\nu)^2+2\hb\,.
$$

The classical optimal transport in this case is obvious: send mass $1/2$ from $-a$ to $-b$, and mass $1/2$ from $+a$ to $+b$.

\begin{figure}
\centering
\begin{tikzpicture}[xscale=2.5,yscale=2.5]
\draw [-] (0,0) -- (2,0);
\draw[-] (0,1) -- (2,1);
\draw[-] (.5,0)--(0.2,1);
\draw[-] (1.5,0)--(1.8,1);
\fill (0.5,0) circle[radius=2pt] node[label=below:$x_1$ ${,}$ $\frac12$,draw]{};
\fill (1.5,0) circle[radius=2pt] node[label=below:$x_2$ ${,}$ $\frac12$,draw]{};
\fill (0.2,1) circle[radius=2pt] node[label=above:$y_1$ ${,}$ $\frac12$,draw]{};
\fill (1.8,1) circle[radius=2pt] node[label=above:$y_2$ ${,}$ $\frac12$,draw]{};
\end{tikzpicture}
\begin{tikzpicture}[xscale=2.5,yscale=2.5]
\draw [-] (0,0) -- (2,0);
\draw[-] (0,1) -- (2,1);
\draw[-] (0.5,0)--(0.5,1);
\draw[-] (1.5,0)--(1.5,1);
\draw[-] (1.5,0)--(0.5,1);
\fill (0.5,0) circle[radius=2pt] node[label=below:$x_1$ ${,}$ $\frac{1-\eps}2$,draw]{};
\fill (1.5,0) circle[radius=2pt] node[label=below:$x_2$ ${,}$ $\frac{1+\eps}2$,draw]{};
\fill (0.5,1) circle[radius=2pt] node[label=above:$y_1$ ${,}$ $\frac12$,draw]{};
\fill (1.5,1) circle[radius=2pt] node[label=above:$y_2$ ${,}$ $\frac12$,draw]{};
\end{tikzpicture}

\caption{Left: equal masses; Right: unequal mass case} \label{f1}
\end{figure}
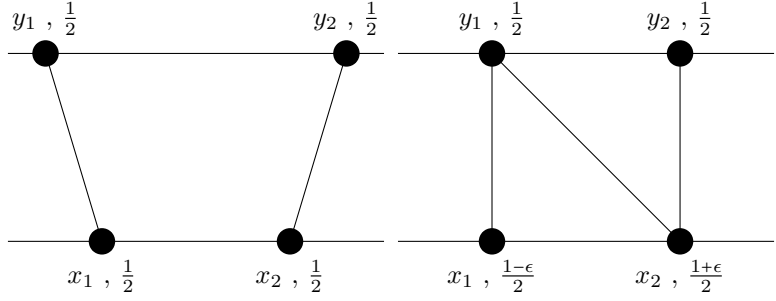

\smallskip
Next we consider the following example, with identical locations, but unequal masses: for $0<\eps<1$, set 
$$
\mu=\tfrac12(\de_{(+a,0)}+\de_{(-a,0)})\,\text{ and }\,\rho_\eps=\tfrac{1+\eps}2\de_{(+a,0)}+\tfrac{1-\eps}2\de_{(-a,0)}\in\cP_2(\bR\times\bR)\,.
$$
\textbf{Proposition 5.} For each $\eps\in(0,1)$, one has
$$
\fd(\cT[\mu],\cT[\rho_\eps])^2<\cW_2(\mu,\rho_\eps)^2+2\hb\,.
$$

\smallskip
We shall not prove Propositions 4 and 5, which rely on rather intricate computations, and refer instead the interested reader to the article \cite{CaglioFGPaul}. However, we shall discuss both results.

\smallskip
Because of Lemma 3, an optimal coupling for $\cT[\mu]$ and $\cT[\nu]$ is $\cT[\lbd]$, where $\lbd$ is the optimal coupling of $\mu$ and $\nu$, i.e.
$$
\lbd:=\tfrac12(\de_{(+a,0)}\otimes\de_{(+b,0)}+\de_{(-a,0)}\otimes\de_{(-b,0)})\,,
$$
so that
$$
\lbd:=\tfrac12(|+a,0,+b,0\ra\la +a,0,+b,0|+|+a,0,+b,0\ra\la +a,0,+b,0|)\,,
$$
with the notation
$$
|q,p,q',p'\ra(x,y)=|q,p\ra\otimes|q',p'\ra(x,y)=|q,p\ra(x)|q',p'\ra(y)\,.
$$

In the unequal mass case, it is proved in \cite{CaglioFGPaul} that there exists a quantum coupling of the form
$$
\begin{aligned}
T=&\sum_{k,l\in\{\pm\}}\tau_{klkl}|ka,0,la,0\ra\la ka,0,la,0|
\\
&+\sum_{(k,l)\not=(m,n)\in\{\pm\}}\tau_{klmn}|ka,0,la,0\ra\la ma,0,na,0|\in\cC(\cT[\mu],\cT[\rho_\eps])\,,
\end{aligned}
$$
with
$$
\sum_{(k,l)\not=(m,n)\in\{\pm\}}|\tau_{klmn}|^2>0\,,
$$
and
$$
\Tr(T^{1/2}C_\hb T^{1/2})<\cW_2(\mu,\rho_\eps)^2+2\hb\,.
$$

Clearly, any coupling of $\mu$ and $\rho_\eps$ must be of the form
$$
\sum_{k,l\in\{\pm\}}q_{kl}\de_{(ka,0)}\otimes\de_{(la,0)}\,,
$$
and therefore belongs to the $4$-dimensional linear space 
$$
\text{span}\{\de_{(\pm a,0)}\otimes\de_{(\pm a,0)}\}\,.
$$
On the contrary, couplings of two rank-$2$ operators with 
$$
R=\tfrac12(|+a,0\ra\la+a,0\ra+|-a,0\ra\la-a,0|)
$$
and
$$
S_\eps=\tfrac{1+\eps}2|+a,0\ra\la+a,0|+\tfrac{1-\eps}2|-a,0\ra\la-a,0|
$$
with $0<\eps<1$ belong to the $16$-dimensional linear space 
$$
\text{span}\{|ka,0,la,0\ra\la m,a,0,na,0|\,:\,k,l,m,n\in\{\pm\}\}\,.
$$

Therefore, one could summarize the results in \cite{CaglioFGPaul} as follows: since there are more degrees of freedom in the set of quantum couplings than in the set of
classical couplings, it is natural to surmise that quantum optimal transport is cheaper than classical optimal transport, since more couplings are allowed in the quantum 
case than in the classical case. However, this simple argument is not sufficient to prove a \textit{strict} inequality as in \cite{CaglioFGPaul}. Putting together Proposition 5 
and Lemma 3 shows that any optimal coupling for $\fd(\cT[\mu],\cT[\rho_\eps])$ must be of the form
$$
\begin{aligned}
T=&\sum_{k,l\in\{\pm\}}\tau_{klkl}|ka,0,la,0\ra\la ka,0,la,0|
\\
&+\sum_{(k,l)\not=(m,n)\in\{\pm\}}\tau_{klmn}|ka,0,la,0\ra\la ma,0,na,0|\in\cC(\cT[\mu],\cT[\rho_\eps])\,,
\end{aligned}
$$
with
$$
\sum_{(k,l)\not=(m,n)\in\{\pm\}}|\tau_{klmn}|^2>0\,.
$$
While the term
$$
\sum_{k,l\in\{\pm\}}\tau_{klkl}|ka,0,la,0\ra\la ka,0,la,0|
$$
has a classical interpretation, since it is the image of a bounded, positive Radon measure on phase space by the Toeplitz map, the term
$$
\sum_{(k,l)\not=(m,n)\in\{\pm\}}\tau_{klmn}|ka,0,la,0\ra\la ma,0,na,0|\not=0
$$
does not have any classical interpretation.

At the time of this writing, the structure of optimal couplings for $\fd(R_1,R_2)$ with $R_1,R_2\in\cD_2(\fH)$ is not very well understood in general, at variance with the 
classical case (see however the discussion of this point in \cite{CagliotiFGPaulSNS}), which is completely clarified by the Knott-Smith and the Brenier theorems. The very 
simple examples discussed in \cite{CaglioFGPaul} show that optimal couplings between quantum densities are much more involved than in the classical case.


\section{Lecture II: Applying the Quantum Wasserstein Pseudometric\\ to Particle Dynamics}


In this lecture, we shall discuss several applications of the quantum Wasserstein pseudometric $\fd$ introduced in Lecture I. These applications include

\smallskip
\noindent
\bu various limits of many-body problems in quantum mechanics (section \ref{SS-MFSCLim}),

\noindent
\bu proofs of the uniform in $\hb$ convergence of some numerical schemes for quantum dynamics (section \ref{SS-Split}), and

\noindent
\bu observation inequalities for the Schr\"odinger and for the von Neumann equations (section \ref{SS-Split}).

\subsection{Basics of Quantum Dynamics.}


As already explained, one does not need to be an expert in quantum mechanics to read these notes. 

However, some familiarity with the most elementary notions of quantum mechanics is required in order to understand the simple ideas behind the computations and the mathematical 
apparatus studied in this course. An excellent reference to learn quantum mechanics is \cite{BasdDalib}; see also \cite{CTDL} for a more detailed, yet equally lucid presentation.
The reference \cite{Hall} is interesting for mathematicians, but contains few physical explanations, at variance with \cite{BasdDalib,CTDL}.

\subsubsection{Classical Mechanics}

We recall the fundamentals of classical dynamics for a point particle of mass $m$, in Hamiltonian form.

The position of this point particle at time $t$ is denoted by $q(t)\in\bR^d$, while its momentum at time $t$ is denoted by $p(t)\in\bR^d$.

The \textbf{Hamiltonian} for the point particle is the total energy of that particle, expressed in terms of its position $q$ and momentum $p$ --- a word of caveat: it is essential at this point
to use these variables, and not other variables, say for instance the velocity instead of the momentum. For a point particle with mass $m$
$$
H(q,p)=\underbrace{|p|^2/2m}_{\text{kinetic}}+\underbrace{V(q)}_{\text{potential}}=\text{total energy.}
$$
With this, we can write \textbf{Newton's 2nd law of motion in Hamiltonian form}:
$$
\dot q(t)=\d H/\d p=p(t)/m\,,\qquad\dot p(t)=-\d H/\d q=-\grad V(q(t))\,.
$$
(In fact, the first equation is of a kinematic nature, since it can be viewed as a definition of the momentum $p(t)$ in terms of the particle mass $m$ and of the velocity $\dot q(t)$.
Only the second equation corresponds to Newton's second law.)

Newton's second law of motion can be viewed as the governing equation for the special phase space probability measure $\de_{(q(t),p(t))}$. But of course, it is equally interesting
to consider the dynamics of phase space Borel probability measures more general than $\delta_{q(t),p(t)}$.

This is precisely the purpose of the \textbf{Liouville equation}: if $f\equiv f(t,x,\xi)$ is the probability density of finding the point particle at the position $x\in\bR^d$ with momentum
$\xi\in\bR^d$ at time $t$, it satisfies the equation
$$
\d_tf(t,x,\xi)+\tfrac1m\xi\cdot\grad_xf(t,x,\xi)-\grad V(x)\cdot\grad_\xi f(t,x,\xi)=0\,.
$$
This equation can be recast in terms of the \textbf{Poisson bracket}:
$$
\{H(x,\xi),f(x,\xi)\}:=\grad_\xi H(x,\xi)\cdot\grad_xf(x,\xi)-\grad_xH(x,\xi)\cdot\grad_\xi f(x,\xi)\,,
$$
as follows:
$$
\d_tf(t,x,\xi)+\{H(x,\xi),f(t,x,\xi)\}=0\,.
$$

Newton's second law of motion in Hamiltonian form is a system of ODEs to which one can apply the Cauchy-Lipschitz theorem. Specifically, the local existence of a unique solution
of the Cauchy problem for the Hamiltonian formulation of Newton's second law of motion is implied by the assumption
$$
V\in C^{1,1}(\bR^d)
$$
--- meaning that $V\in C^1(\bR^d)$ and $\grad V$is Lipschitz continuous on $\bR^d$. Since
$$
H(q(t),p(t))=H(q(0),p(0))
$$
(the verification of this is left to the reader as an exercise), which corresponds to the conservation of total energy by the dynamics deduced from Newton's second law of motion,
it is easily seen that all solutions of the Cauchy problem for Newton's second law of motion are defined for all $t\in\bR$ under the condition
$$
\lim_{|x|\to+\infty}V(x)=+\infty\,.
$$
(Indeed, in that case, the map $(q,p)\mapsto H(q,p)$ is proper on $\bR^d\times\bR^d$, i.e. the inverse image of any compact subset of $\bR$ is compact in $\bR^d\times\bR^d$;
this implies that the Hamiltonian flow generated by Newton's second law of motion is global by the most elementary continuation argument for ODEs; see for instance (10.5.5)
in section 5 of chapter X in \cite{Dieudo}.)

\subsubsection{Quantum Mechanics}

The state of a quantum particle at time $t$ is given by its \textbf{wave function}:
$$
\psi\equiv\psi(t,x)\in L^2(\bR^d;\bC)=:\fH
$$ 
such that
$$
\|\psi(t,\cdot)\|_{\fH}=1\,.
$$

In classical mechanics, the Hamiltonian is a function on the phase space $\bR^d\times\bR^d$; in quantum mechanics, the \textbf{quantum Hamiltonian} is an (unbounded) 
self-adjoint operator on $\fH$:
$$
\bH=-\tfrac{\hb^2}{2m}\Dlt_x+V(x)=\bH^*\,.
$$
(Here, the real-valued potential operator $V$ is to be understood as a multiplication operator, i.e. $\psi(x)\mapsto V(x)\psi(x)$.) 

The reason for considering this specific operator by analogy with the case of a point particle in classical mechanics is 

\noindent
\textbf{The correspondence principle.} 
$$
V(q)\to\text{multiplication by }V(x)\text{ and }p_j\to-i\hb\d_{q_j}=\hb D_{q_j}\,.
$$

\smallskip
With these mathematical objects, the quantum analogue of Newton's second law of motion is 

\noindent
\textbf{The Schr\"odinger equation.}
$$
i\hb\d_t\psi(t,x)=\bH\psi(t,x)\,.
$$
Assuming that $\bH$ is an unbounded self-adjoint operator on $\fH$, it generates a unitary group on $\fH$ by Stone's theorem, denoted by $e^{-it\bH/\hb}$, so that
$$
\psi(t,\cdot)=e^{-it\bH/\hb}\psi(0,\cdot)\,,\qquad t\in\bR\,.
$$

As explained in Lecture I, the quantum analogue of Borel probability measures on $\bR^d\times\bR^d$ are density operators on $\fH=L^2(\bR^d)$. The quantum analogue
of the Liouville equation, defining the dynamics of phase space distribution functions is 

\noindent
\textbf{The von Neumann equation} for $R(t)\in\cD(\fH)$, 
$$
i\hb\d_tR(t)=\underbrace{\bH R(t)-R(t)\bH}_{=:[\bH,R(t)]}\,.
$$
One easily checks that, under the assumption that $\bH$ is an unbounded self-adjoint operator on $\fH$,
$$
R(t)=e^{-it\bH/\hb}R(0)e^{it\bH/\hb}\,.
$$
For instance, one easily checks that, if $\psi(t,\cdot)$ is a solution of the Schr\"odinger equation, then the rank-$1$ density operator $R(t):=|\psi(t,\cdot)\ra\la\psi(t,\cdot)|$
is a solution of the von Neumann equation.

This brings forward a further analogy in the correspondence principle, between the Poisson bracket $\{f,g\}$ of two $C^1$ functions defined on $\bR^d\times\bR^d$, and 
the commutator $[A,B]$ of two operators on $L^2(\bR^d)$:
$$
\{\cdot,\cdot\}\to\tfrac{i}{\hb}[\cdot,\cdot]\,.
$$

\smallskip
We conclude this section with a quick discussion of sufficient conditions on the potential $V$ under which the quantum Hamiltonian $\bH=-\tfrac{\hb^2}{2m}\Dlt+V$ is 
self-adjoint on $\fH=L^2(\bR^d)$. As explained above, if $\bH$ is self-adjoint, by Stone's theorem, it generates a quantum dynamics, namely the unitary group
$e^{-it\bH/\hb}$ on $\fH$.

\noindent
\textbf{Self-adjointness of $\bH$.} A first procedure for generating self-adjoint Hamiltonians of the form $-\Dlt+V$ is based on the associated (sesquilinear) quadratic form $b$
defined as follows:
$$
(\phi,\psi)\mapsto b(\phi,\psi)=\int_{\bR^d}\grad\overline{\phi(x)}\cdot\grad\psi(x)dx+\int_{\bR^d}V(x)\overline\phi(x)\psi(x)dx
$$
for all $\phi,\psi\in\cQ(b)$, where
$$
\cQ(b):=\{\phi\in H^1(\bR^d)\text{ s.t. }|V|^{1/2}\phi\in L^2(\bR^d)\}\,.
$$
Assume that $V$ is real-valued, and satisfies the following condition:
$$
V\in L^\infty_{loc}(\bR^d)\,,\quad\text{ and there exists }M\ge 0\text{ s.t. }V(x)\ge -M\text{ for a.e. }x\in\bR^d\,.
$$
Then $C^\infty_c(\bR^d)\subset\cQ(b)$ which is therefore dense in $\fH=L^2(\bR^d)$. Then, the quadratic form $b$ is semi-bounded, since
$$
\phi\in\cQ(b)\implies b(\phi,\phi)\ge-M\|\phi\|^2_\fH\,.
$$
Besides, the quadratic form is \textbf{closed}, meaning that, for each sequence $\phi_n\in\cQ(b)$ such that
$$
\phi_n\to\phi\text{ in }\fH\text{ as }n\to\infty\,,\quad\text{ and }b(\phi_n-\phi_m,\phi_n-\phi_m)\to 0\text{ as }m,n\to\infty\,,
$$
one has
$$
\phi\in Q(b)\,,\quad\text{ and }b(\phi_n-\phi,\phi_n-\phi)\to 0\text{ as }n\to\infty\,.
$$
(Indeed, one easily checks that $\grad\phi_n$ is a Cauchy sequence in $L^2(\bR^d;\bR^d)$, and therefore converges towards an $L^2$ vector field $\xi$ on $\bR^d$; since
$\grad\phi_n\to\grad\phi$ in the sense of distributions on $\bR^d$, one has $\phi\in H^1(\bR^d)$ and $\xi=\grad\phi$. Similarly, $(1+M+V)^{1/2}\phi_n$ is a Cauchy sequence
in $\fH$, and therefore converges to some limit $\ell\in\fH$; then, one easily checks that $\ell/\sqrt{1+M+V}\in\fH$ and that $\phi_n\to\ell/\sqrt{1+M+V}$ as $n\to\infty$, so that
$\phi=\ell/\sqrt{1+M+V}$ by uniqueness of the limit in $\fH$. Hence $\sqrt{1+M+V}\phi\in\fH$, so that $\phi\in\cQ(b)$. The remaining part of the proof is routine, and left to the
reader.) By Theorem VIII.15 of \cite{RS1}, there exists an unbounded self-adjoint operator $A$ with domain $D(A)\subset\cQ(b)$ such that 
$$
\text{ for all }\phi,\psi\in D(A)\,,\quad b(\phi,\psi)=\la\psi|A\phi\ra\,.
$$
Obviously $C^\infty_c(\bR^d)\subset D(A)$ and $A$ coincides with $-\Dlt+V$ on $C^\infty_c(\bR^d)$.

A second procedure is the \textbf{Kato-Rellich Theorem} (see Theorem X.12 of \cite{RS2}). Assume that $A$ is a self-adjoint operator on $\fH$ with domain $D(A)$, and $B$
is a symmetric\footnote{I.e. $\la\phi|B\psi\ra=\la B\phi|\psi\ra$ for all $\phi,\psi\in D(B)$. In other words, $D(B)\subset D(B^*)$ and $B^*\rstr_{D(B)}=B$.} operator on $\fH$ with 
domain $D(B)$. Assume that $D(A)\subset D(B)$, and that there exists $a\in[0,1)$ and $b\ge 0$ such that
$$
\|B\phi\|\le a\|A\phi\|+b\|\phi\|\,,\quad\phi\in D(A)\,.
$$
Then $A+B$ is self-adjoint on $D(A)$.

With this result, one can prove that $-\Dlt+V$ is self-adjoint on $D(-\Dlt)=H^2(\bR^3)$ provided that
$$
V\in L^2(\bR^3)+L^\infty(\bR^3)\,.
$$
This result is particularly important in atomic physics, since the Coulomb potential 
$$
V(x)=\pm\frac{1}{|x-x_0|}=\pm\left(\frac{\indc_{|x-x_0|\le 1}}{|x-x_0|}+\frac{\indc_{|x-x_0|>1}}{|x-x_0|}\right)\in L^2(\bR^3)+L^\infty(\bR^3)\,.
$$
Here, $V$ is the potential energy of an electron at the position $x$ interacting with a nucleus located at the position $x_0$.

One can also combine these two result to treat the case of $-\Dlt+V+V_\infty$ where $V\in L^2(\bR^3)+L^\infty(\bR^3)$, while $V_0\ge 0$ belongs to $L^\infty_{loc}(\bR^3)$
and $V_0(x)\to+\infty$ as $|x|\to+\infty$ (in other words, $V_0$ is a confining potential).

In all these cases, one can see that much less regularity is required on the potential in order to define the quantum dynamics, than in the case of the classical dynamics.
Perhaps the reason for this difference is that classical mechanics deals with the dynamics of much more singular objects (i.e. $\de_{(q(t),p(t))}$) than the wave function
$\psi(t,\cdot)\in L^2(\bR^d)$ in quantum mechanics.

\smallskip
\noindent
\textbf{Quiz 14.} Explain how one can use the Kato-Rellich theorem to prove that 
$$
V\in L^2(\bR^3)+L^\infty(\bR^3)\implies -\Dlt+V\text{ is self-adjoint with domain }H^2(\bR^3)\,.
$$
(Hint: use the Sobolev embedding.)

In all the situations described above, one starts from the operator $-\tfrac{\hb^2}{2m}\Dlt+V$, which is well defined on $C^2_c(\bR^d)\subset L^2(\bR^d)$. Both methods described 
above (either the method involving a quadratic form, or the Kato-Rellich Theorem) produce an unbounded self-adjoint operator on $L^2(\bR^d)$, which is therefore densely defined 
in $L^2(\bR^d)$. The domain of this self-adjoint operator contains $C^2_c(\bR^d)$, and this self-adjoint operator coincides with the differential operator $-\tfrac{\hb^2}{2m}\Dlt+V$ 
on $C^2_c(\bR^d)$. This self-adjoint operator is therefore an extension of the differential operator $-\tfrac{\hb^2}{2m}\Dlt+V$, defined on $C^2_c(\bR^d)\subset L^2(\bR^d)$, and we 
shall keep the notation $-\tfrac{\hb^2}{2m}\Dlt+V$ to designate this extension.

\subsubsection{The Classical Limit of Quantum Mechanics}

Consider a point particle of mass $m$, moving at a speed $v$; its de Broglie wavelength is the ratio $2\pi\hb/mv$. For instance a dust particle of diameter $1\mu$ with mass 
$m=10^{-6}\mu$g moving at speed $1$mm/s, has a de Broglie wavelength $6.6\cdot 10^{-6}$\AA$\ll 1\mu$, the size of the dust particle. (This example is taken from chapter I,
complement A in \cite{CTDL}.) If the de Broglie wavelength of a particle is negligible when compared to its size, or to the typical length scale of the experiment, one expects that
the laws of classical mechanics should be sufficient to describe its behavior.

There are various ways of describing the classical limit of quantum mechanics. 

As explained in \S 6 of \cite{LL6}, in the quasi-classical regime, the phase of the wave function of a particle is proportional to the mechanical action of that particle, and the constant 
of proportionality is Planck's constant $\hb$. One could therefore study solutions of the Schr\"odinger equation
$$
i\hb\d_t\psi=-(\tfrac{\hb^2}{2m}\Dlt+V)\psi
$$
in the form of a \textbf{WKB ansatz} i.e. a formal series
$$
\psi(t,x)=\sum_{n\ge 0}\hb^na_n(t,x)e^{iS(t,x)/\hb}\,,\quad S(t,x)\text{ and }a_n(t,x)\in\bR\,,
$$
where the phase $S$ and the amplitude coefficients $a_0,a_1,\ldots$ are smooth.

Assuming that $a_0(t,x)\not=0$ for all $(t,x)$, one finds that

\noindent
(a) the phase $S$ is a solution of the \textit{eikonal equation}, that is a Hamilton-Jacobi equation:
$$
\d_tS+H(\grad_xS,x)=0\,;
$$
(b) at leading order, the amplitude $a_0$ is a solution of the \textit{transport equation}:
$$
\d_ta_0^2+\Div_x(a_0^2\grad_xS(t,x))=0\,.
$$
However, this description usually fails after some finite time, for the following reason: the graph of the map $x\mapsto \grad_xS(t,x)$ is the image of the graph of the map
$x\mapsto\grad_xS(0,\cdot)$ by the flow of the classical Hamiltonian $H(x,\xi)$. In general, the graph of $\grad_xS(0,\cdot)$ becomes folded after some finite time in such
a way that it is no longer the graph of map from $\bR^d$ to $\bR^d$. The image of these folds by the projection $\bR^d\times\bR^d\ni(x,\xi)\mapsto x\in\bR^d$ is referred
to as the ``caustic'', by analogy with geometric optics. The appearance of caustics is the reason why the WKB ansatz is in general only local in time.

\smallskip
\noindent
\textbf{Quiz 15.} Write the WKB ansatz for the free Schr\"odinger equation
$$
i\hb\d_t\psi(t,x)=-\tfrac{\hb^2}{2m}\partial_x^2\psi(t,x)\,,\quad\psi(0,x)=a^{in}(x)e^{iS^{in}(x)/\hb}\,,\qquad x\in\bR\,,
$$
where 
$$
S^{in}(x)=\left\{\ba{}&-x^{1/3}&&\text{ for }x\ge 0\,,\\ &+|x|^{1/3}&&\text{ for }x<0\,.\ea\right.
$$
Study the dynamics for all $t\in\bR$ (in the past $t<0$ as well as in the future $t>0$), and describe the caustic in this case.

\smallskip
Another approach to the classical limit of quantum mechanics involves the \textbf{Wigner Transform} already introduced in Quiz 11. We briefly recall the essentials: start from
some integral operator $R\in\cL(\fH)$ of the form
$$
R\phi(x)=\int_{\bR^d}r(x,y)\phi(y)dy\,.
$$
For $R=\text{projection on }\bC\psi$ with $\|\psi\|_\fH=1$, written $R=|\psi\ra\la\psi|$ in Dirac's notation:
$$
R\phi(x)=\underbrace{\left(\int_{\bR^d}\overline{\psi(y)}\phi(y)dy\right)}_{=:\la\psi|\phi\ra}\psi(x)\implies r(x,y)=\psi(x)\overline{\psi(y)}\,.
$$
We consider its Wigner transform at scale $\hb$, given by the formula
$$
W_\hb[R](q,p):=\tfrac1{(2\pi)^d}\int_{\bR^d}r(q+\tfrac12\hb y,q-\tfrac12\hb y)e^{ip\cdot y}dy\,.
$$
Some assumptions are needed on the integral kernel $r$ for the Wigner transform to make sense. For instance, if $R\in\cL^2(\fH)$ is a Hilbert-Schmidt operator, its integral
kernel $r$ belongs to $L^2(\bR^d\times\bR^d)$, so that the map
$$
(q,y)\mapsto r(q+\tfrac12\hb y,q-\tfrac12\hb y)
$$
belongs to $L^2(\bR^d_q\times\bR^d_y)$, since the Jacobian of the transformation 
$$
(q,y)\mapsto(q+\tfrac12\hb y,q-\tfrac12\hb y)
$$ 
is $(-\hb)^d$, which is in particular independent of $(q,y)$. In that case, the Wigner transform $W_\hb[R]$ belongs to $L^2(\bR^d_q\times\bR^d_p)$ by the Plancherel theorem, 
as the partial Fourier transform of a square integrable measurable function.

Observe that
$$
R=R^*\implies W_\hb[R](q,p)\in\bR\,,
$$
but
$$
R\ge 0\text{ does not imply that }W_\hb[R]\ge 0\,.
$$
(For instance, if $\psi$ is an odd wave function, then $W_\hb[|\psi\ra\la\psi|](0,0)<0$.)

We have seen in Quiz 11 that, if a continuous, time-dependent density operator $R(t)$ is a weak solution of the von Neumann equation
$$
i\hb\d_tR(t)=[-\tfrac{\hb^2}{2m}\Dlt+V,R(t)]\,,
$$
its Wigner transform $W_\hb[R(t)]$ is a weak solution of the Wigner equation
$$
(\d_t+q\cdot\grad_q)W_\hb[R(t)](q,p)+\Theta[V]W_\hb[R(t)](q,p)=0\,,
$$
where $\Theta(V)$ is the nonlocal (linear) operator with distribution kernel
$$
\tfrac1{(2\pi)^d}\int_{\bR^d}\tfrac1{i\hb}((V(q+\tfrac12\hb y)-V(q-\tfrac12\hb y))e^{iy\cdot(p'-p)}dy\,.
$$
If $V$ is a polynomial of degree $\le 2$, one easily checks that
$$
\Theta[V]=-\grad V(q)\cdot\grad_p\,.
$$

The classical limit of quantum mechanics can be formulated in terms of the Wigner function, in the following manner.
 
\noindent
\textbf{Theorem (Lions-Paul).} Assume that $V$ is a real-valued function satisfying the assumptions
$$
\inf_{q\in\bR}V(q)>-\infty\,,\,\,V\in C^{1,1}(\bR^d)\text{ and }V(q)=O(|q|^n)\text{ for some }n\ge 0\text{ as }|q|\to\infty\,,
$$
and let 
$$
\bH=-\tfrac{\hb^2}{2m}\Dlt+V
$$
be the quantum Hamiltonian, with domain $H^2(\bR^d)\cap L^2(\bR^d;V(x)_+^2dx)$. Let $R^{in}_\hb$ be a family of $\cD(\fH)$ such that
$$
W_\hb[R_\hb^{in}]\to f^{in}\text{ in }\cS'(\bR^d\times\bR^d)\text{ as }\hb\to 0\,.
$$
Then
$$
f^{in}\text{ is a probability density on }\bR^d\times\bR^d\,,
$$
and
$$
W_\hb[e^{-it\bH/\hb}R_\hb^{in}e^{it\bH/\hb}]\to f(t,\cdot,\cdot)\text{ in }\cS'(\bR^d\times\bR^d)\text{ as }\hb\to 0
$$
uniformly in $t\in[0,T]$ for each $T>0$, where $f$ is the probability density solution to the Liouville equation 
$$
\d_tf(t,q,p)+\{\tfrac1{2m}|p|^2+V(q),f(t,q,p)\}=0\,,\qquad f\rstr_{t=0}=f^{in}\,.
$$

\smallskip
This is Theorem IV.1 in \cite{LionsPaul}, and the interested reader is referred to this article for its proof, together with several other interesting examples involving the Wigner
transform.

\smallskip
The connection between these two approaches to the classical limit of quantum mechanics is made clear by the following example, which is left to the reader as an exercise.

\noindent
\textbf{Quiz 16.} Consider a \textbf{WKB wave function}
$$
\psi_\hb^{in}(x)=a^{in}(x)e^{iS^{in}(x)/\hb}\,,
$$
with $\|a^{in}\|_{L^2}=1$ and $S\in\Lip(\bR^d;\bR)$. Prove that
$$
W_\hb[|\psi^{in}\ra\la\psi^{in}|](q,p)\to |a^{in}(q)|^2\de(p-\grad S^{in}(q))
$$
in the sense of tempered distributions as $\hb\to 0$. Explain how the eikonal and the transport equation predicted in statements (a)-(b) above in this section emerge from the 
dynamics of $f^{in}$ predicted by the Lions-Paul Theorem.

\subsection{Amplification of the Pseudometric $\fd$ by Hamiltonian Dynamics}

This is the core of the present lecture. Our goal is to control amplifications of the pseudometric $\fd$ constructed in Lecture I by Hamiltonian dynamics (classical or quantum).

\subsubsection{Pair Dispersion in Classical Mechanics}

In this brief section, we seek to compare the evolution of two different initial data, with two different potentials, in classical and in quantum mechanics. (The term ``pair
dispersion'' is used in Lagrangian fluid mechanics, and is perhaps not so common in the present setting.) The results obtained here should be thought of as a warm-up 
--- and a motivation --- for the study of the dynamical amplification of $\fd$.

\smallskip
\noindent
\textbf{Classical dynamics.} We seek to compare two solutions of Newton's equations, $t\mapsto(X,\Xi)(t)$ and $t\mapsto(Y,H)(t)$ with two different potentials $V$ and $W$
belonging to $C^{1,1}(\bR^d)$:
$$
\left\{\ba{}&\dot X\!\!=\tfrac1m\Xi\,,\\ &\dot\Xi=-\grad V(X)\,,\ea\right.\qquad\qquad\text{ and }\qquad\qquad\left\{\ba{}&\dot Y=\tfrac1mH\,,\\ &\dot H=-\grad W(Y)\,,\ea\right.
$$
with initial data
$$
(X,\Xi)(0)=(X^{in},\Xi^{in})\,,\qquad (Y,H)(0)=(Y^{in},H^{in})\,.
$$
Setting $L:=\Lip(\grad V)$, we compute
$$
\ba
\tfrac{d}{dt}(|X-Y|^2+|\Xi-H|^2)
\\
=\tfrac2m(\Xi- H)\cdot(X-Y)-2(\grad V(X)-\grad W(Y))\cdot(\Xi-H)
\\
=\tfrac2m(\Xi- H)\cdot(X-Y)-2(\grad V(X)-\grad V(Y))\cdot(\Xi-H)
\\
+2(\grad W(Y)-\grad V(Y))\cdot(\Xi-H)
\\
\le(\tfrac1m+L)(|X-Y|^2+|\Xi-H|^2)
\\
+2\|\grad(V-W)\|_{L^\infty(\bR^d)}(|\Xi|+|H|)
\\
\le(\tfrac1m+L)(|X-Y|^2+|\Xi-H|^2)
\\
+2\|\grad(V-W)\|_{L^\infty(\bR^d)}\sqrt{|\Xi|^2+2V(X)+2\|V\|_{L^\infty(\bR^d)}}
\\
+2\|\grad(V-W)\|_{L^\infty(\bR^d)}\sqrt{|H|^2+2W(Y)+2\|W\|_{L^\infty(\bR^d)}}
\\
=(\tfrac1m+L)(|X-Y|^2+|\Xi-H|^2)
\\
+2\|\grad(V-W)\|_{L^\infty(\bR^d)}\sqrt{|\Xi^{in}|^2+2V(X^{in})+2\|V\|_{L^\infty(\bR^d)}}
\\
+2\|\grad(V-W)\|_{L^\infty(\bR^d)}\sqrt{|H^{in}|^2+2W(Y^{in})+2\|W\|_{L^\infty(\bR^d)}}&\,,
\ea
$$
where the last equality follows from the energy conservation. By Gronwall's inequality
$$
\ba
(|X(t)-Y(t)|^2+|\Xi(t)-H(t)|^2)\le(|X^{in}-Y^{in})|^2+|\Xi^{in}-H^{in}|^2)e^{(\frac1m+L)t}
\\
+M\frac{e^{(\frac1m+L)t}-1}{(\frac1m+L)}\|\grad(V-W)\|_{L^\infty(\bR^d)}&\,,
\ea
$$
where
$$
M:=2\left(\sqrt{|\Xi^{in}|^2+2V(X^{in})+2\|V\|_{L^\infty(\bR^d)}}+\sqrt{|H^{in}|^2+2W(Y^{in})+2\|W\|_{L^\infty(\bR^d)}}\right)\,.
$$

\smallskip
\noindent
\textbf{Quantum dynamics.} Consider the quantum dynamics of two different wave functions, driven by two potentials $V$ and $W$ such that both quantum Hamiltonians
$-\tfrac{\hb^2}{2m}\Dlt+V$ and $-\tfrac{\hb^2}{2m}\Dlt+W$ have self-adjoint extensions to $\fH=L^2(\bR^d)$:
$$
\ba
i\hb\d_t\phi=(-\tfrac{\hb^2}{2m}\Dlt+V)\phi\,,\qquad\phi\rstr_{t=0}=\phi^{in}\,,
\\
i\hb\d_t\psi=(-\tfrac{\hb^2}{2m}\Dlt+W)\psi\,,\qquad\psi\rstr_{t=0}=\psi^{in}\,.
\ea
$$
Thus
$$
\d_t(\phi-\psi)=\tfrac1{i\hb}(-\tfrac{\hb^2}{2m}\Dlt+V)(\phi-\psi)+\tfrac1{i\hb}(V-W)\psi\,,
$$
so that
$$
\ba
\|(\phi-\psi)(t,\cdot)\|_\fH\le\|\phi^{in}-\psi^{in}\|_\fH+\frac1\hb\int_0^t\|V-W\|_{L^\infty(\bR^d)}\|\psi(s,\cdot)\|_{L^2(\bR^d)}ds
\\
=\|\phi^{in}-\psi^{in}\|_\fH+\frac{t}\hb\|V-W\|_{L^\infty(\bR^d)}&\,.
\ea
$$

\smallskip
If we compare these two estimates, we immediately see the following differences:

\noindent
(a) the $L^2$ bound on the difference of wave functions of the quantum particle involves the difference of potentials in sup norm, whereas the difference in the position
and momenta of the classical particle involves the difference of the force fields, i.e. of the gradients of the potentials, in sup norm;

\noindent
(b) there is no exponential amplification of the $L^2$ norm of the difference of wave functions in the quantum case, whereas the bound on the difference in positions
and momenta of the classical particle involves an amplification factor $e^{Lt}$, where $L$ is the Lipschitz constant of one of the force fields; yet

\noindent
(c) the $L^2$ estimate on the difference of wave functions is not uniform in $\hb$ as $\hb\to 0^+$. 

Notice however that the $L^2$ bound on this difference is uniform in $\hb$ in the very special case where both potentials are equal. This is perhaps of limited interest
in the context of numerical analysis, since numerical schemes typically replace the true potential $V$ by an approximation thereof.

Also, the $L^2$ bound on the difference of wave functions is an estimate of the same kind as a bound in trace, or Hilbert-Schmidt norm for the difference of the density
operators, since
$$
\|\,|\phi\ra\la\phi|-|\psi\ra\la\psi|\,\|^2_2\le 2\|\phi-\psi\|^2_\fH\|\phi\|^2_\fH+2\|\psi\|^2_\fH\|\phi-\psi\|^2_\fH=4\|\phi-\psi\|^2_\fH\,.
$$
We have seen in Lecture I, in the case where $\phi=|q_1,p_1\ra$ and $\psi=|q_2,p_2\ra$, that the Hilbert-Schmidt norm for the difference between such density
operators converges to $\|\de_{(q_1,p_1)}-\de_{(q_2,p_2)}\|_{TV}=2$ unless $(q_1,p_1)=(q_2,p_2)$ in the limit as $\hb\to 0^+$. Therefore, in the case where $V=W$, 
the fact that the quantum dynamics is unitary would typically result in the perfectly true, but uninteresting inequality $2\le 4$ in the small $\hb$ limit --- assuming that 
$\phi(t,\cdot)=|q_1,p_1\ra$ and $\psi(t,\cdot)=|q_2,p_2\ra$, corresponding to $\phi(0,\cdot)=U(t)^*|q_1,p_1\ra$ and $\psi(0,\cdot)=U(t)^*|q_2,p_2\ra$, with 
$U(t):=\exp(-it(-\tfrac{\hb^2}{2m}\Dlt+V)\hb)$.

On the contrary, the pair dispersion estimate in the classical setting can be understood as 
$$
\ba
\cW_2(\de_{(X(t),\Xi(t))},\de_{(Y(t),H(t))})^2\le&\cW_2(\de_{(X^{in},\Xi^{in})},\de_{(Y^{in},H^{in})})^2e^{(\frac1m+L)t}
\\
&+M\frac{e^{(\frac1m+L)t}-1}{(\frac1m+L)}\|\grad(V-W)\|_{L^\infty(\bR^d)}&\,.
\ea
$$
It is an easy exercise (left to the reader) to extend this estimate to arbitrary phase space probability measures that are weak solutions of the Liouville equation --- not necessary 
of the form $\de_{(z(t),\zeta(t))}$, where $t\mapsto(z(t),\zeta(t))$ is a solution of Newton's motion equations.

This suggests the following question: is there an estimate analogous to the classical pair dispersion estimate in the quantum setting for $\fd(R_1(t),R_2(t))$, where $R_1$ and
$R_2$ are time-dependent density operators on $\fH=L^2(\bR^d)$ whose dynamics is governed by the von Neumann equation with two different potentials? Should such an
estimate exist, it would be

\noindent
\bu uniform in $\hb$ as $\hb\to 0^+$, and

\noindent
\bu should involve the same amplification factor $e^{(\frac1m+L)t}$ as in the classical case, and a similar error term of order
$$
e^{(\frac1m+L)t}\|\grad(V-W)\|_{L^\infty(\bR^d)}\quad\text{ instead of }\quad\frac1\hb\|V-W\|_{L^\infty(\bR^d)}\,.
$$

Of course, such an estimate, should it exist, would require more regularity on the potentials (typically the same kind of regularity as in the classical setting) than what is needed
for the quantum Hamiltonians
$$
-\tfrac{\hb^2}{2m}\Dlt+V\quad\text{ and }\quad-\tfrac{\hb^2}{2m}\Dlt+W
$$
to have (unbounded) self-adjoint extensions to $\fH=L^2(\bR^d)$.

\subsubsection{Propagation Estimate for $\fd$.}

It will be convenient to consider, instead of the pseudometric $\fd$ introduced in the previous lecture, a deformation thereof, henceforth designated by $\fd_\lbd$, where $\lbd>0$ 
is the deformation parameter.

\smallskip
\noindent
\textbf{Pseudometric $\fd_\lbd$.} For all $R,S\in\cD_2(\fH)$, and all probability density $f$ on the phase space $\bR^d\times\bR^d$ with finite 2nd order moment
$$
\ba
{}&\fd_\lbd(f,R)^2:=\inf_{Q\in\cC(f,R)}\iint_{\bR^{2d}}\Tr_\fH(Q(x,\xi)^\frac12c_{\lbd,\hb}(x,\xi)Q(x,\xi)^\frac12)dxd\xi\,,
\\
&\fd_\lbd(R,S)^2:=\inf_{Q\in\cC(R,S)}\Tr_{\fH\otimes\fH}(Q^{1/2}C_{\lbd,\hb}Q^{1/2})\,,
\ea
$$
where $c_{\lbd,\hb}$ and $C_{\lbd,\hb}$ are the differential operators defined on $\bR^d$ and $\bR^d\times\bR^d$ respectively by the formulas
$$
\ba
c_{\lbd,\hb}(x,\xi)\phi(y):=&(\lbd^2|x-y|^2+|\xi+i\hb\grad_y|^2)\phi(y)\,,
\\
C_{\lbd,\hb}\Phi(x,y):=&(\lbd^2|x-y|^2-\hb^2(\grad_x-\grad_y)\cdot(\grad_x-\grad_y))\Phi(x,y)\,.
\ea
$$
We leave it to the reader as an easy exercise to check that
$$
c_\hb(x,\xi)\ge \lbd d\hb I_\fH\,,\qquad C_\hb\ge 2\lbd d\hb I_{\fH\otimes\fH}\,.
$$

\noindent
\textbf{Theorem 6.}
Assume that $V\in C^{1,1}(\bR^d)$ satisfies
$$
V(y)\to+\infty\text{ as }|y|\to\infty\,,\quad\text{ and }\Lip(\grad V)<\infty\,.
$$
Hence $\bH:=-\tfrac{\hb^2}{2m}\Dlt+$ has a self-adjoint extension to $\fH$, and defines a quantum dynamics via the unitary group $U(t):=e^{-it\bH/\hb}$. 
On the other hand, let $\Phi(t;\cdot,\cdot)$ be the flow of the classical Hamiltonian $H(x,\xi):=\tfrac1{2m}|\xi|^2+V(x)$, which is defined for all $t\in\bR$, 
since $V$ is confining (tends to $+\infty$ at infinity). Then, for each $R_1^{in},R_2^{in}\in\cD_2(\fH)$ and each probability density $f$ on the phase
space $\bR^d\times\bR^d$ with finite 2nd order moment, one has
$$
\ba
\fd_\lbd(f^{in}\circ\Phi(-t,\cdot,\cdot),\,U(t)R_1^{in}U(t)^*)\le&\,\fd_\lbd(f^{in},R_1^{in})e^{L|t|}\,,
\\
\fd_\lbd(U(t)R_1^{in}U(t)^*\!,\!U(t)R_2^{in}U(t)^*\!)\le&\,\fd_\lbd(R_1^{in},R_2^{in})e^{L|t|}\,,
\ea
$$
for all $t\in\mathbf R$, with 
$$
L:=\tfrac12\left(\frac{\lbd}m+\frac{\Lip(\grad V)}{\lbd}\right)\,.
$$

\noindent
\begin{proof}
Let $Q^{in}\in\cC(f^{in},R_1^{in})$; set 
$$
Q(t,x,\xi):=U(t)Q^{in}(\Phi(t,x,\xi))U(t)^*\,.
$$
One easily checks that 
$$
\ba
\Tr_\fH(Q(t,x,\xi))=&\Tr_\fH(U(t)Q^{in}(\Phi(-t,x,\xi))U(t)^*)
\\
=&\Tr_\fH(Q^{in}(\Phi(-t,x,\xi)))=f^{in}(\Phi(-t,x,\xi))\,,
\ea
$$
while
$$
\ba
\iint_{\bR^{2d}}U(t)Q^{in}(\Phi(-t,x,\xi))U(t)^*dxd\xi=U(t)\left(\iint_{\bR^{2d}}Q^{in}(\Phi(-t,x,\xi))dxd\xi\right)U(t)^*
\\
=
U(t)\left(\iint_{\bR^{2d}}Q^{in}(X,\Xi)dXd\Xi\right)U(t)^*=U(t)R^{in}U(t)^*\,,
\ea
$$
so that
$$
Q(t,\cdot,\cdot)\in\cC(f^{in}\circ\Phi(-t,\cdot,\cdot),U(t)R_1^{in}U(t)^*)\,.
$$
Thus
$$
\ba
\fd_\lbd(f^{in}\circ\Phi(-t,\cdot,\cdot),U(t)R_1^{in}U(t)^*)^2
\\
\le\iint_{\bR^{2d}}\Tr_\fH(Q(t,x,\xi)^{1/2}c_{\lbd,\hb}(x,\xi)Q(t,x,\xi)^{1/2})dxd\xi
\\
=\iint_{\bR^{2d}}\Tr_\fH(Q^{in}(X,\Xi)^{1/2}U(t)^*c_{\lbd,\hb}(\Phi(t,X,\Xi)U(t)Q^{in}(X,\Xi)^{1/2})dXd\xi&\,,
\ea
$$
and
$$
\ba
\tfrac{d}{dt}\iint_{\bR^{2d}}\Tr_\fH(Q(t,x,\xi)^{1/2}c_{\lbd,\hb}(x,\xi)Q(t,x,\xi)^{1/2})dxd\xi
\\
=\iint_{\bR^{2d}}\Tr_\fH(Q(t,x,\xi)^{1/2}\{\tfrac1{2m}|\xi|^2+V(x),c_{\lbd,\hb}(x,\xi)\}Q(t,x,\xi)^{1/2})dxd\xi
\\
+\iint_{\bR^{2d}}\Tr_\fH(Q(t,x,\xi)^{1/2}\tfrac{i}\hb[-\tfrac{\hb^2}{2m}\Dlt+V,c_{\lbd,\hb}(x,\xi)]Q(t,x,\xi)^{1/2})dxd\xi&\,.
\ea
$$
Now, we compute
$$
\ba
\{\tfrac1{2m}|\xi|^2+V(x),c_{\lbd,\hb}(x,\xi)\}+\tfrac{i}\hb[-\tfrac{\hb^2}{2m}\Dlt+V,c_{\lbd,\hb}(x,\xi)]
\\
=\tfrac{\lbd^2}{m}((\xi+i\hb\grad_y)\cdot(x-y)+(x-y)\cdot(\xi+i\hb\grad_y))
\\
-(\xi+i\hb\grad_y)\cdot(\grad V(x)-\grad V(y))-(\grad V(x)-\grad V(y))\cdot(\xi+i\hb\grad_y)
\\
=\tfrac{\lbd}{m}((\xi+i\hb\grad_y)\cdot\lbd(x-y)+\lbd(x-y)\cdot(\xi+i\hb\grad_y))
\\
-\tfrac{\Lip(\grad V)}\lbd((\xi+i\hb\grad_y)\cdot\lbd\tfrac{\grad V(x)-\grad V(y)}{\Lip(\grad V)}+\lbd\tfrac{\grad V(x)-\grad V(y)}{\Lip(\grad V)}\cdot(\xi+i\hb\grad_y))
\\
\le(\tfrac{\lbd}{m}+\tfrac{\Lip(\grad V)}\lbd)(\lbd^2|x-y|+|\xi+i\hb\grad_y|^2)&\,.
\ea
$$
Therefore
$$
\ba
\tfrac{d}{dt}\iint_{\bR^{2d}}\Tr_\fH(Q(t,x,\xi)^{1/2}c_{\lbd,\hb}(x,\xi)Q(t,x,\xi)^{1/2})dxd\xi
\\
\le(\tfrac{\lbd}{m}+\tfrac{\Lip(\grad V)}\lbd)\iint_{\bR^{2d}}\Tr_\fH(Q(t,x,\xi)^{1/2}c_{\lbd,\hb}(x,\xi)Q(t,x,\xi)^{1/2})dxd\xi&\,,
\ea
$$
and hence
$$
\ba
\fd_\lbd(f^{in}\circ\Phi(-t,\cdot,\cdot),U(t)R_1^{in}U(t)^*)^2
\\
\le\iint_{\bR^{2d}}\Tr_\fH(Q(t,x,\xi)^{1/2}c_{\lbd,\hb}(x,\xi)Q(t,x,\xi)^{1/2})dxd\xi
\\
\le\exp(|t|(\tfrac{\lbd}{m}+\tfrac{\Lip(\grad V)}\lbd))\iint_{\bR^{2d}}\Tr_\fH(Q^{in}(x,\xi)^{1/2}c_{\lbd,\hb}(x,\xi)Q^{in}(x,\xi)^{1/2})dxd\xi&\,.
\ea
$$
Choosing a minimizing sequence $Q^{in}_n$ of elements of $\cC(f^{in},R^{in}_1)$ so that
$$
\iint_{\bR^{2d}}\Tr_\fH(Q_n^{in}(x,\xi)^{1/2}c_{\lbd,\hb}(x,\xi)Q_n^{in}(x,\xi)^{1/2})dxd\xi\to\fd_\lbd(f^{in},R^{in}_1)^2
$$
as $n\to\infty$ leads to the first inequality.
\end{proof}

\smallskip
\noindent
\textbf{Quiz 17.} Prove the second inequality in Theorem 6 (the argument follows the one presented above for the first inequality).

\smallskip
Theorem 6 will often be used together with Theorem 1 from Lecture I, in the following manner. For instance, one could start from Toeplitz density operators, for which
the pseudometric $\fd_\lbd$ is very well-known, by using Theorem 1 (1). Usually, the quantum dynamics fails to preserve the Toeplitz structure of the density operator,
but at time $t$, one can use the lower bound for $\fd_\lbd$ deduced from Theorem 1 (2) to compare the Husimi transforms of the (quantum) density operators by means
of the classical Wasserstein $\cW_2$ metric. The resulting statement is as follows.

\noindent
\textbf{Corollary 7.} Under the same assumptions as in Theorem 6, let $f^{in}$ be a probability density with finite second order moments on $\bR^d\times\bR^d$, and
let $g_1^{in},g_2^{in}\in\cP_2(\bR^d\times\bR^d)$. Set $R_1^{in}:=\cT[g_1^{in}]$ and $R_2^{in}:=\cT[g_2^{in}]$. Then
$$
\ba
\cW_2(f^{in}\circ\Phi(-t,\cdot,\cdot),\cH[U(t)R^{in}_1U(t)^*])^2\le &e^{L|t|}\frac{\max(1,\lbd^2)}{\min(1,\lbd^2)}\cW_2(f^{in},g_1^{in})^2
\\
&+\frac{(1+\lbd^2)d\hb}{2\min(1,\lbd^2)}(e^{L|t|}+1)\,,
\\
\cW_2(\cH[U(t)R^{in}_1U(t)^*],\cH[U(t)R^{in}_2U(t)^*])^2\le &e^{L|t|}\frac{\max(1,\lbd^2)}{\min(1,\lbd^2)}\cW_2(g_1^{in},g_2^{in})^2
\\
&+\frac{(1+\lbd^2)d\hb}{\min(1,\lbd^2)}(e^{L|t|}+1)\,,
\ea
$$
where $L:=\tfrac12(\tfrac{\lbd}m+\tfrac{\Lip(\grad V)}{\lbd})$.

\smallskip
\noindent
\textbf{Quiz 18.} Prove Corollary 7. (Hint: compute the operators
$$
\iint_{\bR^d\times\bR^d} F(p)|q,p\ra\la q,p|dqdp\quad\text{ and }\quad\iint_{\bR^d\times\bR^d} F(q)|q,p\ra\la q,p|dqdp\,,
$$
where $F$ is a polynomial of degree $2$ on $\bR^d$. See Appendix B of \cite{FGMouPaul}, or the basic properties of the Toeplitz map in Lecture I, and Quiz 8.)

In the remaining part of this lecture, we shall study three (more or less direct) applications of the propagation bound in Theorem 6:

\smallskip
\noindent
\underline{Application 1:} mean-field and classical limits of quantum mechanics (section \ref{SS-MFSCLim});

\smallskip
\noindent
\underline{Application 2:} time-splitting schemes for quantum mechanics (section \ref{SS-Split});

\smallskip
\noindent
\underline{Application 3:} observation inequalities for quantum dynamics (section \ref{SS-Observ}).

\subsection{Mean-Field and Classical Limits of Quantum Mechanics}\lb{SS-MFSCLim}

So far, we have considered the quantum dynamics of a single particle. In this section, we consider the quantum dynamics of $N$ identical particles. We are concerned
with two different limiting regimes:

\bu the large $N$ limit, and

\bu the semiclassical regime.

The large $N$ limit is of special interest: in that case, the problem is set on a space of large dimension (typically $3N$, the number of position variables for $N$ points
in space dimension $3$). Numerical simulations for problems of this type are very often untractable.

For this reason, one seeks to replace the original equation governing the dynamics of $N$-particle systems with ``reduced models'', where the space dimension does
not increase with the number of particles considered. The situation studied in this section is summarized by the following diagram.

\bigskip
\begin{center}
\begin{tabular}{ccc}
{\fbox{\bf von Neumann}}& {$\stackrel{N\to\infty}{\longrightarrow}$}& {\fbox{\bf Hartree}} 
\\ [3mm]
{$\downarrow$} & & {$\downarrow$} 
\\[3mm]
{${\hbar\to0}$}&$\searrow$& {${\hbar\to 0}$} 
\\ [3mm]
$\downarrow$&& $\downarrow$ 
\\ [3mm]
{\fbox{\bf Liouville}}& {$\stackrel{N\to\infty}{\longrightarrow}$}&{\fbox {\bf Vlasov}} 
\end{tabular}
\end{center}

\bigskip
The upper horizontal arrow (quantum mean-field limit) has been proved by Spohn \cite{SpohnRMP} --- see also \cite{BGM} --- in the case of bounded potentials; 
the case of a Coulomb potential was treated subsequently by \cite{ErdosYau} --- see also \cite{BEGMY}, by the method of the ``BBGKY hierarchy'' (see for instance 
\cite{FGTwente} for an elementary introduction to the BBGKY hierarchy). Other approaches involve the method of second quantization \cite{RodSchlein} --- see also 
an original method due to Pickl \cite{Pickl}, which makes use of some notions originating from second quantization without the full machinery of Fock spaces. Both 
\cite{RodSchlein} and \cite{Pickl} include a treatment of potentials with a singularity at the origin of the same type as for the Coulomb potential.

The lower horizontal arrow is the mean-field limit in classical mechanics; it has been proved by Braun-Hepp \cite{BraunHepp} by using the notion of ``Klimontovich''
solutions --- i.e. phase-space empirical measures --- of the Vlasov equation, with a convergence rate obtained by Dobrushin \cite{Dobru}, who used optimal transport
distances for the first time on this kind of problem. Dobrushin's analysis is our first motivation for defining a quantum analogue of the Wasserstein distance, and for the 
analysis in the present section.

The limit corresponding to the left vertical arrow follows from the Lions-Paul Theorem mentioned above (Theorem IV.1 in \cite{LionsPaul}). The limit corresponding to 
the right vertical arrow follows from Theorem IV.2 in \cite{LionsPaul}. It should be mentioned that the case of the classical limit of the Hartree equation for singular 
potentials including the Coulomb case is treated in Theorem IV.4 of \cite{LionsPaul} --- one should however keep in mind that this proof assumes that the Wigner
function of the density operator solution of the Hartree equation is assumed to be bounded in $L^2$, which excludes the case of pure states, such as solutions of
the Schr\"odinger equation with WKB initial data, for instance.

Since we are interested in a situation involving two small parameters (specifically $\tfrac1N$ and $\hbar$), it is natural to investigate the uniformity of the mean-field
(large $N$) limit in the semiclassical (small $\hb$) regime. It seems that the first result in that direction is \cite{GraffiMartPulvi} --- see also \cite{PulviPezzo}.

To conclude this introduction, one should also mention that the mean-field limit in classical mechanics (the lower horizontal arrow) remains an open problem in the
case of the Coulomb potential --- see however \cite{HaurayJabin1,HaurayJabin2} in the case of interactions with a singularity at the origin weaker than that of the
Coulomb potential, and \cite{Serfaty} in the case of the Coulomb potential itself, but for a restricted class of initial data (specifically for monokinetic data). The quantum 
analogue of \cite{Serfaty}, i.e. the joint mean-field and classical limit of the $N$-particle quantum dynamics in the case of monokinetic data, which is based on Serfaty's
remarkable inequality on the Coulomb potential (Proposition 2.3 in \cite{Serfaty}).

\subsubsection{Quantum $N$-Particle Dynamics}

The state of a quantum $N$-particle system at time $t$ is represented by a density operator $R(t)\in\cD(\fH_N)$, where $\fH_N=\fH^{\otimes N}$ is the $N$-particle
Hilbert space. If $\fH=L^2(\bR^d)$, then it is easily seen that $\fH_N=L^2(\bR^{dN})$.

Since the $N$ particles are indistinguishable, their density operator should commute with permutations of the particle labels. More precisely, for each $\si\in\fS_N$,
we define the map
$$
U_\si:\,\fH_N\ni\Psi\equiv\Psi(x_1,\ldots,x_N)\mapsto(U_\si \Psi)\equiv\Psi(x_{\si^{-1}(1)},\ldots,x_{\si^{-1}(N)})\in\fH_N\,.
$$
One easily checks that the map $\si\mapsto U_\si$ is a unitary representation of the symmetric group $\fS_N$ on the set of $N$ elements in $\fH_N$. A density 
operator for a system of $N$ identical particles should satisfy the relation
$$
U_\si R(t)U^*_\si=R(t)\,,\qquad \si\in\fS_N\,,\,\,t\in\bR\,.
$$
Henceforth, we denote by $\cD^s(\fH_N)$ the set of density operators on $\fH_N$ satisfying this symmetry property:
$$
\cD^s(\fH_N):=\{R\in\cD(\fH_N)\text{ s.t. }U_\si RU^*_\si=R\text{ for all }\si\in\fS_N\}\,.
$$
\textbf{Remark.} One should avoid confusing this symmetry, corresponding to indistinguishable particles, with the symmetries corresponding to the Bose-Einstein,
or the Fermi-Dirac statistics. The Bose-Einstein statistics applies to particles with integer spin, referred to as bosons, such as photons, $^4$He nuclei ($\a$ particles).
At low temperature, large numbers of bosons can condense in a a single energy state, thereby forming a Bose-Einstein condensate. The Fermi-Dirac statistics applies
to particles. with half-integer spin, referred to as fermions, such as electrons, protons, neutrons, $^3$He atoms. Fermions satisfy the Pauli exclusion principle: two
(or more than two) fermions in a given quantum system cannot occupy simultaneously the same quantum state. The difference between bosons and fermions can be 
read on their $N$-particle wave functions: for all $\si\in\fS_N$, one has
$$
\ba
U_\si\Psi_N=&\Psi_N\,,&&\qquad\text{ if the }N\text{ particles are bosons,}
\\
U_\si\Psi_N=&(-1)^{\text{sign}(\si)}\Psi_N\,,&&\qquad\text{ if the }N\text{ particles are fermions.}
\ea
$$
Notice that, in both cases, one has
$$
U_\si|\Psi_N\ra\la\Psi_N|U^*_\si=|U_\si\Psi_N\ra\la U_\si\Psi_N|=|\Psi_N\ra\la\Psi_N|\,,\qquad\si\in\fS_N\,,
$$
so that the density operator $|\Psi_N\ra\la\Psi_N|\in\cD^s(\fH_N)$.

\smallskip
\noindent
\textbf{The quantum $N$-particle Hamiltonian} is the unbounded operator on the $N$-particle Hilbert space $\fH_N=L^2(\bR^{dN})$
$$
\sum_{k=1}^N\!-\tfrac{\hb^2}{2m}\Dlt_{x_k}+\sum_{1\le k<l\le N}V(x_k\!-\!x_l)\,.
$$
Of course, it is assumed that $N\ge 2$.

Since the $N$ particles are identical, the total mass of the system is $M=Nm$, so that the ``energy per particle'' is
$$
\tfrac1N\left(\sum_{k=1}^N\!-\tfrac{\hb^2}{2m}\Dlt_{x_k}+\sum_{1\le k<l\le N}V(x_k\!-\!x_l)\right)=\sum_{k=1}^N\!-\tfrac{\hb^2}{2M}\Dlt_{x_k}+\tfrac1N\sum_{1\le k<l\le N}V(x_k\!-\!x_l)\,.
$$
Henceforth we set $M=1$ without loss of generality, and consider as the quantum Hamiltonian the energy per particle, i.e.
$$
\bH_N:=\sum_{k=1}^N\!-\tfrac{\hb^2}{2}\Dlt_{x_k}+\tfrac1N\sum_{1\le k<l\le N}V(x_k\!-\!x_l)\,.
$$

We shall adopt the following \textbf{assumptions on the interaction potential:}
$$
V\in C^{1,1}(\bR^d)\quad\text{ with }V^-\in L^\infty(\bR^d)\quad\text{ and }\quad V(z)=V(-z)\in\bR\,.
$$
Hence 
$$
\bH_N\text{ is self-adjoint on }\fH_N\text{ with domain }\Dom(\bH_N)\supset H^2(\bR^{dN})\,.
$$
By Stone's theorem, the quantum Hamiltonian $\bH_N$ generates a unitary group $\cU_N(t):=e^{-it\bH_N/\hb}$ on $\fH_N$. 

The operator $R_{\hb,N}(t)=\cU_N(t)R_N^{in}\cU_N(t)^*$ solves \textbf{the $N$-particle von Neumann equation}
$$
i\hb\d_tR_{\hb,N}(t)=[\bH_N,R_{\hb,N}(t)]\,,\qquad R_{\hb,N}(0)=R_N^{in}\in\cD(\fH_N)\,.
$$
If $R_N^{in}$ is a pure state, meaning that $R_N^{in}=|\Psi_N^{in}\ra\la\Psi_N^{in}|$, then
$$
R_{\hb,N}(t)=|\Psi_{\hb,N}(t)\ra\la\Psi_{\hb,N}(t)|
$$
where $\Psi_{\hb,N}(t,\cdot):=\cU_N(t)\Psi_N^{in}$ is the solution of the $N$-particle Schr\"odinger equation
$$
i\hb\Psi_{\hb,N}(t,\cdot)=\bH_N\Psi_{\hb,N}(t,\cdot)\,,\qquad\Psi_{\hb,N}(0,\cdot)=\Psi_N^{in}\,.
$$

We leave it to the reader, as an easy exercise, to check that 
$$
U_\si\cU_N(t)=\cU_N(t) U_\si\quad\text{ for all }\si\in\fS_N\text{ and all }t\in\bR\,.
$$
Therefore
$$
R_N^{in}\in\cD^s(\fH_N)\implies R_{\hb,N}(t)=\cU_N(t)R_N^{in}\cU_N(t)^*\in\cD^s(\fH_N)\,\quad\text{ for all }t\in\bR\,.
$$
Henceforth, we are interested in situations where $N\gg 1$ (mean-field regime) and $\hb\ll 1$ (semiclassical regime).

\subsubsection{Mean-Field Equations}

The purpose of the mean-field limit is to replace the description of a $N$-particle system by the quantum Hamiltonian $\bH_N$ --- involving functions of $N$ variable in 
$L^2(\bR^{dN})$, therefore functions of $dN\gg 1$ variables, by an equation, or a system of equations posed on the single-particle phase space $\fH$ instead of $\fH_N$.
The physical idea leading to these mean-field equations can be expressed as follows: one seeks to write an equation governing the evolution of a single, ``typical'' particle.
This typical particle is driven by the interaction with the $N-1$ other particles, approximated as follows: call $\rho(t,x)$ the single-particle \textbf{density function}, i.e.
$$
\rho(t,x)=r(t,x,x)
$$
where $r(t,x,y)$ is an integral kernel of $R(t)\in\cD(\fH)$, the density operator of the typical, single particle, such that $z\mapsto r(t,x+z,x)$ belongs to $C_b(\bR^d;L^1(\bR^d))$ 
for each $t\in\bR$ (see Lecture I, Quiz 4). Then, one expects that
$$
\ba
\tfrac1N\sum_{1\le j\le N}V(x_j-y)\simeq\int_{\bR^d}V(x-y)\rho(t,x)dx=&(V\star\rho(t,\cdot))(y)
\\
=&\Tr(V(\cdot-y)R(t))=:V_{R(t)}(y)
\ea
$$
as $N\to\infty$. This time-dependent potential $V_{R(t)}$ is usually referred to as the mean-field, self-consistent potential. Then one can write the equation governing
the (quantum) evolution of $R(t)$: this is a von Neumann equation where the potential is the mean-field potential, i.e.
$$
i\hb\d_tR(t)=[-\tfrac12\hb^2\Dlt+V_{R(t)},R(t)]\,.
$$
This is a quantum dynamical equation, referred to as the (time-dependent) \textbf{Hartree equation}, set on the single-particle Hilbert space $\fH=L^2(\bR^d)$, instead of the $N$-particle 
Hilbert space $\fH_N=L^2(\bR^{dN})$. The advantage of this description is obvious: one has to manipulate wave functions depending on $d$ space variables ($d\le 3$ in practice) instead 
of $Nd$ space variables, which is untractable. The drawback is that the mean-field equation is nonlinear (but the nonlinearity is relatively mild since it involves a convolution), and that it is 
only an approximation of the true, $N$-particle dynamics.

In the case of pure states, the (time-dependent) Hartree equation takes the form
$$
i\hb\d_t\psi(t,x)=-\tfrac12\hb^2\Dlt_x\psi(t,x)+\psi(t,x)V\star_x|\psi|^2(t,x)\,,\quad x\in\bR^d\,,
$$
and $R(t)=|\psi(t,\cdot)\ra\la\psi(t,\cdot)|$.

\smallskip
In the classical setting, the idea is the same: the classical mean-field potential is defined in terms of the single-particle distribution function $f\equiv f(t,x,\xi)$ by the formula
$$
V_f(t,x):=\iint_{\bR^d\times\bR^d}V(x-y)f(t,y,\eta)dyd\eta=(V\star f(t,\cdot,\cdot))(x)\,.
$$
Then, the mean-field equation, known as the \textbf{Vlasov equation} is written in terms of the mean-field Hamiltonian with the usual single particle Poisson bracket:
$$
\d_tf(t,x,\xi)+\underbrace{\xi\cdot\grad_xf(t,x,\xi)-\grad_xV_f(t,x)\cdot\grad_\xi f(t,x,\xi)}_{=\{\frac12|\xi|^2+V_f(t,x),f(t,x,\xi)\}}=0\,.
$$

\subsubsection{Uniform in $\hb$ Error Bounds for the Quantum Mean-Field Limit}

In this section, we present error bounds comparing the solution of the $N$-particle von Neumann equation and the solution of the quantum (Hartree), or the classical (Vlasov) 
mean-field equations.

We begin with the error bound for the quantum mean-field limit (from the $N$-particle von Neumann equation to the Hartree equation). This is the upper horizontal arrow in the 
diagram above.

\noindent
\textbf{Theorem 8.} Assume that $V\in C^{1,1}(\bR^d)$ is a real-valued function such that 
$$
V(z)=V(-z)\ge -M\quad\text{ for some }M>0\,,
$$
and set
$$
L:=2(1+4\text{Lip}(\grad V)^2)\,.
$$
Choose an initial ($1$-particle) distribution function $f^{in}$ such that $f^{in}dxd\xi\in\cP_2(\bR^{2d})$. Set the Hartree initial data to be $R^{in}=\cT[f^{in}]$, and the $N$-particle
initial data to be $R_N^{in}=(R^{in})^{\otimes N}=\cT[(f^{in})^{\otimes N}]$. Let $t\mapsto R(t)$ be the solution of the time-dependent Hartree equation with initial data $R^{in}$,
while $t\mapsto\cU_N(t)\cT[(f^{in})^{\otimes N}]\cU_N(t)^*$ is the solution of the $N$-particle von Neumann equation with initial data $R_N^{in}$.

Then, for each $t>0$,
$$
\frac{\fd\left(R(t)^{\otimes N},\cU_N(t)\cT[(f^{in})^{\otimes N}]\cU_N(t)^*\right)^2}N\le 2d\hb e^{Lt}+\frac{8\|\grad V\|_{L^\infty}}{N\!-\!1}\frac{e^{Lt}\!-\!1}{L}\,.
$$

\bigskip
Next we study the error bound for the joint mean-field and classical limits (from the $N$-particle von Neumann equation to the Vlasov equation). This is the diagonal arrow in the 
diagram above.

\noindent
\textbf{Theorem 9.} Under the same assumptions as in Theorem 8, choose an initial ($1$-particle) distribution function $f^{in}$ such that $f^{in}dxd\xi\in\cP_2(\bR^{2d})$. Set the 
$N$-particle initial data to be $R_N^{in}=\cT[(f^{in})^{\otimes N}]$. Let $t\mapsto f(t,\cdot,\cdot)$ be the solution of the Vlasov equation with initial data $f^{in}$, while the solution 
of the $N$-particle von Neumann equation with initial data $R_N^{in}$ is $t\mapsto\cU_N(t)\cT[(f^{in})^{\otimes N}]\cU_N(t)^*$. 

Then, for each $t>0$,
$$
\frac{\fd\left(f(t,\cdot,\cdot))^{\otimes N},\cU_N(t)\cT[(f^{in})^{\otimes N}]\cU_N(t)^*\right)^2}N\le d\hb e^{Lt }+\frac{8\|\grad V\|_{L^\infty}}{N\!-\!1}\frac{e^{Lt}\!-\!1}{L}\,.
$$

\bigskip
Several comments are in order before going further.

\noindent
\textbf{Remarks.}

\noindent
(1) In both Theorems 8 and 9, we have restricted our attention to Toeplitz initial density operators. The reason for using such well-prepared initial data comes from statement (1)
in Theorem 1: if $T$
is a Toeplitz operator in $\cD_2(\fH)$, then 
$$
\fd(T,T)=\sqrt{2d\hb}=\min_{R\in\cD(\fH)}\fd(R,R)\,.
$$
Similarly, if $f$ is the symbol of $T$ --- i.e. if $T=\cT[f]$ --- then
$$
\fd(f,T)=\sqrt{d\hb}=\min_{g\in\cP_2(\bR^{2d})\atop R\in\cD_2(\fH)}\fd(g,R)\,.
$$
Therefore, for this choice of initial data, the first terms in the right-hand side of the upper bounds in both theorems are as small as possible. 

\noindent
(2) The reason for considering the expressions
$$
\frac{\fd\left(R(t)^{\otimes N},\cU_N(t)\cT[(f^{in})^{\otimes N}]\cU_N(t)^*\right)^2}N
$$
and
$$
\frac{\fd\left(f(t,\cdot,\cdot))^{\otimes N},\cU_N(t)\cT[(f^{in})^{\otimes N}]\cU_N(t)^*\right)^2}N\,,
$$
instead of
$$
\fd\left(R(t)^{\otimes N},\cU_N(t)\cT[(f^{in})^{\otimes N}]\cU_N(t)^*\right)^2
$$
and
$$
\fd\left(f(t,\cdot,\cdot))^{\otimes N},\cU_N(t)\cT[(f^{in})^{\otimes N}]\cU_N(t)^*\right)^2
$$
comes from the fact that the quantum-to-quantum transport cost and the classical-to-quantum transport cost in $\fH_N$ are sums of $N$ terms, respectively
$$
\sum_{k=1}^NI_{\fH\otimes\fH}^{\otimes (k-1)}\otimes C_\hb\otimes I_{\fH\otimes\fH}^{\otimes (N-k)}\,,
$$
and
$$
\sum_{k=1}^NI_{\fH}^{\otimes (k-1)}\otimes c_\hb(x_k,\xi_k)\otimes I_{\fH}^{\otimes (N-k)}\,.
$$
(3) For all $R_N\in\cD^s(\fH_N)$, we consider its \textbf{$k$-th marginal} $R_{N:k}\in\cD^s(\fH_k)$ defined by the following prescription
$$
\Tr_{\fH_k}(R_{N:k}A)=\Tr_{\fH_N}(R_N(A\otimes I^{N-k}_{\fH_{N-k}}))\,,\quad\text{ for all }A\in\cL(\fH_k)\,.
$$
\textbf{Quiz 19.} Let $R_N,S_N\in\cD^s_2(\fH_N)$ and let $f_N$ be a symmetric probability density on $(\bR^d\times\bR^d)^N$ with finite second order moments.
Prove that
$$
\fd(R_{N:1},S_{N:1})^2\le\ldots\le\frac{\fd(R_{N:k},S_{N:k})^2}{k}\le\ldots\le\frac{\fd(R_N,S_N)^2}{N}\,,\quad 1\le k\le N\,,
$$
and that
$$
\fd(f_{N:1},S_{N:1})^2\le\ldots\le\frac{\fd(f_{N:k},S_{N:k})^2}{k}\le\ldots\le\frac{\fd(f_N,S_N)^2}{N}\,,\quad 1\le k\le N\,,
$$
where
$$
f_{N:k}(x_1,\xi_1,\ldots,x_k,\xi_k):=\int_{(\bR^d\times\bR^d)^{N-k}}f_N(x_1,\xi_1,\ldots,x_N,\xi_N)dx_{k+1}d\xi_{k+1}\ldots dx_Nd\xi_N\,.
$$
(Hint: use the formulas for the quantum-to-quantum and the classical-to-quantum costs in (2)).

In particular, Theorems 8 and 9 imply the following result on the first marginal of the $N$-particle density operator.

\noindent
\textbf{Corollary 10.} Under the same assumptions as in Theorems 8 and 9, one has
$$
\ba
\fd\left(R(t),(\cU_N(t)\cT[(f^{in})^{\otimes N}]\cU_N(t)^*)_{:1}\right)^2\le 2d\hb e^{Lt}+\frac{8\|\grad V\|_{L^\infty}}{N\!-\!1}\frac{e^{Lt}\!-\!1}{L}\,,
\\
\fd\left(f(t,\cdot,\cdot)),(\cU_N(t)\cT[(f^{in})^{\otimes N}]\cU_N(t)^*)_{:1}\right)^2\le d\hb e^{Lt }+\frac{8\|\grad V\|_{L^\infty}}{N\!-\!1}\frac{e^{Lt}\!-\!1}{L}\,.
\ea
$$
(4) One can obviously use Theorems 8 and 9, together with Corollary 10 and Theorem 1 (2) in order to compare the Vlasov solution and the Husimi transform of the Hartree
solution to the Husimi transform of the first marginal of the $N$-particle density operator.

\noindent
\textbf{Corollary 11.} Under the same assumptions as in Theorems 8 and 9, one has
$$
\ba
\cW_2\left(\cH[R(t)],\cH[(\cU_N(t)\cT[(f^{in})^{\otimes N}]\cU_N(t)^*)_{:1}]\right)^2\le 2d\hb(e^{Lt}+1)+\frac{8\|\grad V\|_{L^\infty}}{N\!-\!1}\frac{e^{Lt}\!-\!1}{L}\,,
\\
\cW_2\left(f(t,\cdot,\cdot)),\cH[(\cU_N(t)\cT[(f^{in})^{\otimes N}]\cU_N(t)^*)_{:1}]\right)^2\le d\hb(e^{Lt}+1)+\frac{8\|\grad V\|_{L^\infty}}{N\!-\!1}\frac{e^{Lt}\!-\!1}{L}\,.
\ea
$$
(5) Since density operators are characterized by their Husimi transforms (see Quiz 20 below), and since the Husimi transform of a density operator is a probability density, 
a natural idea to define optimal transport distances between density operators is to use the formula
$$
d_{ZS}(\rho_1,\rho_2)=\cW_2(\cH[\rho_1],\cH[\rho_2])\,,\qquad\rho_1,\rho_2\in\cD_2(\fH)\,.
$$
This approach has been proposed by K. \.Zyczkowski and W. S\l omczy\'nski \cite{ZycSlo}.

This definition has some advantages over the pseudometric $\fd$ discussed in these lectures

\noindent
(a) it is a bona fide metric (see quiz below), and

\noindent
(b) one is always dealing with probability densities, i.e. functions on phase space, which are easier to manipulate than operators.

Similarly, one might prefer considering
$$
\cW_2(f,\cH[R])\quad\text{ instead of }\quad\fd(f,R)
$$
for $R\in\cD_2(\fH)$ and $f$ a probability density on $\cP(\bR^d\times\bR^d)$ with finite second order moments. Indeed, with the first quantity, one transforms the density operator
$R$ into a probability density by means of the Husimi transform, and then compares this probability density with $f$ by means of the Wasserstein distance. On the contrary, the
second quantity compares two very different objects (a probability density and a density operator), which is like comparing apples with pears.

\noindent
\textbf{Quiz 20: the Husimi transform is one-to-one.} Set $\fH=L^2(\bR^d)$.

\noindent
(1) Let $R\in\cD_2(\fH)$. Prove that $\cH[R]$ is a probability density, and compute
$$
\iint_{\bR^d\times\bR^d}(|q|^2+|p|^2)\cH[R](q,p)dqdp
$$
in terms of $R$.

\noindent
(2) Let $R,S\in\cD_2(\fH)$, and assume that $\cH[R]=\cH[S]$. Prove that $R=S$. (Hint: let $r\equiv r(y,y')$ be an integral kernel of $R$. Set
$$
J(x,\xi):=\iint_{\bR^d\times\bR^d}r(y,y')e^{-(|y|^2+|y'|^2)/2\hb}e^{x\cdot(y+y')-i\xi\cdot(y-y')/\hb}dydy'.
$$
Prove that $J$ extends as a holomorphic function on $\bC^d\times\bC^d$, and therefore is uniquely determined by its restriction to $\bR^d\times\bR^d$. Conclude (a) by finding 
the formula relating $\cH[R]$ to $J$, and (b) by computing the integral kernel $r$ of $R$ in terms of $J$.)

\smallskip
However, there is a rather heavy price to pay with this approach, which is that the Husimi transform, and therefore $d_{ZS}$ is not easy to propagate by usual quantum dynamics.
This is easily explained if one returns to Quiz 11 (3) and (5): if $t\mapsto R(t)$ is a time-dependent density operator, solution of the von Neumann equation
$$
i\hb\d_tR(t)=[-\tfrac12\hb^2\Dlt+V,R(t)]\,,\qquad R(0)=R^{in}\,,
$$
we have seen that its Wigner transform $W_\hb[R(t)]$ satisfies the Wigner equation
$$
(\d_t+\xi\cdot\grad_x)W_\hb[R(t)](x,\xi)+\Theta[V]W_\hb[R(t)](x,\xi)=0\,,
$$
and that
$$
\cH[R(t)]=\exp(\tfrac{\hb}4\Dlt_{x,\xi})W_\hb[R(t)]\,.
$$
Therefore, the evolution of $\cH[R(t)]$ can be described as follows:
$$
\cH[R^{in}]\mapsto W[R^{in}]\mapsto W_\hb[R(t)]\mapsto\cH[R(t)]\,.
$$
The second arrow is the value at time $t$ of the group generated by the Wigner equation, and the third arrow is $\exp(\tfrac{\hb}4\Dlt_{x,\xi})$. But the first arrow corresponds to
expressing $W_\hb[R^{in}]$ in terms of $\cH[R^{in}]$ by solving
$$
\exp(\tfrac{\hb}4\Dlt_{x,\xi})W_\hb[R^{in}]=\cH[R^{in}]\,.
$$
This first step corresponds to inverting the heat flow at time $\hb/4$, hence the difficulty of this problem. 

A more thorough discussion of the evolution of the Husimi transform of a solution of the von Neumann equation can be found in \cite{Athanassoulis}.

This discussion explains one of the advantages of the pseudometric $\fd$ over $d_{KS}$, that is the remarkable simplicity of the propagation estimate stated in Theorem 6, and
the fact that the bounds reported in that theorem are exactly the same as the pair dispersion estimate in classical mechanics (with the same potential in both dynamics). In other 
words, when compared by means of the pseudometric $\fd$, quantum particles behave as if they were points moving on classical phase-space trajectories. This picture of quantum
dynamics is known to be wrong in general --- see for instance Young's double-slit experiment and its discussion in chapter I, section 2 of \cite{CTDL} --- but turns out to be true as
far as the pseudodistance $\fd$, or its variants $\fd_\lbd$ discussed in Theorem 6, are concerned. Accordingly, the assumptions on the potential used in Theorem 6 are those used
in the definition of a classical dynamics by means of the Cauchy-Lipschitz theorem, which require in particular that $V\in C^{1,1}(\bR^d)$, which are much more restrictive than 
the conditions on $V$ which imply that $-\frac{\hb^2}{2m}\Dlt+V$ has a self-adjoint extension as an unbounded operator on $L^2(\bR^d)$.

\medskip
After this long list of remarks, here is a quick sketch of the proof of Theorems 8. Theorem 9 is proved in essentially the same way, and the necessary modifications are left to the 
reader as a (relatively) easy exercise.

\begin{proof}[Sketch of the proof of Theorem 8] This proof is split in several steps.

\noindent
\underline{Step 1.} Pick $\cQ_N^{in}\in\cC(R_N^{in},(R^{in})^{\otimes N})$ and solve for $\cQ_N(t)\in\cD(\fH_N\otimes\fH_N)$ the linear von Neumann equation with time-dependent
potential
$$
\left\{
\ba
{}&i\hb\d_t\cQ_N(t)\!=\!\left[\bH_N\!\otimes\! I_{\fH_N}+I_{\fH_N}\otimes\sum_{k=1}^NJ_{k,N}(-\tfrac{\hb^2}2\Dlt\!+\!V_{R(t)}),\cQ_N(t)\right]\,,
\\
&\cQ_N(0)=\cQ_N^{in}\,,\ea
\right.
$$
with the notation
$$
J_{k,N}A:=I_\fH^{\otimes(k-1)}\otimes A\otimes I_\fH^{\otimes(N-k)}\,.
$$
One checks, by taking partial traces and using uniqueness for the solution of the von Neumann equation with time-dependent potential, that
$$
\cQ_N(t)\in\cC(R_N(t),R(t)^{\otimes N})\,,\qquad\text{ for all }t\ge 0\,,
$$
where $R(t)$ is the Hartree solution, while 
$$
R_N(t):=\cU_N(t)R_N^{in}\cU_N(t)^*\,.
$$
\underline{Step 2.} Let 
$$
D_N(t):=\tfrac1N\Tr_{\fH_N\otimes\fH_N}(\cQ_N(t)^\frac12C_\hb\cQ_N(t)^\frac12)\,,
$$
with $C_\hb$ defined for all $\Phi\equiv\Phi(X_N,Y_N)\in\cS(\bR^{2dN})$ by 
$$
C_\hb\Phi\!:\equiv\!\sum_{j=1}^N(|x_j-y_j|^2\Phi-\hb^2(\Div_{x_j}\!-\!\Div_{y_j})((\grad_{x_j}\!-\!\grad_{y_j})\Phi))(X_N,Y_N)\,.
$$
\underline{Step 3.} By definition of $D_N(t)$, Step 1 shows that
$$
D_N(t)\ge\tfrac1N\fd(R_N(t),R(t)^{\otimes N})^2\,,\qquad\text{ for all }t\ge 0\,.
$$
\underline{Step 4.} On the other hand
$$
\ba
i\hb\frac{dD_N}{dt}=\tfrac1N\Tr_{\fH_N\otimes\fH_N}(\cQ_N(t)^\frac12[\bH_N\otimes I_{\fH_N},C_\hb]\cQ_N(t)^\frac12)
\\
+\sum_{k=1}^N\tfrac1N\Tr_{\fH_N\otimes\fH_N}(\cQ_N(t)^\frac12[I_{\fH_N}\otimes J_{k,N}(-\tfrac12\hb^2\Dlt+V_{R(t)}),C_\hb]\cQ_N(t)^\frac12)&\,,
\ea
$$
and it remains to compute
$$
Z_N=-\frac{i}{\hb}\left[\bH_N\otimes I_{\fH_N}+I_{\fH_N}\otimes\sum_{k=1}^NJ_{k,N}(-\tfrac12\hb^2\Dlt+V_{R(t)}),C_\hb\right]\,.
$$
\underline{Step 5.} Using the notation $A\vee B:=AB+BA$, one finds that
$$
\ba
Z_N=\sum_{j=1}^N(x_j-y_j)\vee(-i\hb\grad_{x_j}+i\hb\grad_{y_j})
\\
+\sum_{j=1}^N\tfrac1N\sum_{k=1}^N(-\grad V(x_j-x_k)+\grad V_{R(t)}(y_j))\vee(-i\hb\grad_{x_j}+i\hb\grad_{y_j})&\,,
\ea
$$
and one uses the elementary operator inequality (see Quiz 7 (2))
$$
AB^*+BA^*\le AA^*+BB^*
$$
to prove that
$$
Z_N\le 2C_\hb+\sum_{j=1}^N\left|\grad V_{R(t)}(y_j)-\tfrac1N\sum_{k=1}^N\grad V(x_j-x_k)\right|^2.
$$
Split the summand so as to involve the difference between the $N$-body and the mean-field potentials on the $y_j$ variables only
$$
\ba
Z_N\le&2C_\hb+2\sum_{j=1}^N\left|\tfrac1N\sum_{k=1}^N(\grad V(y_j-y_k)-\grad V(x_j-x_k))\right|^2
\\
&+2\sum_{j=1}^N\left|\grad V_{R(t)}(y_j)-\tfrac1N\sum_{k=1}^N\grad V(y_j-y_k)\right|^2
\\
\le&2C_\hb+\tfrac2N\Lip(\grad V)^2\underbrace{\sum_{j,k=1}^N|(y_j-y_k)-(x_j-x_k)|^2}_{\le 4NC_\hb}
\\
&+2\sum_{j=1}^N\left|\grad V_{R(t)}(y_j)-\tfrac1N\sum_{k=1}^N\grad V(y_j-y_k)\right|^2\,.
\ea
$$
\underline{Step 6.} It remains to bound
$$
\ba
\sum_{j=1}^N\Tr_{\fH_N\otimes\fH_N}\left(\cQ_N(t)^\frac12\left|\grad V_{R(t)}(y_j)-\tfrac1N\sum_{k=1}^N\grad V(y_j-y_k)\right|^2\cQ_N(t)^\frac12\right)
\\
=\sum_{j=1}^N\Tr_{\fH_N}\left(\left|\grad V_{R(t)}(y_j)-\tfrac1N\sum_{k=1}^N\grad V(y_j-y_k)\right|^2R(t)^{\otimes N}\right)&\,.
\ea
$$
This is easy since all the potentials considered involve only the variables $y_j$ for $j=1,\ldots,N$, so that $\cQ_N(t)$ can be replaced with the \textbf{factorized} density 
$R(t)^{\otimes N}$, where $R(t)$ is the Hartree solution. Here is the argument.
$$
\ba
\Tr_{\fH_N}\left(\left|\grad V_{R(t)}(y_1)-\tfrac1N\sum_{k=1}^N\grad V(y_1-y_k)\right|^2R(t)^{\otimes N}\right)
\\
=\Tr_{\fH_N}\left(\left|\tfrac1N\sum_{k=1}^N(\grad V_{R(t)}(y_1)-\grad V(y_1-y_k))\right|^2R(t)^{\otimes N}\right)
\\
=\frac1{N^2}\Tr_{\fH_N}\left(\sum_{k=1}^N|\grad V_{R(t)}(y_1)-\grad V(y_1-y_k)|^2R(t)^{\otimes N}\right)
\\
+2\sum_{1\le k<m\le N}\Tr_{\fH_N}\left(\tfrac{\grad V_{R(t)}(y_1)-\grad V(y_1-y_k)}{N}\cdot\tfrac{\grad V_{R(t)}(y_1)-\grad V(y_1-y_m)}NR(t)^{\otimes N}\right)
\\
=\frac1{N^2}\Tr_{\fH_N}\left(\sum_{k=1}^N|\grad V_{R(t)}(y_1)-\grad V(y_1-y_k)|^2R(t)^{\otimes N}\right)\le\frac4N\|\grad V\|_{L^\infty(\bR^d)}^2&\,,
\ea
$$
since
$$
\Tr_{\fH_N}\left((\grad V_{R(t)}(y_1)-\grad V(y_1-y_k))\cdot(\grad V_{R(t)}(y_1)-\grad V(y_1-y_m))R(t)^{\otimes N}\right)=0
$$
for all $m>k\ge 1$. 

\noindent
\underline{Step 7.} Hence
$$
Z_N\le 2(1+4\Lip(\grad V)^2)C_\hb+2\sum_{j=1}^N\left|\grad V_{R(t)}(y_j)-\tfrac1N\sum_{k=1}^N\grad V(x_j-x_k)\right|^2\,,
$$
and therefore
$$
\ba
\frac{d}{dt}D_N(t)=&\Tr_{\fH_N\otimes\fH_N}(\cQ(t)^\frac12 Z_N\cQ(t)^\frac12)
\\
\le& 2(1+4\Lip(\grad V)^2)D_N(t)+\frac8N\|\grad V\|_{L^\infty(\bR^d)}^2\,.
\ea
$$
By Step 3 and the Gronwall inequality, one has
$$
\tfrac1N\fd(R_N(t),R(t)^{\otimes N})^2\le D_N(t)\le D_N(0)e^{Lt}+\frac8N\|\grad V\|^2_{L^\infty(\bR^d)}\frac{e^{Lt}-1}{L}\,.
$$
This inequality holds for each initial coupling $\cQ_N^{in}$ of $(R^{in})^{\otimes N}$ with $R_N^{in}$, so that
$$
\tfrac1N\fd(R_N(t),R(t)^{\otimes N})^2\le\tfrac1N\fd(R_N^{in},(R^{in})^{\otimes N})^2e^{Lt}+\frac8N\|\grad V\|^2_{L^\infty(\bR^d)}\frac{e^{Lt}-1}{L}\,.
$$
Specializing this inequality to 
$$
R^{in}=\cT[f^{in}]\quad\text{ and }\quad R_N^{in}=(R^{in})^{\otimes N}
$$
and using Theorem 1 (1) leads to the announced result.
\end{proof}

\subsection{Time-Splitting Numerical Schemes for Quantum Mechanics}\lb{SS-Split}

In this section, we study the numerical error relative to some time-discretizations of quantum dynamical equations.

Consider the \textbf{von Neumann equation} with unkown $R(t)\in\cD(\fH)$
$$
i\hbar\d_tR=[-\tfrac12\hbar^2\Dlt+V,R]\,,\qquad R\rstr_{t=0}=R^{in}\,,
$$
where the potential $V$ is chosen so that the operator
$$
\bH_\hb:=-\tfrac12\hbar^2\Dlt+V
$$
has a self-adjoint (unbounded) extension to $\fH=L^2(\bR^d)$.

Our purpose is to obtain uniform in $\hb$ error bounds for various time-splitting schemes for the von Neumann equation. The simplest example of time-splitting scheme for the
von Neumann equation is as follows:

\smallskip
\noindent
\textbf{Time-split Heisenberg equation.} Starting from $R^0=R^{in}\in\cD(\fH)$, we define a sequence $R^m(t)\in\cD(\fH)$ for $m\in\tfrac12\bN$ by the following formulas:
$$
\ba
R^{n+\frac12}=&\exp(\tfrac{i\hbar\Dlt t}2\Dlt)R^n\exp(-\tfrac{i\hbar\Dlt t}2\Dlt)\,,
\\
R^{n+1}=&\exp(\tfrac{\Dlt t}{i\hbar}V)R^{n+\frac12}\exp(-\tfrac{\Dlt t}{i\hbar}V)\,.
\ea
$$
The analysis of time-splitting schemes like this one has been studied in detail, for instance by \cite{DescombesThalhammer}. This particular time-splitting scheme is known as
the \textbf{Lie-Trotter splitting method}.

\noindent
\textbf{Error bound. \cite{DescombesThalhammer}} Pick $R^{in}\in\cD(\fH)$ such that
$$
\|\la\hbar D_x\ra R^{in}\la\hbar D_x\ra\|_1\!=:\!M<\infty\qquad\text{ with }\la\hbar D_x\ra\!:=\!(1\!-\!\hbar^2\Dlt_x)^\frac12\,.
$$
Then, for each integer $n\ge 0$, one has
$$
\|R(n\Dlt t)-R^n\|_1\le C(M,\|V\|_{W^{2,\infty}})\tfrac{\Dlt t}{\hbar}\,.
$$
This error bound is obviously \textbf{not uniform} as $\hbar\to 0$. The convergence of this scheme requires that $\Dlt t\ll\hbar$, which is obviously very costly in the semiclassical
regime, where $\hb\ll 1$.

\smallskip
In the sequel, we seek to obtain an error bound for the Lie-Trotter time-splitting scheme that is uniform in $\hb$ by the methodology of \textbf{asymptotic preserving (AP)} schemes.
The idea is to use the Descombes-Thalhammer error bound for $\hb\ge 1$, and, in the case where $\hb\ll 1$, to use the same time-splitting scheme for the classical Liouville
equation, which is the classical dynamical equation obtained by passing to the $\hb\to 0$ limit in the von Neumann equation, as explained for instance by the Lions-Paul theorem
recalled above. This last error estimate involves comparing the numerical quantum and the classical solutions, and the exact classical and numerical solutions on the other hand.
This leads to two different error bounds depending on $\hb$, and one hopes to get a uniform in $\hb$ error estimate by optimizing in $\hb>0$. This is summarized by the diagram
in Figure 5.

The first result on the uniform in $\hb$ convergence of the Lie-Trotter method for the von Neumann equation was obtained in \cite{BaoJinMarko}, without any error bound. The goal 
of the present section is to provide such uniform in $\hb$ error estimates by means of the $\fd$ pseudometric.

\begin{figure}

\includegraphics[width=11cm]{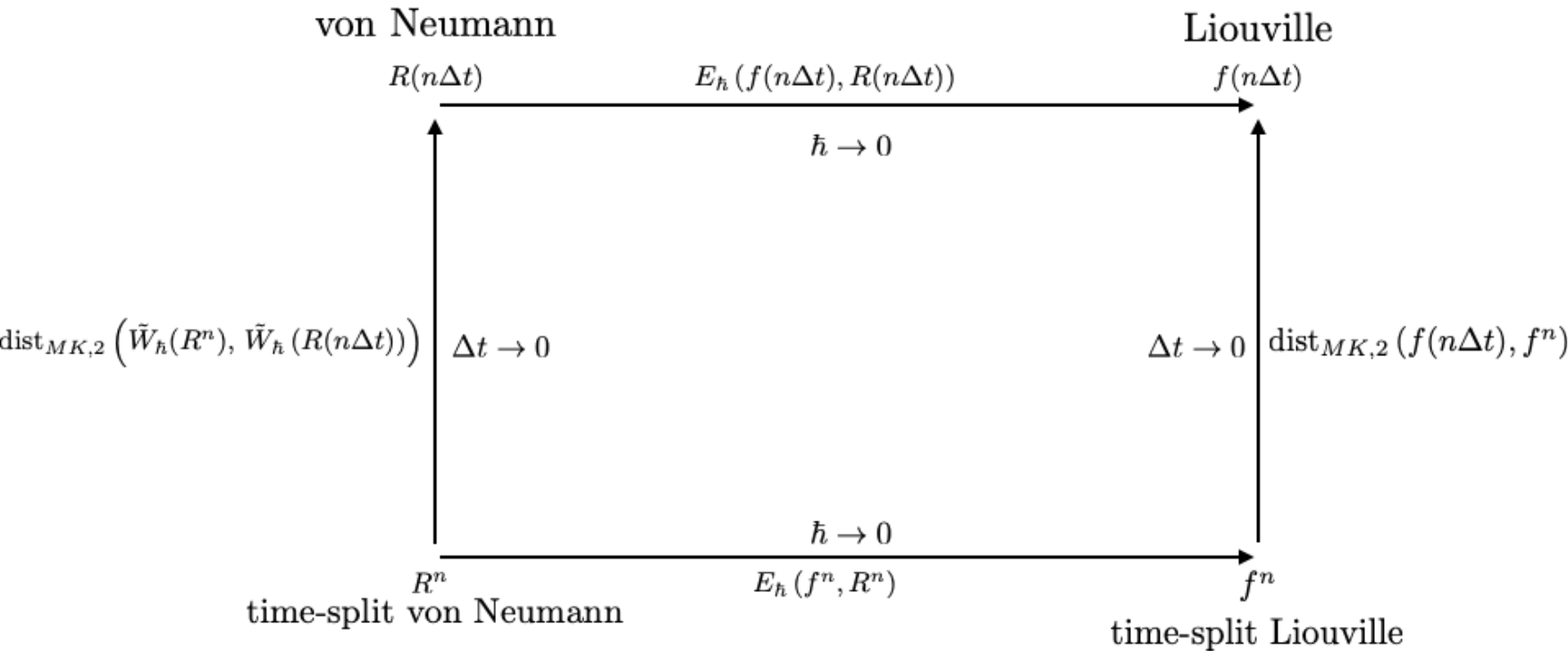}

\caption{The horizontal arrows represent the semiclassical limit $\hbar\ll 1$ and the vertical arrows the convergence of the numerical scheme $\Dlt t\ll 1$.}

\end{figure}

\subsubsection{Lie-Trotter Time Splitting for the Liouville Equation}

As recalled above, the classical dynamical equation corresponding to the von Neumann equation is the \textbf{Liouville equation} with unknown $f\equiv f(t,x,\xi)\ge 0$, written as
$$
\d_tf(t,x,\xi)+\{H(x,\xi),f(t,x,\xi)\}=0\,,\qquad f\rstr_{t=0}=f^{in}\,,
$$
where the classical Hamiltonian is
$$
H(x,\xi):=\tfrac12|\xi|^2+V(x)\,.
$$
We shall assume here that $V\in W^{2,\infty}(\bR^d)$, and that $V\ge -M$ on $\bR^d$ for some constant $M>0$.

This is a first order PDE that is solved by the \textbf{method of characteristics}. If one denotes by $\Phi_t$ the Hamiltonian flow generated by $H$, one finds that
$$
f(t,x,\xi)=f^{in}(\Phi_{-t}(x,\xi))\,,\qquad x,\xi\in\bR^d\,,\,\,t\in\bR\,.
$$
The \textbf{Lie-Trotter time-splitting} method for the Liouville equation is
$$
\ba
f^{n+\frac12}(y,\eta)&=f^n\circ K_{-\Dlt}(y,\eta)\,,
\\
f^{n+1}(y,\eta)&=f^{n+\frac12}\circ P_{-\Dlt t}(y,\eta)\,,
\ea
$$
starting from $f^0:=f^{in}$, where the maps $K_{-\Dlt t}$ and $P_{-\Dlt t}$ are defined by the following formulas:
$$
K_t(y,\eta):=(y+t\eta,\eta)\,,\qquad P_t(y,\eta):=(y,\eta-t\grad V(y))\,.
$$
The \textbf{error bound for the Lie-Trotter splitting scheme for the Liouville equation} is formulated below in terms of the Wasserstein distance of exponent $2$. 
Setting 
$$
(X_t,\Xi_t):=\Phi_t(x,\xi)\quad\text{ and }\quad(Y_t,H_t):=P_t\circ K_t(y,\eta)\,,
$$
one finds that
$$
\ba
|X_t-Y_t|^2+|\Xi_t-H_t|^2\le&(|x-y|^2+|\xi-\eta|^2)e^{(2+\L)|t|}
\\
&+\frac{e^{(2+\L)|t|}-1}{2+\L}\tfrac94\L^2(\tfrac12+\L)^2t^2(1+|y|^2+|\eta|^2)\,,
\ea
$$
with
$$
\L:=\max(1,E,\|\grad^2V\|_{L^\infty})\,,\quad E:=|\grad V(0)|\,.
$$
\textbf{Quiz 21.} Prove the inequality above. (Hint: write a differential system satisfied by $t\mapsto(Y_t,H_t)$, and compare the trajectories of that system with
those of the Hamiltonian system corresponding to $H(x,\xi)$ by means of the Gronwall inequality.)

\smallskip
The inequality above implies the following error estimate for the Lie-Trotter scheme in the case of the Liouville equation.

\noindent
\textbf{Lemma 12.} Let $V\in W^{2,\infty}(\bR^d)$ satisfy $V\ge -M$ for some $M>0$. Assume that $f^{in}$ is a probability density on $\bR^{2d}$ such that
$$
\iint_{\bR^d\times\bR^d}(|x|^2+|\xi|^2)f^{in}(x,\xi)dxd\xi<\infty\,.
$$
Then, for each $\Dlt t\in(0,1)$ and each $n=0,\ldots,[T/\Dlt t]$, one has
$$
\cW_2(f^n,f(n\Dlt t))\le C_T[\L,E,f^{in}]\Dlt t\,,
$$
where $C_T[\L,E,f^{in}]$ is a positive constant depending only on the computing time $T$, on the constants $\L$ and $E$ defined above, and on the initial data $f^{in}$.

\subsubsection{Lie-Trotter Time Splitting for the Liouville Equation}

At this point, we use Theorem 6 first with $\lbd=m=1$ and $V=0$, so that
$$
\fd(f^n\circ K_{-\Dlt t},R^{n+\frac12})\le\fd(f^n,R^n)e^{\frac{\Dlt t}2}\,.
$$
Next we use Theorem 6, this time with $\lbd=1$ and $m\to+\infty$, so that
$$
\fd(f^{n+\frac12}\circ P_{-\Dlt t},R^{n+1})\le\fd(f^{n+\frac12},R^{n+\frac12})e^{\frac{\Dlt t}2\Lip(\grad V)}\,.
$$
Hence, for all integer $n\ge 0$, one has
$$
\fd(f^n,R^n)\le\fd(f^{in},R^{in})\exp\left(\tfrac12n\Dlt t(1+\Lip(\grad V))\right)\,.
$$

On the other hand, applying Theorem 6 with $m=1$ shows that, for all $t\ge 0$,
$$
\ba
\fd(f(t,\cdot,\cdot),R(t))=&\fd(f^{in}\circ\Phi_{-t},e^{-it\bH_\hb/\hb}R^{in}e^{it\bH_\hb/\hb})
\\
\le&\fd(f^{in},R^{in})\exp\left(\tfrac12t(1+\Lip(\grad V))\right)\,.
\ea
$$

Putting together these two bounds, and the error estimate in the previous section, we arrive at the following result.

\noindent
\textbf{Theorem 13.} Let $V\in W^{2,\infty}(\bR^d)$ satisfy $V\ge -M$ for some $M>0$. Let $f^{in}$ be a probability density on $\bR^d\times\bR^d$ such that
$$
\iint_{\bR^d\times\bR^d}(|x|^2+|\xi|^2)f^{in}(x,\xi)dxd\xi<\infty\,.
$$
Set $R^{in}=\cT[f^{in}]$. Then, the Lie-Trotter splitting scheme for the Heisenberg equation satisfies the uniform as $\hb\to 0$ error bound
$$
\fd(R^n,R(n\Dlt t))\le C_T[\L,\|V\|_{W^{2,\infty}},f^{in}]\Dlt t+2\sqrt{d\hbar}e^{\frac{T}2(1+\Lip(\grad V))}\,.
$$
This implies the uniform in $\hbar$ convergence rate
$$
\ba
\sup_{\max(2\|\phi\|_{L^\infty},\Lip(\phi))\le 1}\left|\int_{\bR^{2d}}\phi(x,\xi)(\cH[R^n]-\cH[R(n\Dlt t)])dxd\xi\right|
\\
\le C'_T[\L,\|V\|_{W^{2,\infty}},f^{in}]\Dlt t^{1/3}&\,.
\ea
$$

\smallskip
The uniform as $\hb\to 0$ error estimate follows from the two bounds obtained in this section before Theorem 13, and the error bound for the Lie-Trotter
method for the Liouville equation in the previous section (Lemma 12). One uses also the following triangle inequality, which will be discussed in Lecture III:
$$
\fd(R,S)\le\fd(R,f)+\cW_2(f,g)+\fd(g,S)\,.
$$
The uniform in $\hbar$ convergence rate follows from optimizing between the uniform as $\hbar\to 0$ bound and the Descombes-Thalhammer bound.

\smallskip
There is a similar result with higher order splitting formulas, such as Strang's splitting scheme, leading to a uniform $O\left(\Dlt t^{2/3}\right)$ estimate. 
The interested reader is referred to \cite{FGJinPaul} for these higher order estimates, and for the missing details on the case of the Lie-Trotter splitting 
method.

\subsection{Observation Inequalities for Quantum Dynamics}\lb{SS-Observ}

The observation problem for a general PDE can be formulated as follows: let $P(x,\d_x)$ be a (linear) partial differential operator on some open set $\Om\subset\bR^d$,
and let $\om$ be an open subset of $\Om$. Is a solution $u$ of the PDE $P(x,\d_x)u(x)=0$ for all $x\in\Om$ completely determined by its restriction to $\om$?

\noindent
\textbf{Example.} Here is an elementary example. Set $d=2$, and identify $\bR^2$ to $\bC$ (by sending $(x,y)\in\bR^2$ to $z=x+iy\in\bC$). Set
$$
P(x,y,D_x,D_y)=\bar\d=\tfrac12(\d_x+i\d_y)\,,
$$
(the Cauchy-Riemann operator) and set $\om\subset\Om\subset\bC$ be (nonempty) open sets in $\bC$. Any distribution $u$ on $\Om$ such that $\bar\d u=0$ on $\Om$ is 
a holomorphic function on $\Om$. The restriction of $u$ to the connected component of $\om$ in $\Om$ is uniquely determined by the restriction $u\rstr_{\om}$. In particular, 
if $u\rstr_{\om}=0$, then $u=0$ on $\Om$.

\smallskip
In the context of quantum dynamics, the observation problem can be formulated as follows. Consider the \textbf{von Neumann equation} with unkown $t\mapsto R(t)\in\cD(\fH)$
$$
i\hbar\d_tR=[-\tfrac{\hbar^2}{2m}\Dlt+V,R]\,,\qquad R\rstr_{t=0}=R^{in}\,.
$$
The real-valued potential $V$ is chosen so that the quantum Hamiltonian
$$
\bH:=-\tfrac{\hbar^2}{2m}\Dlt+V
$$
has a self-adjoint (unbounded) extension to $L^2(\bR^d)$.

Specialists of control usually consider $R(t)=|\psi(t,\cdot)\ra\la\psi(t,\cdot)|$ with
$$
i\hb\d_t\psi(t,x)=-\tfrac12\hb^2\Dlt_x\psi(t,x)+V(x)\psi(t,x)\,.
$$
A class $\cK$ of solutions $t\mapsto R(t)\in\cD(\fH)$ of the von Neumann equation can be \textbf{observed} on a domain $\Om\subset\bR^d$ during time $T$ if there exists a 
constant $C_{OBS}>0$, which may depend on $\cK$, on $T$ and on $\Om$, but not on the specific solution $R\in\cK$, such that
$$
1(=\|R(t)\|_1)\le C_{OBS}\int_0^T\Tr_\fH(\indc_\Om R(t))dt\quad\text{ for all }R\in\cK\,.
$$
The reason for the interest in this notion is explained by the HUM method (Hilbert Uniqueness Method) introduced by J.-L. Lions: if $\Om$ and $T$ are such that all solutions of the 
Schr\"odinger equations --- meaning that $\cK=C_b(\bR;L^2(\bR^d))$ --- can be observed on $(0,T)\times\Om$, one can control the Schr\"odinger equation by acting only on the 
domain $\Om$ and on the time interval $(0,T)$ via a source term in the Schr\"odinger equation. (As a matter of fact, Lions' initial interest was with the wave equation, or with the 
equations of linear elasticity \cite{JLLions}, and the problem was to eliminate vibrations on large solid structures by acting with motors on small regions of the structure; the question 
was obviously to find the optimal way of placing the stabilizing motors.)

\noindent
\textbf{Quiz 22: The Lions HUM method.} Let $H$ be an unbounded self-adjoint operator on $L^2(\bR^d)$ --- for instance the Hamiltonian $H=-\tfrac{\hb^2}{2m}\Dlt+V$ under some 
appropriate conditions on the real-valued potential $V$. Consider the control problem
$$
\left\{
\ba
{}&i\d_t\phi=H\phi+\indc_{(0,T)\times\Om}f\,,
\\
&\phi\rstr_{t=T}=0\,.
\ea
\right.
$$
Can one drive any solution $\phi\in C_b(\bR;L^2(\bR^d))$ to $0$ at time $T$ by acting on the domain $\Om$ only via the control $f$? In other words, is the control operator
$$
\cC:\,L^2((0,T)\times\Om)\ni f\mapsto-i\phi\rstr_{t=0}\in L^2(\bR^d)
$$
onto? 

To answer this question, one considers instead the observation problem
$$
\left\{
\ba
{}&i\d_t\psi=H\psi\,,
\\
&\psi\rstr_{t=0}=\psi^{in}\,,
\ea
\right.
$$
and the observation operator
$$
\cO:\,L^2(\bR^d)\ni\psi^{in}\mapsto\psi\rstr_{(0,T)\times\Om}\in L^2((0,T)\times\Om)\,.
$$
If the full class $C_b(\bR;L^2(\bR^d))$ of solutions of the Schr\"odinger equation is observable on $(0,T)\times\Om$ in the sense of the definition above, the operator $\cO$ is 
one-to-one.

\noindent
(1) Prove that $\cC$ and $\cO$ are bounded operators.

\noindent
(2) Prove that $\cO$ is the adjoint of $\cC$.

\noindent
(3) Which statement about the control operator can be deduced from the fact that the class $C_b(\bR;L^2(\bR^d))$ of solutions of the Schr\"odinger equation is observable on 
$(0,T)\times\Om$?

\smallskip
The controllability problem for the wave (or elasticity) equation has been studied by several authors under various specific conditions, until Bardos, Lebeau and Rauch \cite{BLR}
came up with a very satisfying answer based on the propagation of high frequency waves. As is well known, high frequency waves propagate according to the laws of geometric 
optics, and one approach to relating geometric optics to the wave equation is through the consideration of the wave front set in microlocal analysis. (There was a serious additional
difficulty in the Bardos-Lebeau-Rauch paper, namely the fact that they sought to control the solution only at the boundary of the domain, which involved using the Melrose-Sj\"ostrand
theory of propagation of the wave front set along ``broken bicharacteristics'', in other words, optical rays from the theory of geometric optics interacting with the boundary.) However,
the key idea in the Bardos-Lebeau-Rauch approach was a ``geometric condition'' saying that all the rays of geometric optics should hit the region of control at the boundary at least
once in the time interval $(0,T)$. Under this condition, all solutions of the wave equation can be driven to $0$ after time $T$ by some appropriate action on the region of control at the 
boundary of the domain.

In view of the analogy between geometric optics and the classical limit of quantum mechanics (see for instance \S\S 6 and 46 in \cite{LL6}), we shall consider the following geometric
condition, which can be formulated in terms of the trajectories of the classical mechanics for the same potential as in the quantum problem for which the observation problem is posed.

Thus, to the quantum Hamiltonian $\bH$ above, we associate the classical Hamiltonian 
$$
H(x,\xi):=\tfrac1{2m}|\xi|^2+V(x)
$$
generating a flow $\Phi(t;\cdot,\cdot)=(X(t;\cdot,\cdot),\Xi(t;\cdot,\cdot))$ on $\bR^d\times\bR^d$ defined by the following prescription:
$$
\left\{
\ba
{}&\dot X=\tfrac1m\Xi\,,\qquad\qquad&&X(0;x,\xi)\!=x\,,
\\
&\dot\Xi=-\grad V(X)\,,&&\,\Xi(0;x,\xi)=\xi\,.
\ea
\right.
$$
\textbf{Bardos-Lebeau-Rauch Geometric Condition.} Let $K\subset\bR^{2d}$ be compact, consider a domain $\Om\subset\bR^d$ and let $T>0$; the triple $(K,\Om,T)$ is said to satisfy 
the Bardos-Lebeau-Rauch (BLR) geometric condition if
$$
\left\{
\ba
{}&\text{for each }(x,\xi)\in K\,,\text{ there exists }
\\
&t\!\in\!(0,T)\text{ such that }X(t;x,\xi)\!\in\!\Om.
\ea\right.
\leqno{(GC)}
$$
\textbf{Lemma 14.} Assume that $V\in C^{1,1}(\bR^d)$ is real-valued and satisfies  $V\ge -M$ on $\bR^d$ for some $M>0$. Let $K$ be a compact subset of $\bR^d\times\bR^d$, let
$\Om$ be an open set in $\bR^d$ and let $T>0$. Assume that the triple $(K,\Om,T)$ satisfy (GC). Then
$$
C[K,\Om,T]:=\inf_{(x,\xi)\in K}\int_0^T\indc_\Om(X(t;x,\xi))dt>0\,.
$$
\begin{proof} Since $\Om$ is open in $\bR^d$, its indicator function $\indc_\Om$ is lower semicontinuous (l.s.c.) on $\bR^d$. By Fatou's lemma, the function
$$
(x,\xi)\mapsto\int_0^T\indc_\Om(X(t;x,\xi))dt\in(0,+\infty)\text{ is l.s.c. on }\bR^d\times\bR^d\,.
$$
Since $K$ is compact in $\bR^d\times\bR^d$, there exists $(x^*,\xi^*)\in K$ such that
$$
C[K,\Om,T]=\int_0^T\indc_\Om(X(t;x^*,\xi^*))dt\,.
$$
By the geometric condition, there exists $t^*\in(0,T)$ such that
$$
\indc_\Om(X(t^*;x^*,\xi^*))=1\,.
$$
Since $t\mapsto\indc_\Om(X(t;x^*,\xi^*))$ is l.s.c. on $(0,T)$, there exists $\eta\in(0,T)$ such that
$$
|t-t^*|<\eta\implies\indc_\Om(X(t;x^*,\xi^*))=1\,.
$$
Hence
$$
C[K,\Om,T]=\int_0^T\indc_\Om(X(t;x^*,\xi^*))dt\ge\eta>0\,.
$$
\end{proof}

\begin{figure}

\includegraphics[width=8cm]{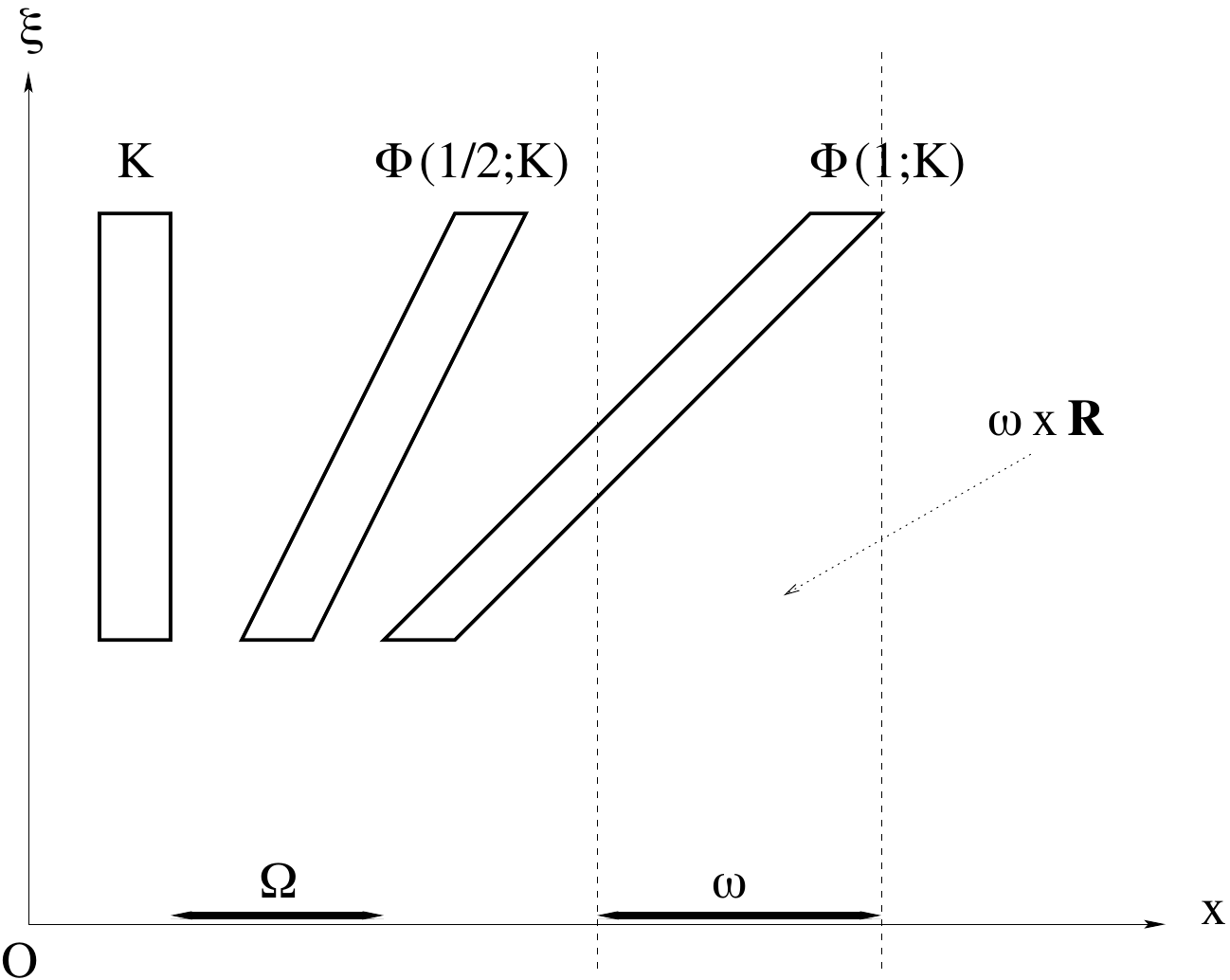}

\caption{The geometric condition in space dimension $d=1$, with $V\equiv 0$. The classical free flow is $\Phi(t;x,\xi)\!:=(X(t;x,\xi),\Xi(t;x,\xi))\!=\!(x+t\xi,\xi)$. The picture 
represents the image of the closed phase-space rectangle $K$ by the map $(x,\xi)\mapsto\Phi(t;x,\xi)$ at time $t=\tfrac12$ and $t=1$. The interval $\Om$ satisfies the 
geometric condition with $T=1$, at variance with $\om$. Indeed, phase-space points on the bottom side of $K$ stay out of the strip $\om\times\bR$ for all $t\in[0,1]$.}

\end{figure}

\smallskip
With the propagation estimate for $\fd$ presented in Theorem 6 above, one can formulate a quantitative observation inequality for the von Neumann equation, with
rather explicit observation constants defined in terms of the classical dynamics. The following result has been obtained in \cite{FGPaulM3AS}.

\noindent
\textbf{Theorem 15.} Let $V\in C^{1,1}(\bR^d)$ be a real-valued potential satisfying $V\ge -M$ for some $M>0$. Let $K$ be a compact subset of $\bR^d\times\bR^d$, 
let $\Om$ be an open set in $\bR^d$ and let $T>0$. Assume that the triple $(K,\Om,T)$ satisfy (GC). Then, for all initial density operator $R^{in}\in\cD_2(\fH)$ and all 
$\de>0$, one has the ``observation'' inequality on the open set $\Om_\de=\Om+B(0,\de)$
$$
\ba
\int_0^T\Tr_\fH(\indc_{\Omega_\de}U(t)R^{in}U(t)^*)dt\ge&C[K,\Omega,T]
\\
-\frac1\de\inf_{\lbd>0}\frac{1}{\l}&\tfrac{\exp\left(\frac12T\left(\frac{\lbd}m+\frac{\Lip(\grad V)}{\lbd}\right)\right)-1}{\tfrac12\left(\frac{\lbd}m+\frac{\Lip(\grad V)}{\lbd}\right)}
\inf_{\Supp(f^{in})\subset K}\fd_\lbd(f^{in},R^{in})\,.
\ea
$$

We have put the word observation between quotes in Theorem 15, since the inequality in that theorem is a bona fide observation inequality only if there exists
$\lbd>0$ and a probability density $f^{in}$ with support in $K$ such that
\[
\fd_\lbd(f^{in},R^{in})<C[K,\Om,T]\frac{\frac{\de\lbd}2\left(\frac{\lbd}m+\frac{\Lip(\grad V)}{\lbd}\right)}{\exp\left(\frac{T}2\left(\frac{\lbd}m+\frac{\Lip(\grad V)}{\lbd}\right)\right)-1}\,.
\]
Here are two examples where this inequality is known to be true.

\noindent
\textbf{Example 1: Toeplitz initial data.} Assume that the initial density operator $R^{in}$ is of the form
$$
R^{in}:=\cT[\mu^{in}]\,,\qquad\text{ where }\mu^{in}\in\cP(\bR^{2d})\text{ satisfies }\Supp(\mu^{in})\subset K\,.
$$
By Theorem 1 (1) (see Lecture I)
$$
\lbd d\hb\le\fd_\hb(f^{in},R^{in})^2\le\max(1,\lbd^2)W_2(f^{in},\mu^{in})^2+d\hb\,,
$$
so that
$$
\Supp(\mu^{in})\subset K\implies\inf_{\Supp(f^{in})\subset K}\fd_\lbd(f^{in},R^{in})=\sqrt{\tfrac12(\lbd^2+1)d\hb}\,.
$$
Indeed, 
$$
\Supp(\mu^{in})\subset K\implies\inf_{\Supp(f^{in})\subset K}\cW_2(f^{in},\mu^{in})=0\,,
$$
since any probability measure $\mu^{in}$ supported in $K$ is the weak limit of a sequence of absolutely continuous (with respect to the Lebesgue measure of $\bR^{2d}$)
probability measures supported in $K$. (To see this, let $\zeta_\eps$ be a regularizing sequence on $\bR^{2d}$, and set 
$$
f_\eps(x,\xi):=Z_\eps^{-1}(\mu^{in}\star\zeta_\eps)(x,\xi)\indc_K(x,\xi)\,,\quad\text{ with }\quad Z_\eps:=\iint_K\mu^{in}\star\zeta_\eps(x,\xi)dxd\xi\,.
$$
Then $f_\eps$ is a probability density on $\bR^{2d}$ with support in $K$, and one easily check that
$$
f_\eps\to\mu^{in}\text{ weakly in the sense of probability measures as }\eps\to 0\,,
$$
and since $\Supp(f_\eps)\subset K$ for each $\eps>0$, this implies that $\cW_2(f_\eps,\mu^{in})\to 0$ as $\eps\to 0^+$.)

\noindent
\textbf{Example 2: Pure state.} Assume now  that $R(t)=|U(t)\psi^{in}\ra\la U(t)\psi^{in}|$, where $U(t)=e^{-it\bH/\hb}$ is the Schr\"odinger group. (We recall that
$$
\bH:=-\tfrac{\hb^2}{2m}\Dlt+V\,,
$$
and the assumptions on $V$ imply that $\bH$ has a self-adjoint extension to $L^2(\bR^d)$.)

\bigskip
Choosing 
$$
f^{in}(q,p):=\frac{|\la q,p|\psi^{in}\ra|^2}{(2\pi\hb)^d}=\cH[|\psi^{in}\ra\la\psi^{in}|]
$$
and setting $\lbd=1$ leads to
$$
\frac1{C_{OBS}}=C[K,\Omega,T]\iint_K|\la q,p|\psi^{in}\ra|^2\tfrac{dqdp}{(2\pi\hb)^d}-D[T,\Lip(\grad V)]\frac{\Sigma[\psi^{in}]}\de\,,
$$
where
$$
\ba
D[T,L]:=&4\frac{e^{(\frac1m+L)T/2}-1}{\frac1m+L}\,,
\\
\Sigma[\psi^{in}]^2:=&\la\psi^{in}|\,|x|^2|\psi^{in}\ra-|\la\psi^{in}|x|\psi^{in}\ra|^2
\\
&+\la\psi^{in}|-\hb^2\Dlt_x|\psi^{in}\ra-|\la\psi^{in}|-i\hb\grad_x|\psi^{in}\ra|^2\,.
\ea
$$
\textbf{Quiz 23.} The missing argument to arrive at the formula for $C_{OBS}$ in the pure state case is a good opportunity to revise some elementary facts about $\fd$.

\noindent
(1) Find $\cC(\cH[|\psi^{in}\ra\la\psi^{in}|],|\psi^{in}\ra\la\psi^{in}|)$.

\noindent
(2) Compute $\fd(\cH[|\psi^{in}\ra\la\psi^{in}|],|\psi^{in}\ra\la\psi^{in}|)$. (Hint: observe that $x_j$, $-i\hb\d_{x_j}$, $x_j^2$ and $-\hb^2\d_{x_j}^2$ are Toeplitz operators,
and compute their symbols.)

\smallskip
Here is a sketch of the proof of Theorem 15.

\begin{proof}
Call $f(t,\cdot,\cdot):=f^{in}\circ\Phi(-t;\cdot,\cdot)$ and $R(t):=U(t)R^{in}U(t)^*$. For each coupling $Q(t)\in\cC(f(t,\cdot,\cdot),R(t))$, one has
$$
\ba
\left|\Tr_\fH(\chi R(t))-\iint_{\bR^{2d}}\chi(x)f(t,x,\xi)dxd\xi\right|
\\
=\left|\iint_{\bR^{2d}}\Tr_\fH((\chi(x)-\chi(y))Q(t,x,\xi)dxd\xi\right|
\\
\le\tfrac{\Lip(\chi)}{\lbd}\left(\iint_{\bR^{2d}}\Tr_\fH(Q_t^\frac12(\lbd^2|x\!-\!y|^2\!+\!|\xi+i\hb\grad_y|^2)Q_t^\frac12)dxd\xi\right)^\frac12&\,,
\ea
$$
so that, by the propagation estimate in Theorem 6, 
$$
\ba
\left|\Tr_\fH(\chi R(t))-\iint_{\bR^{2d}}\chi(x)f(t,x,\xi)dxd\xi\right|\le\tfrac{\Lip(\chi)}{\lbd}\fd_\lbd(f(t,\cdot,\cdot),R(t))
\\
\le\tfrac{\Lip(\chi)}{\lbd}\fd_\lbd(f^{in},R^{in})\exp\left(\tfrac12t\left(\frac{\lbd}m+\frac{\Lip(\grad V)}{\lbd}\right)\right)&\,.
\ea
$$
Since
$$
\iint_{\bR^{2d}}\chi(x)f(t,x,\xi)dxd\xi=\iint_{\bR^{2d}}\chi(X(t;x,\xi))f^{in}(x,\xi)dxd\xi\,,
$$
one has
$$
\ba
\int_0^T\Tr_\fH(\chi R(t))dt\ge&\inf_{(x,\xi)\in K}\int_0^T\chi(X(t;x,\xi))dt\iint_Kf^{in}(x,\xi)dxd\xi
\\
&-\frac{\Lip(\chi)}{\lbd}\frac{\exp\left(\tfrac12T\left(\frac{\lbd}m+\frac{\Lip(\grad V)}{\lbd}\right)\right)-1}{\tfrac12\left(\frac{\lbd}m+\frac{\Lip(\grad V)}{\lbd}\right)}\fd_\lbd(f^{in},R^{in})\,.
\ea
$$
Conclude by choosing $\chi(x):=\left(1-\frac{\text{dist}(x,\Omega)}{\delta}\right)_+$, so that $\Lip(\chi)=\frac1\delta$.
\end{proof}

\medskip
The main result in this lecture is obviously Theorem 6, which explains how the pseudometric $\fd$, or its variant $\fd_\lbd$ are propagated by self-adjoint quantum Hamiltonians 
of the form
$$
-\tfrac{\hb^2}{2m}\Dlt+V\,.
$$
However, there are other quantum dynamics for which it may be useful to know how $\fd$ is propagated. The example in the following exercise was communicated to us by
E. Carlen.

\noindent
\textbf{Quiz 24: Quantum Heat Equation.}

\noindent
(1) Let $\mu,\nu\in\cP_2(\bR^n)$. Prove that
$$
\cW_2(e^{t\Dlt}\mu,e^{t\Dlt}\nu)\le\cW_2(\mu,\nu)\,.
$$
(Hint: represent $e^{t\Dlt}\mu(x)$ by means of the Brownian motion, and consider the process $(x+B_t,y+B_t)\in\bR^n\times\bR^n$, with the \textbf{same} Brownian motion $B_t$.)

\noindent
(2) Find another proof of (1) without appealing to the representation of the solution by means of the Brownian motion. (Hint: pick $\rho_{in}\in\cC(\mu,\nu)$ and propagate $\rho_{in}$ 
by a degenerate diffusion operator $A^*A$, i.e. set
$$
\d_t\rho_t+A^*A\rho_t=0\,,\qquad\rho_0=\rho_{in}\,,
$$
where $A$ is a 1st order differential operator such that $A|x-y|^2=0$.)

\noindent
(3) Set $\fH:=L^2(\bR)$ and $q\psi(y):=y\psi(y)$ while $p\psi(y):=-i\hb\tfrac{d\psi}{dy}(y)$. Consider the Quantum Heat Equation
$$
\d_tR=-\tfrac1{\hb^2}[p,[p,R]]-\tfrac1{\hb^2}[q,[q,R]]\,,\qquad R(0)=R^{in}\in\cD_2(\fH)\,.
$$
Prove that the Cauchy problem above is solved by a contraction semigroup on $\cL^2(\fH)$, and that $R(t)\in\cD_2(\fH)$ for all $t\ge 0$.

\noindent
(4) Let $R_1,R_2$ be the solutions of
$$
\ba
\d_tR_1=-\tfrac1{\hb^2}[p,[p,R_1]]-\tfrac1{\hb^2}[q,[q,R_1]]\,,\qquad R_1(0)=R_1^{in}\in\cD_2(\fH)\,,
\\
\d_tR_2=-\tfrac1{\hb^2}[p,[p,R_2]]-\tfrac1{\hb^2}[q,[q,R_2]]\,,\qquad R_2(0)=R_2^{in}\in\cD_2(\fH)\,.
\ea
$$
Prove that 
$$
\fd(R_1(t),R_2(t))\le\fd(R_1^{in},R_2^{in})\,,\qquad t\ge 0\,.
$$
(Hint: consider the operators $[p\otimes I+I\otimes p,q\otimes I-I\otimes q]$ and $[q\otimes I+I\otimes q,p\otimes I-I\otimes p]$.)


\section{Lecture III: Triangle Inequalities and Optimal Transport\\ in the Quantum Setting}


We have seen in Lecture II that the pseudometric $\fd$ defined in Lecture I behaves very satisfyingly under propagation by the most fundamental quantum dynamics
(i.e. the unitary group generated by self-adjoint Hamiltonians of the form $-\frac{\hb^2}{2m}\Dlt+V$, and also by the quantum heat equation). We have not treated the
interesting case of the Schr\"odinger equation with a magnetic field, leading to Hamiltonians of the form
$$
\tfrac1m|-i\hb\grad+A|^2+V
$$
where $A$ is the vector potential (so that the magnetic field is $B=\text{curl}A$): see \cite{BenPorath} for a comprehensive treatment of this case.

Although Lecture II gives (hopefully) convincing arguments, mostly based on applications, in favor of the pseudometric $\fd$, we have already seen in Lecture I, that
$\fd$ is not a bona fide metric, in particular because\footnote{One might think of changing the definition of $\fd$, for instance by replacing the quantum-to-quantum cost 
operator $C_\hb$ in Lecture I by $C_\hb-2d\hb I_{\fH\otimes\fH}$. However, if you have understood the essence of the proof of Theorem 6, this is not going to help.
Indeed, this proof is based on the operator inequality
$$
[H\otimes I_\fH+I_\fH\otimes H,C_\hb]\le LC_\hb
$$
for some positive constant $L$, where $H$ is a Hamiltonian of the form $-\tfrac{\hb^2}{2m}\Dlt+V$. Changing $C_\hb$ into $C_\hb-\a I_{\fH\otimes\fH}$ will obviously 
not change the commutator on the left-hand side, but changes the right-hand side.} $\fd(T,T)>0$ for all $T\in\cD_2(\fH)$.

In Lecture I, we have postponed the necessary task of exploring the properties of the pseudometric $\fd$, except those reported in Theorem 1, which have proved
already very useful for applications of the pseudometric $\fd$ to quantum dynamical problems. For instance, we have not discussed the triangle inequality for $\fd$, 
in spite of the fact that we used it in Lecture II in deriving a uniform in $\hb$ error estimate for the Lie-Trotter splitting scheme for the von Neumann equation.

In this last lecture we shall return in particular to this question. We shall also discuss further properties of the pseudometric $\fd$ by analogy with the Wasserstein
distance $\cW_2$. Specifically, we shall study the following topics:

\noindent
\bu a Kantorovich-type duality for quantum optimal transport (section \ref{SS-QDual})

\noindent
\bu two kinds of triangle inequalities for the quantum pseudometric $\fd$ on $\fD$ (presented in section \ref{SS-RestTriang}), and

\noindent
\bu the structure of optimal couplings for the pseudometric $\fd$ (section \ref{SS-QOT}).

\subsection{Restricted Triangle Inequalities}\lb{SS-RestTriang}


In this section, we shall prove the triangle inequality for $\fd$ in some special cases --- hence the terminology of ``restricted'' triangle inequality used here. Our goal
is the following statement.

\noindent
\textbf{Theorem 16.} For all $\rho_1,\rho_2,\rho_3\in\fD=\cP_2(\bR^d\times\bR^d)\cup\cD_2(\fH)$, one has
$$
\fd(\rho_1,\rho_3)\le\fd(\rho_1,\rho_2)+\fd(\rho_2,\rho_3)\,,
$$
provided that $\rho_2$ is a probability density in $\bR^d\times\bR^d$ or one of the $\rho_j$s is a rank-$1$ density operator on $\fH$.

\smallskip
See Theorems A and 3.1 in \cite{FGPaulJFA}; see also Theorem 3.5 in \cite{FGPaulJMPA}.

\subsubsection{Operator Inequalities}


As a preparation to the proof of Theorem 16, we prove several inequalities involving the quantum-to-quantum and the classical-to-quantum transport cost operators.
These inequalities are of the same type as triangle inequalities, with some arbitrary parameter $\a>0$.

\smallskip
First, we discuss inequalities where the intermediate ``point'' is an operator.

\noindent
\textbf{Lemma 17.} For all $\a>0$, one has
$$
\ba
|x-z|^2+|\xi-\zeta|^2\le&(1+\a)c_\hb(x,\xi;y,\hb\grad_y)
\\
&+(1+\tfrac1\a)c_\hb(z,\zeta;y,\hb\grad_y)\,,
\\
c_\hb(x,\xi;z,\hb\grad_z)\le&(1+\a)c_\hb(x,\xi;y,\hb\grad_y)
\\
&+(1+\tfrac1\a)C_\hb(y,\hb\grad_y,z,\hb\grad_z)\,,
\\
C_\hb(x,\hb\grad_x,z,\hb\grad_z)\le&(1+\a)C_\hb(x,\hb\grad_x,y,\hb\grad_y)
\\
+&(1+\tfrac1\a)C_\hb(y,\hb\grad_y,z,\hb\grad_z)\,.
\ea
$$

\medskip
These operator inequalities mean that, for all $\phi\in\cS(\bR^{2d}_{x,\xi}\times\bR^d_y\times\bR^{2d}_{z,\zeta})$, or all $\phi\in\cS(\bR^{2d}_{x,\xi}\times\bR^d_y\times\bR^d_z)$,
or all $\phi\in\cS(\bR^d_x\times\bR^d_y\times\bR^d_z)$,
$$
\la\phi|r.h.s.-l.h.s.|\phi\ra\ge 0\,.
$$

\begin{proof} Write
$$
\ba
C_\hb(x,\hb\grad_x,z,\hb\grad_z)=&|x-y+y-z|^2-\hb^2|\grad_x-\grad_y+\grad_y-\grad_z|^2
\\
=&C_\hb(x,\hb\grad_x,y,\hb\grad_y)+C_\hb(y,\hb\grad_y,z,\hb\grad_z)
\\
&+2(x-y)\cdot(y-z)-2\hb^2(\grad_x-\grad_y)\cdot(\grad_y-\grad_z)\,.
\ea
$$
Use the Peter-Paul elementary inequality
$$
2(x-y)\cdot(y-z)\le\a|x-y|^2+\tfrac1\a|y-z|^2\,,
$$
and, for operators $A,B$, the analogous inequality
$$
A^*B+B^*A\le\a|A|^2+\tfrac1\a|B|^2
$$
with $A=A^*=-i\hb(\d_{x_j}-\d_{y_j})$ and $B=B^*=-i\hb(\d_{y_j}-\d_{z_j})$ for all indices $j=1,\ldots,d$. (Observe that these operators commute, which is inessential here). 
The operator inequality comes from expanding
$$
0\le\left|\a^\frac12 A-\a^{-\frac12}B\right|^2=\a|A|^2+\tfrac1\a|B|^2-A^*B-B^*A\,.
$$
See also Quiz 7 in Lecture I.

Hence
$$
\ba
2(x-y)\cdot(y-z)-2\hb^2(\grad_x-\grad_y)\cdot(\grad_y-\grad_z)
\\
\le\a C_\hb(x,\hb\grad_x,y,\hb\grad_y)+\tfrac1\a C_\hb(y,\hb\grad_y,z,\hb\grad_z)
\ea
$$
With the previous equality involving $C_\hb(x,\hb\grad_x,z,\hb\grad_z)$, we arrive at the 3rd inequality of Lemma 17. 
\end{proof}

\noindent
\textbf{Quiz 25.} Prove the first two inequalities in Lemma 17.

\smallskip
Next, we discuss inequalities similar to those of Lemma 17, but with classical phase-space point as intermediate term.

\noindent
\textbf{Lemma 18.} For all $\a>0$, one has
$$
\ba
c_\hb(x,\xi;z,\hb\grad_z)\le&(1+\a)(|x-y|^2+|\xi-\eta|^2)
\\
&+(1+\tfrac1\a)c_\hb(y,\eta;z,\hb\grad_z)
\\
C_\hb(x,\hb\grad_x,z,\hb\grad_z)\le&(1+\a)c_\hb(x,\hb\grad_x,y,\eta)
\\
+&(1+\tfrac1\a)c_\hb(y,\eta,z,\hb\grad_z)\,.
\ea
$$

\medskip
These operator inequalities mean that
$$
\la\phi|r.h.s.-l.h.s.|\phi\ra\ge 0
$$
for all $\phi\in\cS(\bR^{2d}_{x,\xi}\times\bR^d_{y,\eta}\times\bR^{2d}_{z,\zeta})$, or all $\phi\in\cS(\bR^d_x\times\bR^d_{y,\eta}\times\bR^d_z)$.

\medskip
\noindent
\textbf{Quiz 26.} Prove Lemma 18 (by the same method as in the proof of Lemma 17).

\subsubsection{The Rank-$1$ Case}


At this point, we can prove the part of Theorem 16 involving a rank-$1$ density operator.

\begin{proof}[Proof of Theorem 16: sketch for the rank-$1$ case]
Assume for example that $\rho_1$ and $\rho_2\in\cD_2(\fH)$ while $\rho_3$ is a rank-$1$ density operator, and let $Q\in\cC(\rho_1,\rho_2)$. Set
$$
T:=Q\otimes\rho_3\,,\quad T_{13}=\Tr_2T\in\cC(\rho_1,\rho_3)\,.
$$
Hence, by the 3rd inequality in Lemma 17,
$$
\ba
\fd(\rho_1,\rho_3)^2\le\Tr_{\fH^{\otimes 2}}(T_{13}^\frac12 C_\hb(x,\hb\grad_x,z,\hb\grad_z)T_{13}^\frac12)
\\
=\Tr_{\fH^{\otimes 3}}(T^\frac12 C_\hb(x,\hb\grad_x,z,\hb\grad_z)T^\frac12)
\\
\le(1+\a)\Tr_{\fH^{\otimes 3}}(T^\frac12 C_\hb(x,\hb\grad_x,y,\hb\grad_y)T^\frac12)
\\
+(1+\tfrac1\a)\Tr_{\fH^{\otimes 3}}(T^\frac12 C_\hb(y,\hb\grad_y,z,\hb\grad_z)T^\frac12)
\\
=(1+\a)\Tr_{\fH^{\otimes 2}}(Q^\frac12 C_\hb(x,\hb\grad_x,y,\hb\grad_y)Q^\frac12)
\\
+(1+\tfrac1\a)\underbrace{\Tr_{\fH^{\otimes 2}}((\rho_2\otimes\rho_3)^\frac12 C_\hb(y,\hb\grad_y,z,\hb\grad_z)(\rho_2\otimes\rho_3)^\frac12)}_{=\fd(\rho_2,\rho_3)^2}&\,.
\ea
$$
That the last term on the r.h.s. is equal to $\fd(\rho_2,\rho_3)^2$ follows from the fact that $\cC(\rho_2,\rho_3)=\{\rho_2\otimes\rho_3\}$ since $\rho_3$ has rank $1$ 
(see Lemma 2 in Lecture I). Minimizing the last r.h.s. in $Q\in\cC(\rho_1,\rho_2)$ shows that
$$
\fd(\rho_1,\rho_3)^2\le(1+\a)\fd(\rho_1,\rho_2)^2+(1+\tfrac1\a)\fd(\rho_2,\rho_3)^2\,.
$$
Minimizing the r.h.s. in $\a>0$, i.e. setting
$$
\a:=\frac{\fd(\rho_2,\rho_3)}{\fd(\rho_1,\rho_2)}\quad\text{ assuming }\fd(\rho_1,\rho_2)>0\,,
$$
leads to
$$
\fd(\rho_1,\rho_3)^2\le\fd(\rho_1,\rho_2)^2+\fd(\rho_2,\rho_3)^2+2\fd(\rho_1,\rho_2)\fd(\rho_2,\rho_3)\,.
$$
Conclude by taking the square root of both sides of this inequality.
\end{proof}

\noindent
\textbf{Quiz 27.} Complete the missing argument in the proof by justifying the equality
$$
\ba
\Tr_{\fH^{\otimes 2}}(T_{13}^\frac12 C_\hb(x,\hb\grad_x,z,\hb\grad_z)T_{13}^\frac12)
\\
=\Tr_{\fH^{\otimes 3}}(T^\frac12 C_\hb(x,\hb\grad_x,z,\hb\grad_z)T^\frac12)&\,.
\ea
$$
(1) Prove this identity when $C_\hb$ is replaced with $(I_{\fH^{\otimes 2}}+\tfrac1nC_\hb)^{-1}C_\hb$.

\noindent
(2) Using the Fatou lemma for trace-class operators, prove that
$$
\ba
\lim_{n\to\infty}\Tr_{\fH^{\otimes 2}}\left(T_{13}^\frac12 \tfrac{C_\hb(x,\hb\grad_x,z,\hb\grad_z)}{I_{\fH^{\otimes 2}}+\frac1nC_\hb(x,\hb\grad_x,z,\hb\grad_z)}T_{13}^\frac12\right)
\\
=\Tr_{\fH^{\otimes 2}}\left(T_{13}^\frac12 C_\hb(x,\hb\grad_x,z,\hb\grad_z)T_{13}^\frac12\right)&\,,
\\
\lim_{n\to\infty}\Tr_{\fH^{\otimes 3}}\left(T^\frac12 \tfrac{C_\hb(x,\hb\grad_x,z,\hb\grad_z)}{I_{\fH^{\otimes 3}}+\frac1nC_\hb(x,\hb\grad_x,z,\hb\grad_z)}T^\frac12\right)
\\
=\Tr_{\fH^{\otimes 3}}\left(T^\frac12 C_\hb(x,\hb\grad_x,z,\hb\grad_z)T^\frac12\right)&\,.
\ea
$$
\textbf{Quiz 28.} Complete the proof of Theorem 16 by treating the missing cases where one of the $\rho_j$'s is a rank-$1$ density operator.

\subsubsection{The Case of a Classical Intermediate Density}


All the cases of the triangle inequality involving a rank-$1$ operator are easy, because of Lemma 2. Indeed, the set of couplings of any density (quantum or classical) with 
a rank-$1$ density operator is a singleton, and therefore the pseudometric $\fd$ is easily computed explicitly in such a case.

In the present section, we shall discuss all the cases of the triangle inequality where the intermediate point is a classical density on phase-space. The triangle inequality in
such cases is much more involved, but fortunately, the proof can be modelled on one of the proofs of the triangle inequality for $\cW_2$ (See the proof of Theorem 7.3 in 
\cite{VillaniAMS}.)

The key step is the following lemma, which explains how to \textit{disintegrate} a coupling between a classical probability density on phase-space and a density operator.

\noindent
\textbf{Lemma 19.} Let $f$ be a probability density on $\bR^d\times\bR^d$, let $R\in\cD(\fH)$ and let $Q\in\cC(f,R)$.  There exists a weakly measurable map 
$$
\bR^d\times\bR^d\ni(x,\xi)\mapsto Q_f(x,\xi)\in\cL^1(\fH)
$$
defined a.e. on $\bR^d\times\bR^d$, which satisfies
$$
Q_f(x,\xi)=Q_f(x,\xi)^*\ge 0\,,\quad\Tr(Q(x,\xi))=1\,,
$$
and
$$
Q(x,\xi)=f(x,\xi)Q_f(x,\xi)\quad\text{ a.e. in }(x,\xi)\in\bR^d\times\bR^d\,.
$$

\begin{proof} First replace $f$ with a Borel representative, and consider the set 
$$
\cN:=f^{-1}(\{0\})\,,
$$
which is Borel measurable. Pick $u\in\fH$ such that $\|u\|_\fH=1$, and set
$$
Q_f(x,\xi):=\frac{Q(x,\xi)+\indc_\cN(x,\xi)|u\ra\la u|}{f(x,\xi)+\indc_\cN(x,\xi)}\in\cL(\fH)\,.
$$
Obviously
$$
Q(x,\xi)=Q(x,\xi)^*\ge 0\,\,\text{ and }\,\, f(x,\xi)\ge 0
$$
and hence
$$
Q_f(x,\xi)=Q_f(x,\xi)^*\ge 0\,,\quad\text{ for a.e. }(x,\xi)\in\bR^d\times\bR^d\,.
$$
Moreover
$$
\Tr_\fH(Q(x,\xi)+\indc_\cN(x,\xi)|u\ra\la u|)=f(x,\xi)+\indc_\cN(x,\xi)\,,
$$
so that
$$
\Tr_\fH(Q_f(x,\xi))=1\,,\quad\text{ for a.e. }(x,\xi)\in\bR^d\times\bR^d\,.
$$
Finally
$$
f(x,\xi)Q_f(x,\xi)=\frac{f(x,\xi)Q(x,\xi)}{f(x,\xi)+\indc_\cN(x,\xi)}=Q(x,\xi)
$$
for a.e. $(x,\xi)\in\bR^d\times\bR^d$. Indeed, whenever $f(x,\xi)>0$, one has $\indc_\cN(x,\xi)=0$ and the claimed equality is obvious. On the other hand, since 
$$
Q(x,\xi)=Q(x,\xi)^*\ge 0\quad\text{ and }\Tr_\fH(Q(x,\xi))=f(x,\xi)
$$
for a.e. $(x,\xi)\in\bR^d\times\bR^d$, it follows that
$$
f(x,\xi)=0\implies Q(x,\xi)=0=f(x,\xi)Q_f(x,\xi)\,.
$$
\end{proof}

\smallskip
With Lemma 19 at our disposal, the proof of Theorem 16 in the case where the intermediate point is a classical density follows the proof of Theorem 7.3 in \cite{VillaniAMS}.

\begin{proof}[Proof of Theorem 16: the case of a classical intermediate density]
Consider for example the case where both $\rho_1$ and $\rho_3\in\cD_2(\fH)$, and assume that $\rho_2=f(y,\eta)dyd\eta\in\cP_2(\bR^d\times\bR^d)$. Choose couplings 
$Q^1\in\cC(\rho_1,f)$ while $Q^3\in\cC(f,\rho_3)$. Call $Q^3_f$ the disintegration of $Q^3$ with respect to $f$ as in Lemma 19. Set
$$
T(y,\eta):=Q^1(y,\eta)\otimes Q^3_f(y,\eta)\,.
$$
By construction
$$
T(y,\eta)=T(y,\eta)^*\ge 0\,,
$$
and 
$$
\ba
\Tr_1(T(y,\eta))=&f(y,\eta)Q^3_f(y,\eta)=Q^3(y,\eta)\,,
\\
\Tr_3(T(y,\eta))=&Q^1(y,\eta)\Tr_\fH(Q^3_f(y,\eta))=Q^1(y,\eta)\,.
\ea
$$
In particular
$$
\int_{\bR^{2d}}T(y,\eta)dyd\eta=:\cQ\in\cC(\rho_1,\rho_3)\,.
$$
By the second inequality in Lemma 17
$$
\ba
\fd(\rho_1,\rho_3)^2\le\Tr_{\fH^{\otimes 2}}(\cQ^\frac12C_\hb(x,\hb\grad_x,z,\hb\grad_z)\cQ^\frac12)
\\
=\int_{\bR^{2d}}\Tr_{\fH^{\otimes 2}}(T(y,\eta)^\frac12C_\hb(x,\hb\grad_x,z,\hb\grad_z)T(y,\eta)^\frac12)dyd\eta
\\
\le(1+\a)\int_{\bR^{2d}}\Tr_1\left(\Tr_3(T(y,\eta)^\frac12c_\hb(x,\hb\grad_x,y,\eta)T(y,\eta)^\frac12)\right)dyd\eta
\\
+(1+\tfrac1\a)\int_{\bR^{2d}}\Tr_3\left(\Tr_1(T(y,\eta)^\frac12c_\hb(x,\hb\grad_x,y,\eta)T(y,\eta)^\frac12)\right)dyd\eta
\\
\le(1+\a)\int_{\bR^{2d}}\Tr_\fH(Q^1(y,\eta)^\frac12c_\hb(x,\hb\grad_x,y,\eta)Q^1(y,\eta)^\frac12)dyd\eta
\\
+(1+\tfrac1\a)\int_{\bR^{2d}}\Tr_\fH\left(Q^3_f(y,\eta)^\frac12c_\hb(x,\hb\grad_x,y,\eta)Q^3_f(y,\eta)^\frac12\right)f(y,\eta)dyd\eta&\,.
\ea
$$
Minimizing the last right-hand side in $Q^1\in\cC(\rho_1,\rho_2)$ and in $Q^3\in\cC(\rho_2,\rho_3)$ leads to the inequality
$$
\fd(\rho_1,\rho_3)^2\le(1+\a)\fd(\rho_1,\rho_2)^2+(1+\tfrac1\a)\fd(\rho_2,\rho_3)^2\,,
$$
and we conclude as in the rank-1 case. 
\end{proof}

\smallskip
\noindent
\textbf{Quiz 28.}

\noindent
(1) Complete the missing details in the proof of Theorem 16 in the case where $\rho_1,\rho_3\in\cD_2(\fH)$ and $\rho_2=f(y,\eta)dyd\eta$. In particular, prove the identity
$$
\ba
\Tr_{\fH^{\otimes 2}}(\cQ^\frac12C_\hb(x,\hb\grad_x,z,\hb\grad_z)\cQ^\frac12)
\\
=\int_{\bR^{2d}}\Tr_{\fH^{\otimes 2}}(T(y,\eta)^\frac12C_\hb(x,\hb\grad_x,z,\hb\grad_z)T(y,\eta)^\frac12)dyd\eta&\,.
\ea
$$
(2) Write the proof of Theorem 16 in the missing cases.

\subsection{Applications of the Restricted Triangle Inequalities}


Before going further in our discussion of the triangle inequality for the pseudometric $\fd$, we shall present two easy applications of the (restricted) triangle inequalities already
established in the previous section.

\subsubsection{Definition of $\fd$ on $\cP_2(\bR^d\times\bR^d)\times\cD_2(\fH)$}


Our first application is of a quite fundamental nature, since it completes our definition of $\fd$ in Lecture I. So far we have defined $\fd(\mu,R)=\fd(R,\mu)$ for 
$\mu\in\cP_2(\bR^d\times\bR^d)$ and $R\in\cD_2(\fH)$ only when $\mu=f(x,\xi)dxd\xi$, with  $f$ a probability density on $\bR^d\times\bR^d$ --- i.e. only when 
$\mu$ is absolutely continuous with respect to the phase-space Lebesgue measure, by the Radon-Nikodym theorem.

\noindent
\textbf{Theorem 20.}
For each $R\in\cD_2(\fH)$, the map $f\mapsto\fd(f,R)$, defined for all probability density $f$ with finite second order moments on $\bR^d\times\bR^d$ has a 
unique extension to $\cP_2(\bR^d\times\bR^d)$ satisfying
$$
|\fd(\mu,R)-\fd(\nu,R)|\le\cW_2(f,g)\,,\qquad\mu,\nu\in\cP_2(\bR^d\times\bR^d)\,.
$$

See section 3 in \cite{FGPaulJFA}.

\smallskip
\noindent
\textbf{Remark.} This extension of $\fd$ obviously satisfies the restricted triangle inequality
$$
\fd(\rho_1,\rho_3)\le\fd(\rho_1,\rho_2)+\fd(\rho_2,\rho_3)
$$
for all $\rho_1,\rho_2,\rho_3\in\fD$ provided that $\rho_2\in\cP_2(\bR^d\times\bR^d)$ or if one of the $\rho_j$s is a rank-$1$ density operator on $\fH=L^2(\bR^d)$.

\begin{proof} For all $f,g$ probability densities with finite 2nd order moments on $\bR^d\times\bR^d$, one has the triangle inequality
$$
\fd(f,R)\le\fd(f,g)+\fd(g,R)
$$
by Theorem 16, so that 
$$
\fd(f,R)-\fd(g,R)\le\fd(f,g)=\cW_2(f,g)\,.
$$
Exchanging $f$ and $g$ in the inequality above implies that
$$
|\fd(f,R)-\fd(g,R)|\le\cW_2(f,g)\,.
$$
Thus the function $f\mapsto\fd(f,R)$ is Lipschitz-continuous for the metric $\cW_2$. It has therefore a unique Lipschitz-continuous extension to $\cP_2(\bR^d\times\bR^d)$ by the density 
argument recalled in Lemma 21 below.
\end{proof}

\noindent
\textbf{Lemma 21.}
Let $\mu\in\cP_2(\bR^n)$ and let $\chi_\eps(x)=\chi(x/\eps)/\eps^n$ be an even $C^\infty$ mollifier with support in $B_\eps(0)$. Then $f_\eps:=\chi_\eps\star\mu$ is a $C^\infty$ 
probability density on $\bR^n$ and
$$
\cW_2(f_\eps,\mu)\to 0\quad\text{ as }\eps\to 0\,.
$$

This is Lemma 3.2 in \cite{FGPaulJFA}.

\begin{proof}
For all $\phi\in C_0(\bR^n)$, one has
$$
\int_{\bR^n}\phi(x)\mu(dx)-\int_{\bR^n}f_\eps(x)\phi(x)dx=\int_{\bR^n}(\phi(x)-\chi_\eps\star \phi(x))\mu(dx)
$$
since $\chi_\eps$ is even, so that
$$
\left|\int_{\bR^n}\phi(x)\mu(dx)-\int_{\bR^n}f_\eps(x)\phi(x)dx\right|\le\|\phi-\phi\star\chi_\eps\|_{L^\infty(\bR^n)}\to 0\,.
$$
Hence $f_\eps\to\mu$ weakly in $\cP(\bR^n)$ as $\eps\to 0$.

It remains to establish the tightness property (see the properties of $\cW_2$ recalled in Lecture I). Since $\chi_\eps$ is even,
$$
\int_{\bR^n}\indc_{|x|>R}|x|^2\chi_\eps\star\mu(x)dx=\int_{\bR^n}\chi_\eps\star(\indc_{|x|>R}|x|^2)\mu(dx)\,.
$$
On the other hand, for all $\eps\in(0,1)$
$$
\ba
\chi_\eps\star(\indc_{|x|>R}|x|^2)\le&\indc_{|x|+1\ge R}\int_{\bR^n}|x-\eps y|^2\chi(y)dy
\\
\le& 2\indc_{|x|+1\ge R}\left(|x|^2+\eps^2\underbrace{\int_{\bR^n}|y|^2\chi(y)dy}_{\le 1}\right)
\ea
$$
Hence
$$
\sup_{0<\eps<1}\int_{\bR^n}\indc_{|x|>R}|x|^2\chi_\eps\star\mu(x)dx\le 2\int_{\bR^n}\indc_{|x|+1>R}(|x|^2+1)\mu(dx)\to 0
$$
as $R\to\infty$, by dominated convergence.

Therefore, by Theorem 7.12 of \cite{VillaniAMS}
$$
\cW_2(\chi_\eps\star\mu,\mu)\to 0\qquad\text{ as }\eps\to 0\,.
$$
\end{proof}

\subsubsection{$\cW_2$ is the Classical Limit of $\fd$}


Our next application of the restricted triangle inequalities presented in Theorem 16 can be thought of as a confirmation of the geometric picture proposed in Lecture I.

We recall the idea of considering $\cP_2(\bR^d\times\bR^d)$ as a limit set, or boundary set, of $\cD_2(\fH)$ in $\fD$. The next result completes this picture by showing 
that $\cW_2$ is the limiting metric deduced from the pseudometric $\fd$ on $\cD(\fH)$.

\noindent
\textbf{Theorem 22.}
Let $R_\hb,S_\hb\in\cD_2(\fH)$ and $\mu,\nu\in\cP_2(\bR^d\times\bR^d)$. Assume that $\mu,\nu$ are the classical limits of $R_\hb,S_\hb$ respectively, i.e. 
$$
\fd(\mu,R_\hb)+\fd(\nu,S_\hb)\to 0\quad\text{ as }\hb\to 0\,.
$$
Then
$$
\lim_{\hb\to 0}\fd(R_\hb,S_\hb)=\fd(\mu,\nu)\,.
$$

This is Theorem C (see also Theorem 5.5) in \cite{FGPaulJFA}.

\begin{figure}

\includegraphics[width=10cm]{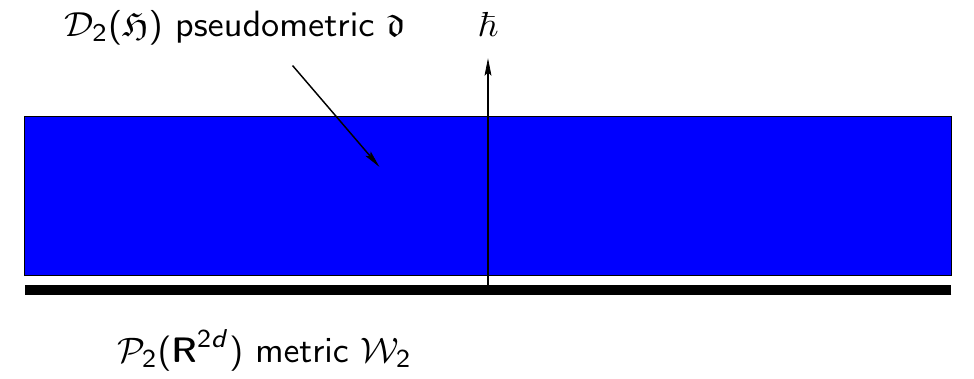}

\caption{The Wasserstein distance $\cW_2$ is the classical limit of the pseudometric $\fd$ as $\hb\to 0$.}

\end{figure}

\begin{proof} By the restricted triangle inequality
$$
\fd(R_\hb,S_\hb)\le\fd(R_\hb,\mu)+\fd(\mu,\nu)+\fd(\nu,S_\hb)\,,
$$
so that
$$
\varlimsup_{\hb\to 0}\fd(R_\hb,S_\hb)\le\fd(\mu,\nu)=\cW_2(\mu,\nu)\,.
$$
On the other hand, by Theorem 1 (2) of Lecture I,
$$
\ba
\fd(R_\hb,\mu)^2\ge\!\cW_2(\cH[R_\hb],\mu)^2-d\hb\implies\lim_{\hb\to 0}\cW_2(\cH[R_\hb],\mu)=0\,,
\\
\fd(S_\hb,\nu)^2\ge\cW_2(\cH[S_\hb],\nu)^2-d\hb\implies\lim_{\hb\to 0}\cW_2(\cH[S_\hb],\nu)=0\,.
\ea
$$
Hence
$$
\fd(R_\hb,S_\hb)^2\ge\cW_2(\cH[R_\hb],\cH[S_\hb])^2-2d\hb\,,
$$
implies that
$$
\varliminf_{\hb\to 0}\fd(R_\hb,S_\hb)\ge\lim_{\hb\to 0}\cW_2(\cH[R_\hb],\cH[S_\hb])=\cW_2(\mu,\nu)\,.
$$
(This part of the argument does not appeal to the restricted triangle inequality, and had been already established in Theorem 2.3 (2) of \cite{FGMouPaul}, albeit in a slightly 
different form). Summarizing
$$
\cW_2(\mu,\nu)\le\varliminf_{\hb\to 0}\fd(R_\hb,S_\hb)\le\varlimsup_{\hb\to 0}\fd(R_\hb,S_\hb)\le\cW_2(\mu,\nu)\,.
$$
\end{proof}

\subsection{Quantum Kantorovich Duality}\lb{SS-QDual}


In the classical setting, Kantorovich duality has several important consequences. The Knott-Smith and Brenier theorems, giving invaluable information on the structure 
of optimal couplings between two (Borel) probability measures on $\bR^n$ are among the most important applications of duality --- at least in the case of the Wasserstein
distance $\cW_2$.

It is therefore natural to seek extensions of Kantorovich duality to the quantum setting. However, the examples studied at the end of Lecture I (in the section showing that
``quantum optimal transport is cheaper'') suggest that the structure of quantum optimal couplings might differ significantly from that of classical optimal couplings.

First we consider the case of $\fd(f,R)$, where $f$ is a probability density on $\bR^d\times\bR^d$ with finite 2nd order moments and $R\in\cD_2(\fH)$.

\medskip
Define the set $\fk$ of test Kantorovich potentials, as follows:
$$
\fk:=\{(a,B)\,:\,a\in C_b(\bR^d\times\bR^d)\text{ and }B=B^*\in\cL(\fH)\text{ s.t. }a(x,\xi)I_\fH+B\le c_\hb(x,\xi)\}\,.
$$
The operator inequality in the definition of $\fk$ means that
$$
a(x,\xi)\|\phi\|^2_\fH+\la\phi|B|\phi\ra\le\la\phi|c_\hb(x,\xi)|\phi\ra\,,\qquad x,\xi\in\bR^d\,,
$$
for all $\phi\in H^1(\bR^d)\cap L^2(\bR^d;|y|^2dy)$.

\smallskip
\noindent
\textbf{Theorem 23.} Let $f$ be a probability density on $\bR^d\times\bR^d$ with finite second order moments and let $R\in\cD_2(\fH)$. Then
$$
\ba
\fd(f,R)^2=&\sup_{(a,B)\in\fk}\left(\int_{\bR^{2d}}a(x,\xi)f(x,\xi)dxd\xi+\Tr_\fH(BR)\right)\,.
\ea
$$

The proof of Theorem 23 is somewhat involved technically, and we shall not repeat it in these notes. We refer instead the interested reader to 
Theorem 4.1 and section 4 of \cite{FGPaulJFA}. 

However, it is a good idea to keep in mind the core argument in that proof, which is based on convex duality, exactly as in the classical setting.

Set $\cE:=C_b(\bR^{2d};\cL(\fH))$ with 
$$
\|T\|_\cE:=\sup_{x,\xi\in\bR^d}\|T(x,\xi)\|_{\cL(\fH)}\,.
$$
Define
$$
G(T):=\left\{\ba{}&0 &&\text{ if }T(x,\xi)=T(x,\xi)^*\ge-c_\hb(x,\xi),\\ &+\infty &&\text{ otherwise, }\ea\right.
$$
and
$$
H(T):=\left\{\ba{}&\int_{\bR^{2d}}\tilde af(x,\xi)dxd\xi+\Tr_\fH(\tilde BR)&&\text{ if }\left\{\ba{}&T(x,\xi)\!=\!T(x,\xi)^*\\&=\tilde a(x,\xi)I_\fH+\tilde B,\ea\right. 
\\ \\ &+\infty &&\text{ otherwise. }\ea\right.
$$
Theorem 23 follows from the Fenchel-Rockafellar duality formula
$$
\inf_{T\in\cE}\left(G(T)+H(T)\right)=\max_{\Lambda\in\cE'}\left(-G^*(-\Lambda)-H^*(\Lambda)\right)\,.
$$
The role of the functional $G$ in this formulation is obviously to penalize the inequality constraint in the definition of $\fk$. (Sometimes, in convex 
analysis, the functional equal to $0$ if a constraint is satisfied, and to $+\infty$ otherwise is referred to as the ``indicator function'' of that constraint. 
This is not to be confused with the classical notion of indicator function of a set, equal to one on the set, and to zero outside.)

\smallskip
Next we consider the case of $\fd(R,S)$ where $R,S\in\cD_2(\fH)$. Here again, there is an analogue of the classical Kantorovich duality theorem.

\medskip
Define the set $\fK$ of test Kantorovich potentials as follows
$$
\fK:=\{(A,B)\,:\,A=A^*\text{ and }B=B^*\in\cL(\fH)\text{ s.t. }A\otimes I_\fH+I_\fH\otimes B\le C_\hb\}\,.
$$
The operator inequality means that for all $\Phi\equiv\Phi(x,y)\in\fH\otimes\fH$ such that
$$
(\grad_x-\grad_y)\Phi\in L^2(\bR^d\times\bR^d)\quad\text{ and }\quad\Phi\in L^2(\bR^d\times\bR^d;|x-y|^2dxdy)\,,
$$
it holds
$$
\la\Phi|A\otimes I_\fH+I_\fH\otimes B|\Phi\ra\le\la\Phi|C_\hb|\Phi\ra\,.
$$
\textbf{Theorem 24.} For all $R,S\in\cD_2(\fH)$, one has
$$
\fd(R,S)^2=\sup_{(A,B)\in\fK}\Tr_\fH(AR+BS)\,.
$$

Here again, the proof of Theorem 24 is too technically involved to be of immediate interest for these lecture notes. We refer the interested reader to
\cite{CagliotiFGPaulSNS} for a detailed exposition of duality in the quantum-to-quantum setting, together with various applications thereof.

\smallskip
Although we have chosen to avoid reproducing the proofs of Theorems 23 and 24, we believe that both results are of key importance in the study
of the pseudometric $\fd$, and in our approach of quantum optimal transport. As a matter of fact, the end of Lecture III will be focussed on two
applications of Theorem 23, which we believe are of some importance.

\subsection{Generalized Triangle Inequalities}\lb{ss-GenTriang}


In this section, we return to the question of the triangle inequality for $\fd$ for $\rho_1,\rho_2,\rho_3\in \fD$, and consider the case where the intermediate 
point $\rho_2$ is a density operator and none of the points $\rho_1,\rho_2,\rho_3$ are rank-$1$ density operators.

In that case, we do not know of any analogue of Lemma 19, and at the time of this writing we do not know how to ``glue'' $Q_{12}\in\cC(\rho_1,\rho_2)$ 
and $Q_{23}\in\cC(\rho_2,\rho_3)$ along $\rho_2$ as in Lemma 7.6 in \cite{VillaniAMS}, to mimic the proof of Theorem 7.3 in \cite{VillaniAMS} in the
classical setting.

At the beginning of this lecture, we have proved ``restricted'' triangle inequalities for $\fd$, i.e. the triangle inequality under addditional assumptions on 
$\rho_1,\rho_2,\rho_3$ --- specifically if one of the points $\rho_1,\rho_2,\rho_3$ is a rank-$1$ density operator, or if the intermediate point $\rho_2$ is
an element of $\cP_2(\bR^d\times\bR^d)$.

In the sequel, we shall prove a ``generalized triangle inequality'', which holds for all $\rho_1,\rho_2,\rho_3\in \fD$, but includes a correction term of order 
$\sqrt{\hb}$ in the right-hand side.

\noindent
\textbf{Theorem 25.} 
For all $\rho_1,\rho_2,\rho_3\in\fD$, one has
$$
\fd(\rho_1,\rho_3)<\fd(\rho_1,\rho_2)+\fd(\rho_2,\rho_3)+\sqrt{d\hb}\,.
$$
In particular
$$
\fd(\rho_1,\rho_3)<\fd(\rho_1,\rho_2)+\fd(\rho_2,\rho_3)+\tfrac1{\sqrt{2}}\fd(\rho_2,\rho_2)\,.
$$

\smallskip
\noindent
\textbf{Remark.} One should compare this result with the De Palma-Trevisan generalized triangle inequality for their distance --- which is reminiscent
of our generalization of the Wasserstein distance $\cW_2$ to $\cD_2(\fH)$ in \cite{FGMouPaul} (the restriction of $\fd$ to $\cD_2(\fH)\times\cD_2(\fH)$).
They arrive at the inequality
$$
d_{dPT}(R,T)\le d_{dPT}(R,S)+d_{dPT}(R,T)+d_{dPT}(S,S)
$$
for all $R,S,T\in\cD_2(\fH)$ for $d_{dPT}$ defined in formula (38) of Definition 8, on p. 3208 in \cite{DePalmaTrev}: see formula (51) in Theorem 2 of 
\cite{DePalmaTrev} on p. 3210. Notice however that $d_{dPT}$ differs from the restriction of $\fd$ to $\cD_2(\fH)\times\cD_2(\fH)$ --- in particular
the definition of couplings in \cite{DePalmaTrev} uses the notion of quantum channel and is very different from the definition used in these lectures, 
which comes from \cite{FGMouPaul}.

\smallskip
\noindent
\textbf{Quiz 29.} Can one slightly modify the definition of $\fd$ so as to obtain a functional $\tilde\fd$ defined on $\fD\times\fD$ that is symmetric and
satisfies the (genuine) triangle inequality? Same question for the DePalma-Trevisan functional $d_{dPT}$. (Of course, the functional $\tilde\fd$ so
obtained satisfies $\tilde\fd>\fd>0$ and therefore is not a bona fide metric either, since $\tilde\fd(\rho,\rho)>0$ for each $\rho\in\fD$. The functional
$\tilde\fd$ obtained in this exercise is therefore of limited interest.)

\smallskip
The proof of Theorem 25 is very different from the proof of the triangle inequality for the original Wasserstein metric $\cW_2$ --- although it uses at
some point the restricted triangle inequality in Theorem 16, whose proof is modelled on the proof of Theorem 7.3 in \cite{VillaniAMS}.

\smallskip
A key step in the proof of Theorem 25 is the following lemma, which can be seen as a consequence of duality for the classical-to-quantum pseudometric
presented in Theorem 23.

\noindent
\textbf{Lemma 26.} For each $R,S\in\cD_2(\fH)$, one has
$$
\fd(R,S)^2\ge\fd(R,\cH[S])^2-d\hb\,.
$$

\smallskip
Taking this lemma for granted, we give a quick proof of Theorem 25.

\noindent
\textbf{Proof of Theorem 25.} Using $\cH[\rho_2]$ as intermediate point, the restricted triangle inequality implies that
$$
\fd(\rho_1,\rho_3)\le\fd(\rho_1,\cH[\rho_2])+\fd(\cH[\rho_2],\rho_3)\,.
$$
Then, Lemma 26 implies that
$$
\ba
\fd(\rho_1,\cH[\rho_2])\le\sqrt{\fd(\rho_1,\rho_2)^2+d\hb}<\fd(\rho_1,\rho_2)+\tfrac12\sqrt{d\hb}\,,
\\
\fd(\cH[\rho_2],\rho_3)\le\sqrt{\fd(\rho_2,\rho_3)^2+d\hb}<\fd(\rho_2,\rho_3)+\tfrac12\sqrt{d\hb}\,.
\ea
$$
The second inequalities above result from the following elementary observation
$$
X>Y>0\implies\sqrt{X^2+Y^2}\le X+\tfrac12Y\,,
$$
whose proof is left to the reader as an (easy) exercise.

With the restricted triangle inequality above (Theorem 16), this implies the first generalized triangle inequality in Theorem 25.

To get the second inequality, observe that
$$
\rho_2\in\cD_2(\fH)\implies\fd(\rho_2,\rho_2)\ge\sqrt{2d\hb}\,.
$$
\rightline{$\Box$}

\smallskip
\noindent
\textbf{Remark.} In fact, we have proved the slightly more precise inequality
$$
\fd(\rho_1,\rho_3)\le\sqrt{\fd(\rho_1,\rho_2)^2+d\hb}+\sqrt{\fd(\rho_2,\rho_3)^2+d\hb}\,.
$$

\begin{proof}[Proof of Lemma 26] For all $a\in C_b(\bR^d\times\bR^d)$ and all $B=B^*\in\cL(\fH)$ satisfying
$$
a(x,\xi)I_\fH+B\le c_\hb(x,\xi)\,,
$$
one applies the Toeplitz map to both sides of the inequality above in the variables $x,\xi$, to find
$$
\ba
\cT[a]\otimes I_\fH+(2\pi\hb)^dI_\fH\otimes B\le&(2\pi\hb)^d\int|q,p\ra\la q,p|c_\hb(q,p)dqdp
\\
\le&(2\pi\hb)^d\left(C_\hb+d\hb I_{\fH\otimes\fH}\right)
\ea
$$
(see the basic properties of the Toeplitz map and Quiz 8 in Lecture I). Thus, for all $T\in\cC(R,S)$, one has
$$
\ba
(2\pi\hb)^d\left(\Tr_{\fH\otimes\fH}(T^\frac12C_\hb T^\frac12)+d\hb\right)
\\
\ge\Tr_{\fH\otimes\fH}\left(T^\frac12(\cT[a]\otimes I_\fH+(2\pi\hb)^dI_\fH\otimes B\right)T^\frac12)
\\
=\Tr_{\fH\otimes\fH}\left(T(\cT[a]\otimes I_\fH+(2\pi\hb)^dI_\fH\otimes B\right))
\\
=\Tr_\fH(R\cT[a])+(2\pi\hb)^d\Tr_\fH(SB)&\,.
\ea
$$

Transforming $\Tr_\fH(R\cT[a])$ into an integral involving the functions $a$ and $\cH[R]$, i.e. (see Lecture I, formula (4) on Husimi transforms)
$$
\Tr_\fH(R\cT[a])=(2\pi\hb)^d\int_{\bR^{2d}}\cH[R](q,p)a(q,p)dqdp\,,
$$
we arrive at the formula
$$
\ba
(2\pi\hb)^d\left(\Tr_{\fH\otimes\fH}(T^\frac12C_\hb T^\frac12)+d\hb\right)
\\
\ge(2\pi\hb)^d\left(\int_{\bR^{2d}}\cH[R](q,p)a(q,p)dqdp+\Tr_{\fH}(SB)\right)\,.
\ea
$$
Maximizing the right-hand side above in $a\in C_b(\bR^d\times\bR^d)$ and $B=B^*\in\cL(\fH)$ s.t.
$$
a(x,\xi)I_\fH+B\le c_\hb(x,\xi)\,,
$$
and applying the duality formula in Theorem 16 shows that
$$
(2\pi\hb)^d\left(\Tr_{\fH\otimes\fH}(T^\frac12C_\hb T^\frac12)+d\hb\right)\ge(2\pi\hb)^d\fd(\cH[R],S)^2
$$
i.e.
$$
\Tr_{\fH\otimes\fH}(T^\frac12C_\hb T^\frac12)\ge \fd(\cH[R],S)^2-d\,.
$$
Minimizing the left-hand side of the inequality above in $T\in\cC(R,S)$ leads to the desired inequality.
\end{proof}

\smallskip
\noindent
\textbf{Quiz 30.} Use Lemma 26 to recover the following result (already known as statement (2) in Theorem 1 of Lecture I)
$$
\fd(R,S)^2\ge\fd(\cH[R],\cH[S])^2-2d\hb\,.
$$
\textbf{Remarks.} 

\noindent
(1) If you include the proof of the duality formula in Theorem 16, this is the longest and most difficult proof of the inequality above...  On the other hand, 
Lemma 26 is a (much) stronger statement  --- it is the key to the generalized triangle inequality. That its proof is more involved than the proof of the 
inequality (2) in Theorem 1 is only natural.

\noindent
(2) Summarizing, in order to prove the triangle inequality for $\fd$ when the intermediate point is not a classical density and none of the density operators
involved are rank-$1$ projections, you 

\noindent
(i) first use the exact (restricted) triangle inequality from Theorem 16
$$
\fd(\rho_1,\rho_3)\le\fd(\rho_1,\cH[\rho_2])+\fd(\cH[\rho_2],\rho_3)\,,
$$
(ii) and then pay the price for replacing $\rho_2$ with its Husimi function 
$$
\fd(\rho_1,\rho_3)\le\sqrt{\fd(\rho_1,\rho_2)^2+d\hb}+\sqrt{\fd(\rho_2,\rho_3)^2+d\hb}
$$ 
which is the result in Lemma 26, based on the Kantorovich duality for the the classical-to-quantum distance (Theorem 23). The end of the proof is 
Kindergarten analysis.

\medskip
The reason for the detour through $\cH[\rho_2]$ instead of $\rho_2$ is due to the fact that question (5) in the following quiz is answered in the 
negative\footnote{While preparing the final version of these lecture notes, I showed the problem to Prof. Denis Serre, who found a counterexample.
At first sight, Serre's counterexample does not seem to suggest that the triangle inequality itself (without the extra $\sqrt{d\hbar}$ term on the 
right-hand side) should be wrong. However, it shows that the procedure of ``glueing'' classical couplings described in Lemma 7.6 of \cite{VillaniAMS}, 
which is key to proving the triangle inequality for $\cW_p$ with $1<p<\infty$, does not have an analogue for general quantum couplings. This
is a rather fundamental difference between classical and quantum optimal transport.}. 

Before working on this exercise, it is a good idea to review the proofs of Theorem 7.3 (the triangle inequality for $\cW_p$) and of Lemma 7.6 (i.e. the
disintegration and the glueing of couplings) in \cite{VillaniAMS}.

\smallskip
\noindent
\textbf{Quiz 31.}
Pick $\rho_1,\rho_2,\rho_3\in\cD_2(\fH)$, all of them or rank $\ge 2$ --- otherwise, there is nothing to prove. Pick $R_{12}$ and $R_{23}$ to be optimal 
couplings of $\rho_1,\rho_2$ and $\rho_2,\rho_3$ (recall briefly why such couplings exist).

\noindent
(1) Assume there exists $T\in\cD(\fH\otimes\fH\otimes\fH)$ such that
$$
\Tr_1(T)=R_{23}\quad\text{ and }\quad\Tr_3(T)=R_{12}\,.
$$
Prove that
$$
\fd(\rho_1,\rho_3)\le\fd(\rho_1,\rho_2)+\fd(\rho_2,\rho_3)\,.
$$
(Hint: observe that $\Tr_2(T)\in\cC(\rho_1,\rho_3)$.)

\medskip
Therefore, proving the triangle inequality boils down to proving the existence of such a $T$. The classical analogue of this is precisely the content of 
Lemma 7.6 in Villani's book \cite{VillaniAMS}.

Let us consider this problem in finite dimension: $\fH=\bC^2$ --- notice that $2$ is the first interesting dimension, because if one of the densities $\rho_j$ 
for $j=1,2,3$  has rank $1$, the triangle inequality is already known (see Theorem 16 on the restricted triangle inequality). 

\noindent
(2) Let $R,R'\in M_2(\bC)$. Find a necessary and sufficient condition on $R,R'$ such that there exists $A,B,C\in M_2(\bC)$ for which the block-wise matrix 
$$
T:=\left(\begin{matrix} A & B\\ B^* & C\end{matrix}\right)\,,\qquad A=A^*\,,\,\, C=C^*\,,
$$
satisfies
$$
\tau'(T):=A+C=R\quad\text{ and }\quad\tau(T):=\left(\begin{matrix} \Tr(A) & \Tr(B)\\ \Tr(B^*) & \Tr(C)\end{matrix}\right)=R'\,.
$$
(3) Assume now that $R,R'\in M_2(M_2(\bC))$. Find a necessary and sufficient condition on $R,R'$ such that there exists $A,B,C\in M_2(M_2(\bC))$ for 
which the block-wise matrix
$$
T:=\left(\begin{matrix} A & B\\ B^* & C\end{matrix}\right)\,,\qquad A=A^*\,,\,\, C=C^*
$$
satisfies
$$
\tau'(T)=A+C=R\quad\text{ and }\quad\left(\begin{matrix} \Tr_{M_2(\bC)}(A) &\Tr_{M_2(\bC)}(B)\\ \Tr_{M_2(\bC)}(B^*) &\Tr_{M_2(\bC)}(C)\end{matrix}\right)=R'.
$$
The notation needs being explained. An element of $B\in M_2(M_2(\bC))$ is of the form
$$
B=\left(\begin{matrix} B_{11} & B_{12}\\ B_{21} & B_{22}\end{matrix}\right)\quad\text{ with }B_{kl}\in M_2(\bC)\,.
$$
Then
$$
B^*:=\left(\begin{matrix} \overline{B^T_{11}} & \overline{B^T_{21}}\\ \overline{B_{12}^T} & \overline{B_{22}^T}\end{matrix}\right)
$$
while
$$
\Tr_{M_2(\bC)}(B):=B_{11}+B_{22}\,.
$$
(4) Explain how (3) is related to the problem of finding $T$ as in (1), in the case where $\rho_1,\rho_2,\rho_3\in\cD(\bC^2)$.

\smallskip
Now, there's the rub...

\noindent
(5) Assuming that $R,R'\in\cD(\bC^2)$, does (the) block-wise matrix (matrices) $T$ obtained in (3) satisfy $T=T^*\ge 0$?

\smallskip
\noindent
\textbf{Remark.} A final observation on the generalized triangle inequality is in order. The presence of the additional term $\sqrt{d\hb}$ 
in the right-hand side of the generalized triangle inequality may be related to the fact that some points in $\fD$ have positive thickness,
in the sense that $\fd(R,R)\ge\sqrt{2d\hb}$ for all $R\in\cD_2(\fH)$. However, the idea of relating this additional term to the ``thickness''
of the intermediate point $\rho_2$ in the triangle inequality could be misleading. In the first place, the genuine triangle inequality holds
if $\rho_2$ is a rank-$1$ density operator, although $\fd(\rho_2,\rho_2)\ge\sqrt{2d\hb}$ in that case.

Besides, the extra term $\sqrt{d\hb}$ on the right-hand side of the generalized triangle inequality seems more related to the method of 
proof of Theorem 25 (viz. the idea of replacing the intermediate point $\rho_2$ with its Husimi transform $\cH[\rho_2]$) than to some
intrinsic feature of $\fd$.

\smallskip
Ultimately, it could be that question (5) in Quiz 31 is answered in the negative, and yet the genuine triangle inequality holds on $\fD$
because there always exists an optimal coupling with a special structure for each pair of finite energy density operators on $\fH$, and 
this special structure acts in favor of the genuine triangle inequality.

This suggests investigating the structure of optimal couplings for the ``pseudometric'' $\fd$ --- however, the example discussed in 
Proposition 5 suggests that this is not an easy task.

\subsection{Towards Quantum Optimal Transport}\lb{SS-QOT}


Until now, we have not said much about quantum optimal transport \textit{per se}, although it is the topic of this school. 

We shall conclude these lectures with some (partial) remarks in that direction. The material in this section mostly comes from \cite{FGPaulJFA}.

We recall that the proofs of the Brenier, or of the Knott-Smith theorems use some form of Kantorovich duality to obtain some additional information
on the structure of optimal couplings. That an optimal coupling for $\cW_2$ is supported in the graph of the subdifferential of some appropriate
function comes from the specific structure of the inequality constraint in the Kantorovich dual formulation of the $\cW_2$ metric, and from the
specific structure of the quadratic transport cost.

It is therefore natural to study the quantum analogue of Kantorovich duality in order to obtain some information on the structure of optimal couplings 
for $\fd$. In the sequel, we shall follow this approach with the classical-to-quantum duality formula in Theorem 23.

Our first task is to give a systematic procedure for constructing elements of the class $\fk$ of test Kantorovich potentials defined before the statement
of Theorem 23.

Set $z:=(x,\xi)\in\bR^d\times\bR^d$ and $Z:=(y,-i\hb\grad_y)$, with the notation
$$
z\cdot Z:=x\cdot y-i\hb\xi\cdot\grad_y\,.
$$
Thus $c_\hb(x,\xi)=|Z|^2+|z|^2I_\fH-2z\cdot Z\ge d\hb I_\fH$ and by Weyl's theorem (see Corollary 2 of Theorem XIII.14 in \cite{RS4})
$$
\tilde B\in\cL(\fH)\implies c_\hb(z)^{-1}\tilde B\in\cK(\fH)\implies\text{ess-spec}(c_\hb(z)-\tilde B)=\varnothing\,.
$$

Assume that $\tilde B=\tilde B^*$ is such that $c_\hb(z)-\tilde B$ has nondegenerate ground state (i.e. with geometric multiplicity $1$) for each $z\in\bR^{2d}$.
(For instance, choose for $\tilde B$ to be a bounded multiplication operator, and apply Theorem XIII.47 in \cite{RS4}). 

Define next
$$
\ba
\tilde a(z):=\min\text{spec}(c_\hb(z)-\tilde B)=&\inf_{\|\phi\|_\fH=1}\la\phi|c_\hb(z)-\tilde B|\phi\ra
\\
&\implies c_\hb(z)-\tilde B\ge\tilde a(z)I_\fH\,.
\ea
$$
Besides, $z\mapsto\tilde a(z)$ is continuous (even real-analytic) by the Kato-Rellich theorem (Theorem XII.8 of \cite{RS4}), and
$$
\ba
\tilde a(z)&\le\la z|c_\hb(z)-\tilde B|z\ra=d\hb-\la z|\tilde B|z\ra\le d\hb+\|\tilde B\|\,,
\\
\tilde a(z)&\ge d\hb+\inf_{\|\phi\|_\fH=1}\la\phi|-\tilde B|\phi\ra\ge d\hb-\|\tilde B\|\,.
\ea
$$
Hence $\tilde a\in C_b(\bR^d\times\bR^d)$, and we have obtained in this way
$$
(\tilde a,\tilde B)\in\fk\,.
$$
The Kato-Rellich theorem recalled above also implies the existence of a continuous (even real-analytic) map 
$$
\bR^d\times\bR^d\ni z\mapsto\psi_z\in\fH\quad\text{ s.t. }\left\{\ba{}&(c_\hb(z)-\tilde B)\psi_z=\tilde a(z)\psi_z\,,\\ &\text{and }\|\psi_z\|_\fH=1,\,\,\, z\in\bR^{2d}\,.\ea\right.
$$

With this, we can define a notion of quantum optimal transport from $\cP_2(\bR^d\times\bR^d)$ to $\cD_2(\fH)$. 

\noindent
\textbf{Theorem 27.}
Let $\tilde B=\tilde B^*$ be such that $c_\hb(z)-\tilde B$ has nondegenerate ground state, set $\tilde a(z):=\min\text{spec}(c_\hb(z)-\tilde B)$, and let $z\mapsto\psi_z$
be a continuous map from $\bR^d\times\bR^d$ to $\fH$ such that $\|\psi_z\|_\fH=1$ and $\psi_z\in\Ker(c_\hb(z)-\tilde B-\tilde a(z)I_\fH)$.

Then, for each probability density $f$ with finite 2nd order moments, the map $z\mapsto f(z)|\psi_z\ra\la\psi_z|$ is an optimal coupling for 
the pseudometric $\fd$ between $f$ and the operator
$$
\cT^{\tilde B}[f]:=\int_{\bR^{2d}}f(z)|\psi_z\ra\la\psi_z|dz\in\cD_2(\fH)\,.
$$

\smallskip
\noindent
\textbf{Example.} Take for example $\tilde B=0$; then, one easily checks that 
$$
\tilde a(z)=d\hb\,,\quad\Ker(c_\hb(z)-d\hb I_\fH)=\bC|z\ra\,,
$$
where $|z\ra$ is the Schr\"odinger coherent state centered at $z:=q+ip$, so that 
$$
\cT^0[f]=\cT[f]
$$
is the Toeplitz operator of symbol $f$. 

We already knew from Theorem 1 (1) in Lecture I that 
$$
\fd(f,\cT[f])=\sqrt{d\hb}=\inf_{\cP_2(\bR^{2d})\times\cD_2(\fH)}\fd\,.
$$

\smallskip
\noindent
\textit{Proof of Theorem 27.}
Set $Q(z):=f(z)|\psi_z\ra\la\psi_z|$, so that $Q(z)^\frac12=\sqrt{f(z)}|\psi_z\ra\la\psi_z|$, and
$$
\ba
\int_{\bR^{2d}}\Tr_\fH\left(Q(z)^\frac12 c_\hb(z)Q(z)^\frac12\right)dz=\int_{\bR^{2d}}\la\psi_z|c_\hb(z)|\psi_z\ra f(z)dz
\\
=\int_{\bR^{2d}}(\tilde a(z)\la\psi_z|\psi_z\ra+\la\psi_z|B|\psi_z\ra) f(z)dz
\\
=\int_{\bR^{2d}}\tilde a(z)f(z)dz+\Tr_\fH\left(\tilde B\cT^{\tilde B}[f]\right)&\,.
\ea
$$
Since $(\tilde a,\tilde B)\in\fk$ and $Q\in\cC(f,\cT^{\tilde B}[f])$, this implies that
$$
\ba
\int_{\bR^{2d}}\Tr_\fH\left(Q(z)^\frac12 c_\hb(z)Q(z)^\frac12\right)dz
\\
=\min_{T\in\cC(f,\cT^{\tilde B}[f])}\int_{\bR^{2d}}\Tr_\fH\left(T(z)^\frac12 c_\hb(z)T(z)^\frac12\right)dz&=\fd\left(f,\cT^{\tilde B}[f]\right)^2&\,.
\ea
$$
This also implies that
$$
\ba
\int_{\bR^{2d}}\tilde a(z)f(z)dz+\Tr_\fH\left(\tilde B\cT^{\tilde B}[f]\right)
\\
=\sup_{(a,B)\in\fk}\int_{\bR^{2d}}a(z)f(z)dz+\Tr_\fH\left(B\cT^{\tilde B}[f]\right)&\,.
\ea
$$
Therefore, in this case, the sup is attained in $\fk$ (this is not true in general).
\rightline{$\Box$}

\smallskip
\noindent
\textbf{Remarks.} 

\noindent
(1) Thus the optimal transport map for $\fd$ between $\cP_2(\bR^d\times\bR^d)$ and $\cD_2(\fH)$ can be thought of as a \textit{deformation} 
of the Toeplitz quantization, at least when $\tilde B$ is such that $c_\hb(z)-\tilde B$ has a ground state of geometric multiplicity $1$.

\noindent
(2) Notice that the starting point in Theorem 27 is the pair $(\tilde a,\tilde B)\in\fk$, and not the pair consisting of $f$ (the probability density) and
the density operator which are the arguments of $\fd$. This approach is vaguely reminiscent of the notion of geodesic in Riemannian geometry:
the original definition of a geodesic curve on a Riemannian manifold is that of the shortest path between two points on the manifold. The calculus
of variations shows that geodesics define local solutions of an ODE system set on the tangent, or cotangent bundle of the manifold. Conversely,
solutions of this ODE system define \textit{local} geodesic curves, which may differ from prescribing arbitrary end points and finding a shortest 
path between these points. There is a very loose analogy between the local theory of geodesics through the ODE system, and the definition of
optimal couplings starting from elements of $\fk$ as explained in Theorem 27. In other words, the transport map $\cT^{\tilde B}$ is independent
of the choice of the endpoints --- or more precisely the endpoints $f$ and $\cT^{\tilde B}[f]$ follow from the optimal transport map, instead of
the other way around.

\smallskip
This raises the following question: in Brenier's theorem, the (classical) optimal transport map is the gradient of a convex function. Is there some analogous 
property in the quantum setting?

\subsubsection*{Operator Legendre Duality and Quantum Optimal Transport}


If $(\tilde a,\tilde B)\in\fk$, one has
$$
\underbrace{|Z|^2+|z|^2I_\fH-2z\cdot Z}_{=c_\hb(z)}\ge\tilde a(z)I_\fH+\tilde B\iff a(z)+B\ge z\cdot Z\,,
$$
with
$$
a(z):=\tfrac12(|z|^2-\tilde a(z))\,,\qquad B=\tfrac12(|Z|^2-\tilde B)\,.
$$
Besides, one has 
$$
\Dom(c_\hb(z))\!=\!\Dom(|Z|^2)\!=H^2(\bR^d)\cap L^2(\bR^d,|y|^4dy)\!=:\!D\,.
$$

After these preliminaries, we define a notion of Legendre transform of an (unbounded) operator on $\fH$.

\noindent
\textbf{Definition.} Let $B$ satisfy $|Z|^2-2B\in\cL(\fH)$. The Legendre dual of $B$ is the convex function (upper envelope of affine functions)
$$
B^L(z):=\sup_{\phi\in D,\,\|\phi\|_\fH=1}\left(z\cdot\la\phi|Z|\phi\ra-\la\phi|B|\phi\ra\right)\,.
$$

Indeed, we recall that, if $T$ is an operator on $\fH$, one can think of $\psi\mapsto\la\psi|T|\psi\ra$ as the noncommutative analogue of 
the evaluation at $x\in\bR^n$ of a real-valued function $f$ defined on $\bR^n$. Therefore, the definition above is analogous to the usual 
definition of the Legendre(-Fenchel) transform
$$
\phi^*(\xi)=\sup_{x\in E}(\la\xi,x\ra_{E',E}-\phi(x))\,,\qquad\xi\in E'\,,
$$
for all $\phi:\,E\to(-\infty,+\infty]$ where $E$ is a normed linear space on $\bR$, and $\phi$ is not identically equal to $+\infty$, and $E'$ is
the topological dual of $E$, i.e. the space of linear functionals on $E$ that are continuous for the norm topology.

\noindent
\textbf{Theorem 28.} 
Let $\tilde B=\tilde B^*$ be such that $c_\hb(z)-\tilde B$ has nondegenerate ground state, set $\tilde a(z):=\min\text{spec}(c_\hb(z)-\tilde B)$, and let $z\mapsto\psi_z$
be a continuous map from $\bR^d\times\bR^d$ to $\fH$ such that $\|\psi_z\|_\fH=1$ and $\psi_z\in\Ker(c_\hb(z)-\tilde B-\tilde a(z)I_\fH)$.

\noindent
(1) Setting 
$$
a(z):=\tfrac12(|z|^2-\tilde a(z))\,,\qquad B:=\tfrac12(|Z|^2-\tilde B)\,,
$$
one has
$$
a=B^L\,.
$$
(2) Besides
$$
\grad a(z)=z-\grad\tilde a(z)=\la\psi_z|Z|\psi_z\ra\,.
$$

\smallskip
\noindent
\textit{Proof.} Statement (1) follows from the definition and the variational formula for the ground state. As for (2), differentiate in $z$ the identity
$$
B\psi_z-z\cdot Z\psi_z+a(z)\psi_z=0\,,
$$
and take the inner product with $\psi_z$ to get
$$
\underbrace{\la\psi_z|B-z\cdot Z+a(z)|\dot\psi_z\ra}_{=0\text{ since }B=B^*,\,Z=Z^*,\,a(z)\in\bR}+\la\psi_z|-Z+\grad a(z)|\psi_z\ra=0\,.
$$
\rightline{$\Box$}

\smallskip
\noindent
\textbf{Remarks.}

\noindent
(1) In the Knott-Smith theorem recalled in Lecture I, optimal couplings for $\cW_2$ are supported in the graph of the subdifferential of a l.s.c. convex function, 
while, in the Brenier theorem, the optimal transport map is the gradient of a convex function. In both results, the function is obtained from an optimal Kantorovich 
potential by the same transformation as $\tilde a\mapsto a$. Theorem 28 (2) is a partial analogue of this crucial piece of information, except that, in the quantum 
setting, density operators are not ``functions of $Z$''.

\noindent
(2) In classical optimal transport, there exist an optimal pair $(a,b)$ of Kantorovich potentials; they are l.s.c. proper convex functions and are Legendre duals of 
each other, so that $\nabla a\circ\nabla b=\text{Id}$; besides $a\in L^1_\mu$ and $b\in L^1_\nu$. If one tries to proceed by analogy, in the present case, one 
should define some notion of ``quantum gradient'' of the operator $B$.

One idea to do so is to use the phase space symplectic structure. For a smooth function $\a\equiv\a(x,\xi)$ on $\bR^d\times\bR^d$, one has
$$
\d_{x_j}\alpha=\{\xi_j,\alpha\}\,,\quad\d_{\xi_j}\alpha=-\{x_j,\alpha\}\,,\qquad j=1,\ldots,d\,.
$$
This suggests to define ``quantum derivatives'' as follows,
$$
\d^Q_{y_j}B:=\tfrac{i}\hb[-i\hb\d_{y_j},B]\,,\quad\d^Q_{\eta_j}B:=-\tfrac{i}\hb[y_j,B],,\qquad j=1,\ldots,d\,,
$$
by using the correspondence principle and the analogy between commutator and Poisson bracket
$$
\tfrac{i}{\hb}[\cdot,\cdot]\to\{\cdot,\cdot\}
$$
recalled in Lecture II. Since
$$
B\psi_z=z\cdot Z\psi_z+a(z)\psi_z\,,\quad B=B^*\,,\quad Z=Z^*\text{ and }a(z)\in\bR\,,
$$
one easily checks that
$$
\left\{\ba{}&x_j=\la\psi_z|\d^Q_{y_j}B|\psi_z\ra\\ &\xi_j=\la\psi_z|\d^Q_{\eta_j}B|\psi_z\ra\ea\right.
$$
This formula can be viewed as the inverse transform of Theorem 28 (2).

\noindent
(3) Analogous ideas on a definition of an optimal transport ``map'' between elements of $\cD_2(\fH)$ can be found in \cite{CagliotiFGPaulSNS}.
Partial results analogous to Theorem 28 have been obtained there, but much remains to be done.

\smallskip
Following the proof of Theorems 27--28 suggests viewing the operator 
$$
-\tfrac12(|x|^2-\hb^2\Dlt_x-A)
$$ 
as the ``smallest eigenvalue'' of the operator 
$$
\tfrac12(|y|^2-\hb^2\Dlt_y-B)-x\cdot y+\hb^2\grad_x\cdot\grad_y\,,
$$
viewed as a ``matrix'' whose entries are operators in the $x$-variables. However, inequalities between operators do not define a total order relation,
so that even the notion of ground state in this setting does not seem to make much sense.

New ideas on this problem are obviously needed.


\end{document}